\PassOptionsToPackage{unicode}{hyperref}
\PassOptionsToPackage{hyphens}{url}
\PassOptionsToPackage{dvipsnames,svgnames,x11names}{xcolor}
\documentclass[
]{article}

\usepackage{amsmath,amssymb}
\usepackage{iftex}
\ifPDFTeX
  \usepackage[T1]{fontenc}
  \usepackage[utf8]{inputenc}
  \usepackage{textcomp} 
\else 
  \usepackage{unicode-math}
  \defaultfontfeatures{Scale=MatchLowercase}
  \defaultfontfeatures[\rmfamily]{Ligatures=TeX,Scale=1}
\fi
\usepackage{lmodern}
\ifPDFTeX\else  
\fi
\IfFileExists{upquote.sty}{\usepackage{upquote}}{}
\IfFileExists{microtype.sty}{
  \usepackage[]{microtype}
  \UseMicrotypeSet[protrusion]{basicmath} 
}{}
\makeatletter
\@ifundefined{KOMAClassName}{
  \IfFileExists{parskip.sty}{%
    \usepackage{parskip}
  }{
    \setlength{\parindent}{0pt}
    \setlength{\parskip}{6pt plus 2pt minus 1pt}}
}{
  \KOMAoptions{parskip=half}}
\makeatother
\usepackage{xcolor}
\makeatletter
\ifx\paragraph\undefined\else
  \let\oldparagraph\paragraph
  \renewcommand{\paragraph}{
    \@ifstar
      \xxxParagraphStar
      \xxxParagraphNoStar
  }
  \newcommand{\xxxParagraphStar}[1]{\oldparagraph*{#1}\mbox{}}
  \newcommand{\xxxParagraphNoStar}[1]{\oldparagraph{#1}\mbox{}}
\fi
\ifx\subparagraph\undefined\else
  \let\oldsubparagraph\subparagraph
  \renewcommand{\subparagraph}{
    \@ifstar
      \xxxSubParagraphStar
      \xxxSubParagraphNoStar
  }
  \newcommand{\xxxSubParagraphStar}[1]{\oldsubparagraph*{#1}\mbox{}}
  \newcommand{\xxxSubParagraphNoStar}[1]{\oldsubparagraph{#1}\mbox{}}
\fi
\makeatother

\usepackage{longtable,booktabs,array}
\usepackage{calc} 
\usepackage{etoolbox}
\makeatletter
\patchcmd\longtable{\par}{\if@noskipsec\mbox{}\fi\par}{}{}
\makeatother
\IfFileExists{footnotehyper.sty}{\usepackage{footnotehyper}}{\usepackage{footnote}}
\makesavenoteenv{longtable}
\usepackage{graphicx}
\makeatletter
\def\maxwidth{\ifdim\Gin@nat@width>\linewidth\linewidth\else\Gin@nat@width\fi}
\def\maxheight{\ifdim\Gin@nat@height>\textheight\textheight\else\Gin@nat@height\fi}
\makeatother
\setkeys{Gin}{width=\maxwidth,height=\maxheight,keepaspectratio}
\makeatletter
\def\fps@figure{htbp}
\makeatother
\NewDocumentCommand\citeproctext{}{}

\makeatletter
 \let\@cite@ofmt\@firstofone
 \def\@biblabel#1{}
 \def\@cite#1#2{{#1\if@tempswa , #2\fi}}
\makeatother
\newlength{\cslhangindent}
\newlength{\csllabelwidth}
\newenvironment{CSLReferences}[2] 
 {\begin{list}{}{%
  \setlength{\itemindent}{0pt}
  \setlength{\leftmargin}{0pt}
  \setlength{\parsep}{0pt}
  \ifodd #1
   \setlength{\leftmargin}{\cslhangindent}
   \setlength{\itemindent}{-1\cslhangindent}
  \fi
  \setlength{\itemsep}{#2\baselineskip}}}
 {\end{list}}
\usepackage{calc}

\usepackage{arxiv}
\usepackage{orcidlink}
\usepackage{amsmath}
\usepackage[T1]{fontenc}
\usepackage{gensymb}
\makeatletter
\@ifpackageloaded{caption}{}{\usepackage{caption}}
\AtBeginDocument{%
\ifdefined\contentsname
  \renewcommand*\contentsname{Table of contents}
\else
  \newcommand\contentsname{Table of contents}
\fi
\ifdefined\listfigurename
  \renewcommand*\listfigurename{List of Figures}
\else
  \newcommand\listfigurename{List of Figures}
\fi
\ifdefined\listtablename
  \renewcommand*\listtablename{List of Tables}
\else
  \newcommand\listtablename{List of Tables}
\fi
\ifdefined\figurename
  \renewcommand*\figurename{Figure}
\else
  \newcommand\figurename{Figure}
\fi
\ifdefined\tablename
  \renewcommand*\tablename{Table}
\else
  \newcommand\tablename{Table}
\fi
}
\@ifpackageloaded{float}{}{\usepackage{float}}
\floatstyle{ruled}
\@ifundefined{c@chapter}{\newfloat{codelisting}{h}{lop}}{\newfloat{codelisting}{h}{lop}[chapter]}
\floatname{codelisting}{Listing}

\makeatother
\makeatletter
\@ifpackageloaded{caption}{}{\usepackage{caption}}
\@ifpackageloaded{subcaption}{}{\usepackage{subcaption}}
\makeatother
\IfFileExists{pdfcomment.sty}
{
  \usepackage{pdfcomment}

}{

}
\makeatletter
\@ifpackageloaded{algorithm}{}{\usepackage{algorithm}}
\makeatother
\makeatletter
\@ifpackageloaded{algpseudocode}{}{\usepackage{algpseudocode}}
\makeatother
\makeatletter
\@ifpackageloaded{caption}{}{\usepackage{caption}}
\makeatother

\ifLuaTeX
  \usepackage{selnolig}  
\fi
\usepackage{bookmark}

\IfFileExists{xurl.sty}{\usepackage{xurl}}{} 
\hypersetup{
  pdftitle={An uncertainty-aware Digital Shadow for underground multimodal CO2 storage monitoring},
  pdfauthor={Abhinav Prakash Gahlot; Rafael Orozco; Ziyi Yin; Felix J. Herrmann},
  colorlinks=true,
  linkcolor={blue},
  filecolor={Maroon},
  citecolor={Blue},
  urlcolor={Blue},
  pdfcreator={LaTeX via pandoc}}

\renewcommand{\today}{2024-10-01}
\newcommand{\runninghead}{A Preprint }
\renewcommand{\runninghead}{A Preprint }
\title{An uncertainty-aware Digital Shadow for underground multimodal
CO\textsubscript{2} storage monitoring}
\def\asep{\\\\\\ } 
\author{\textbf{Abhinav Prakash
Gahlot}~\orcidlink{0009-0008-7580-1220}\\\\Georgia Institute of
Technology\\\\\href{mailto:agahlot8@gatech.edu}{agahlot8@gatech.edu}\asep\textbf{Rafael
Orozco}~\orcidlink{0000-0003-0917-2442}\\\\Georgia Institute of
Technology\\\\\href{mailto:rorozco@gatech.edu}{rorozco@gatech.edu}\asep\textbf{Ziyi
Yin}~\orcidlink{0000-0002-5024-8771}\\\\Georgia Institute of
Technology\\\\\href{mailto:ziyi.yin@gatech.edu}{ziyi.yin@gatech.edu}\asep\textbf{Felix
J. Herrmann}~\orcidlink{0000-0003-1180-2167}\\\\Georgia Institute of
Technology\\\\\href{mailto:felix.herrmann@gatech.edu}{felix.herrmann@gatech.edu}}
\date{2024-10-01}
\begin{document}
\maketitle
\begin{abstract}
As a society, we are faced with important challenges to combat climate
change. Geological Carbon Storage (GCS), during which gigatonnes of
super-critical CO\textsubscript{2} are stored underground, is arguably
the only scalable net-negative CO\textsubscript{2}-emission technology
that is available. While promising, subsurface complexities and
heterogeneity of reservoir properties demand a systematic approach to
quantify uncertainty when optimizing production and mitigating storage
risks, which include assurances of Containment and Conformance of
injected supercritical CO\textsubscript{2}. As a first step towards the
design and implementation of a Digital Twin for monitoring and control
of underground storage operations, a machine-learning-based
data-assimilation framework is introduced and validated on carefully
designed realistic numerical simulations. Because our implementation is
based on Bayesian inference, but does not yet support control and
decision-making, we coin our approach an uncertainty-aware Digital
Shadow. To characterize the posterior distribution for the state of
CO\textsubscript{2} plumes (its CO\textsubscript{2} concentration and
pressure), conditioned on multi-modal time-lapse data, the envisioned
Shadow combines techniques from Simulation-Based Inference (SBI) and
Ensemble Bayesian Filtering to establish probabilistic baselines and
assimilate multi-modal data for GCS problems that are challenged by
large degrees of freedom, nonlinear multiphysics, non-Gaussianity, and
computationally expensive to evaluate fluid-flow and seismic
simulations. To enable SBI for dynamic systems, a recursive scheme is
proposed where the Digital Shadow's neural networks are trained on
simulated ensembles for their state and observed data (well and/or
seismic). Once training is completed, the system's state is inferred
when time-lapse field data becomes available. Contrary to ensemble
Kalman filtering, corrections to the predicted simulated states are not
based on linear updates, but instead follow during the Analysis step of
Bayesian filtering from a prior-to-posterior mapping through the latent
space of a nonlinear transform. Starting from a probabilistic model for
the permeability field, derived from a baseline surface-seismic survey,
the proposed Digital Shadow is validated on unseen simulated
ground-truth time-lapse data. In this computational study, we observe
that a lack of knowledge on the permeability field can be factored into
the Digital Shadow's uncertainty quantification. Our results also
indicate that the highest reconstruction quality is achieved when the
state of the CO\textsubscript{2} plume is conditioned on both time-lapse
seismic data and wellbore measurements. Despite the incomplete knowledge
of the permeability field, the proposed Digital Shadow was able to
accurately track the unseen physical state of the subsurface throughout
the duration of a realistic CO\textsubscript{2} injection project. To
the best of our knowledge, this work represents the first
proof-of-concept of an uncertainty-aware, in-principle scalable, Digital
Shadow that captures the uncertainty arising from unknown reservoir
properties and noisy observations. This framework provides a foundation
for the development of a Digital Twin aimed at mitigating risks and
optimizing the management of underground storage projects.
\end{abstract}

\floatname{algorithm}{Algorithm}

\renewcommand*\contentsname{Table of contents}
{
\hypersetup{linkcolor=}
\setcounter{tocdepth}{3}
\tableofcontents
}

\newpage

\newcommand{\argmin}{\mathop{\mathrm{argmin}\,}\limits}
\newcommand{\argmax}{\mathop{\mathrm{argmax}\,}\limits}

\[
\def\textsc#1{\dosc#1\csod} 
\def\dosc#1#2\csod{{\rm #1{\small #2}}} 
\] \newcommand{\To}{\textbf{to}~}
\newcommand{\Output}{\textbf{Output:}~}
\renewcommand{\Return}{\State \textbf{return}~}
\renewcommand{\Require}{\textbf{Require:}~}

\section{Introduction}\label{introduction}

\subsection{The need for Geological Carbon
Storage}\label{the-need-for-geological-carbon-storage}

Geological carbon storage (GCS) constitutes capturing carbon dioxide
(CO\textsubscript{2}) emissions from industrial processes followed by
storage in deep geological formations, such as depleted oil and gas
reservoirs, deep saline aquifers, or coal seams (Page et al. 2020).
Because of its ability to scale (Ringrose 2020, 2023), this net-negative
CO\textsubscript{2}-emission technology is widely considered as one of
the most promising technologies to help mitigate the effects of
anthropogenic CO\textsubscript{2} greenhouse gas emissions, which are
major contributors to global warming (Bui et al. 2018; IPCC special
report 2005). According to the Intergovernmental Panel on Climate Change
(IPCC, Panel on Climate Change) (2018)), it is imperative to contain
atmospheric warming to \(<2\degree\)C by the year 2100 to avoid serious
consequences of global warming. To achieve this goal, scalable
net-negative emission technologies that can remove tens of gigatonnes of
CO\textsubscript{2} per year (IPCC special report 2018) are needed in
addition to reducing emissions. These needs call for a transition
towards increased use of renewable energy in combination with the global
deployment of GCS technology to store significant amounts of
supercritical CO\textsubscript{2} into geological formations to mitigate
the effects of climate change (Metz et al. 2005; Orr Jr 2009). When
supercritical CO\textsubscript{2}---captured at anthropogenic sources
(fossil-fuel power plants, steel, cement, and chemical industries), is
stored securely and permanently in underground geological storage
reservoirs, it can be scaled to mitigate upto \(20\%\) of the total
CO\textsubscript{2} emission problem (Ketzer, Iglesias, and Einloft
2012). This work is a first step towards an uncertainty-aware Digital
Twin aimed at optimizing underground operations while mitigating risks.

\subsection{\texorpdfstring{The challenge of monitoring subsurfcae
CO\textsubscript{2} injection
projects}{The challenge of monitoring subsurfcae CO2 injection projects}}\label{the-challenge-of-monitoring-subsurfcae-co2-injection-projects}

Our main technical objective is directed towards monitoring
CO\textsubscript{2}-storage projects, so that assurances can be made
whether storage proceeds as expected and whether CO\textsubscript{2}
plumes remain within the storage complex. These assurances are referred
to as Conformance and Containment by the regulators and operators of GCS
(Ringrose 2020, 2023). Even though the proposed methodology could be
applied to other settings, our focus is on CO\textsubscript{2} storage
in offshore saline aquifers in a marine setup as illustrated in
figure~\ref{fig-acquisition}. Compared to storage projects on land, this
choice for marine settings circumvents potential complications and risks
associated with future liability and geomechanically-induced pressure
changes that may trigger earthquakes (Zoback and Gorelick 2012; He et
al. 2011; Raza et al. 2016). While this approach simplifies the
complexity of underground CO\textsubscript{2} storage projects to some
degree, simulations for the state of CO\textsubscript{2} plumes are on
balance more challenging than conventional reservoir simulations because
the fluid-flow properties are less well-known. Besides, the process of
displacing brine with supercritical CO\textsubscript{2} is intricate,
taxing multi-phase flow simulators (Ringrose 2020, 2023). To illustrate
the possible impact of lack of knowledge on the reservoir properties
(the permeability), figure~\ref{fig-acquisition} also includes a
juxtaposition between multi-phase flow simulations for two different
realizations of permeability fields drawn from a stochastic baseline
established from active-source surface seismic data (Yin, Orozco, et al.
2024). While both realizations for the permeability are consistent with
this baseline, they yield vastly different outcomes where in one case
the CO\textsubscript{2} plume remains well within
(figure~\ref{fig-acquisition} bottom-left) the storage complex (denoted
by the white dashed line in figure~\ref{fig-acquisition}) whereas the
other simulation obviously breaches Containment
(figure~\ref{fig-acquisition} bottom-right), due to the presence of an
``unknown'' high-permeability streak. By maximally leveraging
information derived from multimodal time-lapse monitoring data, we plan
to guide uncertain reservoir simulations, so outlying ``unpredicted''
geological events, such as the one caused by the high-permeability
streak, can be accounted for.

\subsection{Proposed approach with Digital
Shadows}\label{proposed-approach-with-digital-shadows}

As stated eloquently by Ringrose (2020); Ringrose (2023),
CO\textsubscript{2}-storage projects can not rely on static reservoir
models alone (Ringrose and Bentley 2016). Instead, reservoir models and
simulations need to be updated continuously from multimodal sources
(e.g.~active-source surface seismic and/or pressure/saturation
measurements at wells) of time-lapse data. By bringing this insight
together with the latest developments in generative Artificial AI, which
includes the sub-discipline of Simulation-Based Inference (SBI, Cranmer,
Brehmer, and Louppe (2020)), where conditional neural networks are
trained to act as posterior density estimators, and nonlinear ensemble
filtering (Spantini, Baptista, and Marzouk 2022), a Digital Shadow (Asch
2022) is coined as a framework for monitoring underground
CO\textsubscript{2}-storage projects. Digital Shadows differ from
Digital Twins (Asch 2022) in the sense that they integrate information
on the state of a physical object or system --- e.g.~the temporal
behavior of CO\textsubscript{2}-injection induced changes in
pressure/saturation for GCS --- by capturing information
unidirectionally from real-time, possibly multimodal, observed
time-lapse data without feedback loops. While our framework is designed
to allow for future optimization, control, and decision-making to
mitigate risks of GCS that would turn our Shadow into a Twin, our main
focus in this work is to introduce a scalable machine-learning assisted
data-assimilation framework (Asch 2022; Spantini, Baptista, and Marzouk
2022). To be relevant to realistic CO\textsubscript{2}-storage projects,
use is made of scalable state-of-the-art reservoir simulation,
wave-equation based seismic inversion (Virieux and Operto 2009; Baysal,
Kosloff, and Sherwood 1983), ensemble Bayesian filtering (Asch 2022),
and the advanced machine-learning technique of neural posterior density
estimation (Cranmer, Brehmer, and Louppe 2020; Radev et al. 2020), to
capture uncertainties due to our lack of knowledge on the reservoir
properties, the permeability field, in a principled way. Because the
envisioned Digital Shadow is aimed at monitoring
CO\textsubscript{2}-injection projects over many decades, its
performance is validated \emph{in silico}---i.e.~via a well-designed
computer-simulation study. As with the development of climate models,
this approach is taken by necessity because suitable time-lapse datasets
spanning over long periods of time are not yet available.

\subsection{Relation to exisiting
work}\label{relation-to-exisiting-work}

The central approach taken in this work finds its basis in ensemble
Kalman Filtering (EnKF, Evensen (1994); Asch, Bocquet, and Nodet (2016);
Carrassi et al. (2018)), a widely employed and scalable Bayesian
data-assimilation (DA) technique. Here, DA refers to a statistical
approach where indirect and/or sparse real-world time-lapse measurements
are integrated into a model that describes the dynamics of the hidden
state of CO\textsubscript{2} plumes. While other time-lapse
data-collection modalities exist, including controlled-source
electromagnetics (Ayani, Grana, and Liu 2020; Grana, Liu, and Ayani
2021), gravity (Appriou and Bonneville 2022), history matching from data
collected at wells (G. Zhang, Lu, and Zhu 2014; Strandli, Mehnert, and
Benson 2014), and even tomography with muons (Gluyas et al. 2019), we
employ active-source time-lapse surface-seismic surveys (D. E. Lumley
2001) in combination with data collected at wells. Seismic surveys are
known for their relatively high attainable resolution, sensitivity to
small CO\textsubscript{2}-induced changes in seismic rock properties,
and spatial coverage in both the vertical and horizontal coordinate
directions (Ringrose 2020, 2023).

By including crucial multi-phase fluid-flow simulations (Nordbotten and
Celia 2011) for the state, and by exchanging the extended Kalman filter
(Asch, Bocquet, and Nodet 2016; Asch 2022) for better scalable EnKF
(Bruer et al. 2024) type of approach, our work extends early seismic
monitoring work by C. Huang and Zhu (2020), where full-waveform
inversion (FWI, Virieux and Operto (2009)) is combined with DA. Our
choice for a scalable ensemble filtering type of approach (Herrmann
2023; Gahlot et al. 2023; Spantini, Baptista, and Marzouk 2022) is
supported by recent work of Bruer et al. (2024), who conducted a careful
study on the computational complexity of realistic CO\textsubscript{2}
projects. This work is aimed at further improving several aspects of
ensemble-filter approaches including \emph{(i)} incorporation of
stochasticity in the state dynamics by treating the permeability
distribution as a random variable newly drawn after each time-lapse
timestep; \emph{(ii)} removal of reliance on limiting Gaussian
assumptions and linearizations of the observation operator; \emph{(iii)}
replacement of the crucial linear Kalman correction step by a learned
nonlinear mapping similar to the coupling approach recently proposed by
Spantini, Baptista, and Marzouk (2022); and \emph{(iv)} extending the
approach to multimodal time-lapse data.

To meet these additional complexities and evident nonlinearities, we
develop a new neural sequential Bayesian inference framework based on
Variational Inference (Jordan et al. 1998) with Conditional Normalizing
Flows (CNFs, Rezende and Mohamed (2015); Papamakarios et al. (2021);
Radev et al. (2020)). As in the work by Spantini, Baptista, and Marzouk
(2022), nonlinearities in both the state and observation operators are
accounted for by a prior-to-posterior mapping through the latent space
of a nonlinear transform, given by a CNF in this work. However, by
allowing the permeability field, which emerges in the state for the
fluid flow, to appear as a stochastic nuisance parameter, our proposed
formulation sets itself apart from existing work. By using the fact that
CNFs implicitly marginalize over nuisance parameters during training
(Cranmer, Brehmer, and Louppe 2020; Radev et al. 2020), stochasticity
due to the lack of knowledge on the permeability field is factored into
the uncertainty captured by our neural approximation for the posterior
distribution of the CO\textsubscript{2} plume's state.

Finally, the presented approach provides an alternative to ``pure''
machine-learning approaches, where neural networks are either trained on
pairs saturation/seismic shot records (Um et al. 2023), on reservoir
models and saturation plumes (Tang et al. 2021), or on ``time-lapse
movies'' (X. Huang, Wang, and Alkhalifah 2024), to produce estimates for
the dynamic state with uncertainties based on Mont-Carlo dropout. Aside
from the argument that this type of uncertainty quantification may not
be Bayesian (Folgoc et al. 2021), our proposed approach is solidly
anchored by the fields of DA, where high-fidelity simulations for the
state and observations are used to recursively create forecast
ensembles, and by SBI (Cranmer, Brehmer, and Louppe 2020; Radev et al.
2020), where advantage is taken of the important concept of
\emph{summary statistics} (Orozco et al. 2023; Orozco, Siahkoohi, et al.
2024) to simplify the complex relationship between the state
(saturation/pressure) and seismic observations (shot records). Instead
of relying on the network to provide the uncertainty, our deployment of
CNFs aims to learn the Bayesian posterior distribution directly that
includes both aleatoric, by adding additive noise to the observed
time-lapse data, and epistemic uncertainty, by treating the permeability
field as a stochastic nuisance parameter. While our approach can be
extended to include network-related epistemic uncertainty (Qu,
Araya-Polo, and Demanet 2024), we leave this to future work to keep our
approach relatively simple.

\subsection{Outline}\label{outline}

The Digital Shadow for underground CO\textsubscript{2}-storage
monitoring is aimed to demonstrate how learned nonlinear Bayesian
ensemble filtering techniques can be used to characterize the state from
multimodal time-lapse monitoring data. Our work is organized as follows.
In Section~\ref{sec-problem}, we start by introducing the main
ingredients of nonlinear ensemble filtering, followed in
Section~\ref{sec-method} by an exposition on how techniques from
amortized Bayesian inference (Cranmer, Brehmer, and Louppe 2020; Radev
et al. 2020) can be used to derive our nonlinear extension of ensemble
filtering based on CNFs. After specifying algorithms for Sequential
Simulation-based Bayesian Inference and \(\textsc{Forecast}\),
\(\textsc{Training}\), and \(\textsc{Analysis}\), which includes an
introduction of the important aspect of summary statistics, we proceed
by illustrating the ``lifecycle'' of the proposed Digital Shadow, which
includes the important prior-to-posterior mapping that corrects the
predictions for the state from time-lapse observations. Next, we outline
a detailed procedure to validate the proposed Digital Shadow prototype
in a realistic off-shore setting representative of the South-Western
North Sea. In addition to detailing this setting, which includes
descriptions of the probabilistic baseline, ground-truth simulations,
and monitoring with the Digital Shadow, Section~\ref{sec-shadow} also
introduces certain performance metrics. Guided by these metrics, a case
study is presented in Section~\ref{sec-case} with an emphasis on
examining the advantages of working with multimodal data. This case
study is followed by Section~\ref{sec-discussion} in which the Digital
Shadow is put in a broader context, followed by conclusions in
Section~\ref{sec-conclusion}.

\begin{figure}

\begin{minipage}{\linewidth}
\includegraphics{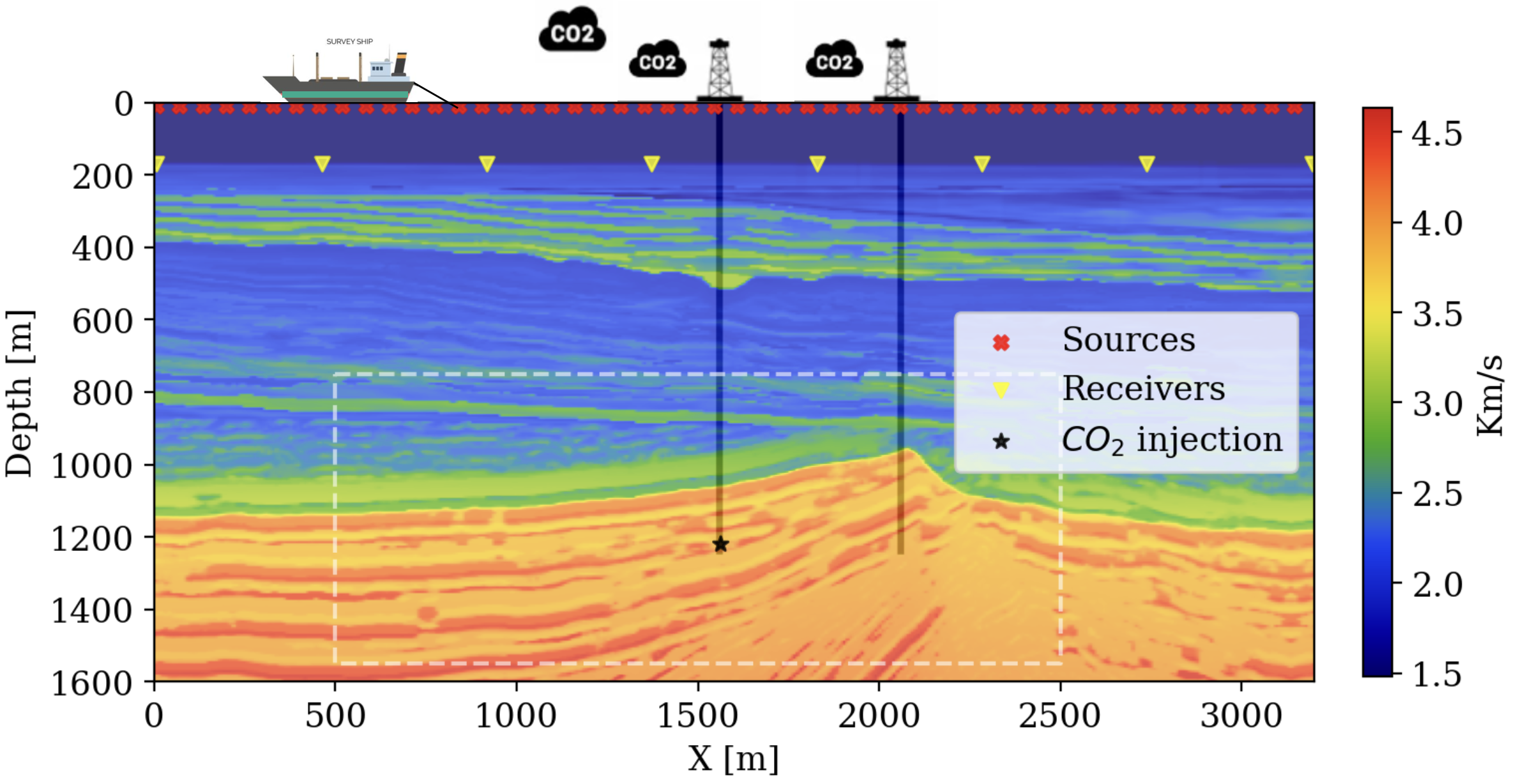}\end{minipage}%
\newline
\begin{minipage}{\linewidth}
\includegraphics{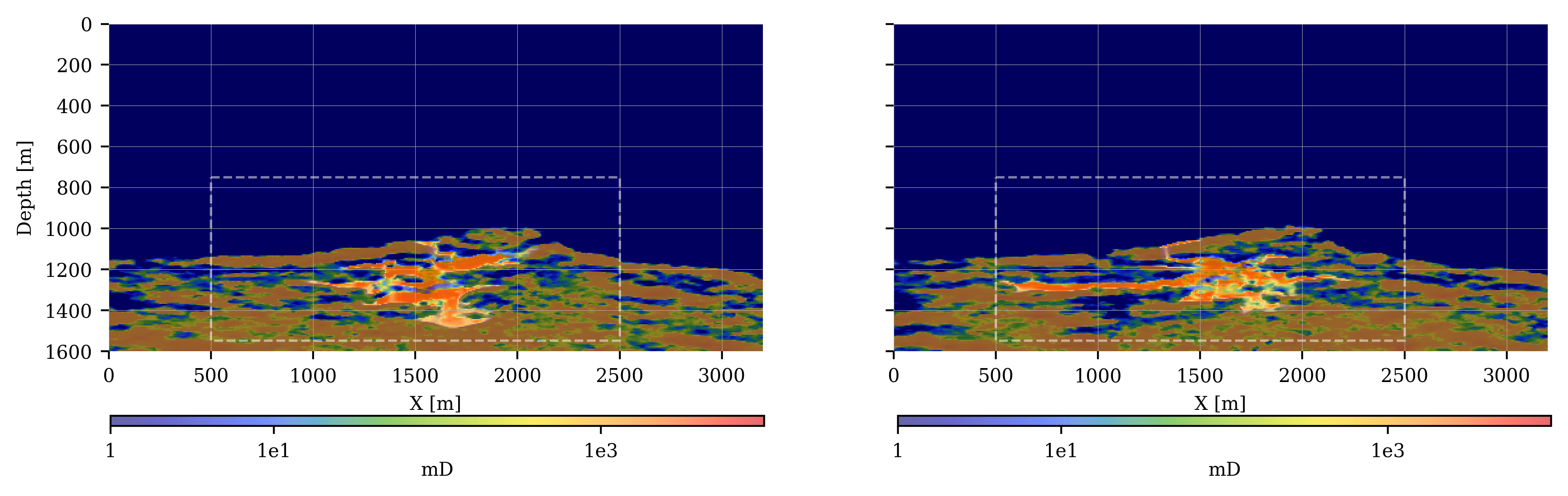}\end{minipage}%

\caption{\label{fig-acquisition}Example of a setup of Geological Carbon
Storage in a marine setting. Top: Schematic diagram showing the velocity
model, the seismic acquisition setup with sources (denoted by red \(X\)
symbol), and receivers (denoted by the yellow \(\nabla\) symbol). The
setup also includes injection (left) and monitoring (right) wells.
Bottom: the juxtaposition of multi-phase flow simulations for two
different realizations of a permeability field drawn from a stochastic
baseline established from active-source surface seismic data (Yin,
Orozco, et al. 2024). Both realizations for the permeability are
consistent with this baseline seismic survey but yield vastly different
outcomes. Bottom-left CO\textsubscript{2} plume remains well within the
storage complex (denoted by the white dashed box). Bottom-right: plume
obviously breaches Containment, due to the presence of an ``unknown''
high-permeability streak.}

\end{figure}%

\section{Problem formulation}\label{sec-problem}

Monitoring CO\textsubscript{2} plumes from multimodal time-lapse
measurements from well and/or surface-seismic data calls for an approach
where nonlinear causes of epistemic uncertainty, reflecting lack of
knowledge on the reservoir properties determining the dynamics, and
aleatoric uncertainty, due to noise in the data, are accounted for in a
sequential Bayesian inference framework (Tatsis, Dertimanis, and Chatzi
2022; Rubio, Marzouk, and Parno 2023). As a lead into the definition of
our Digital Shadow based on amortized neural posterior estimation
(Papamakarios and Murray 2016; Lueckmann et al. 2017; Cranmer, Brehmer,
and Louppe 2020), we begin by defining general models for nonlinear
state dynamics and observations, followed by formulating a general
approach to nonlinear Bayesian filtering (Asch 2022).

\subsection{Uncertain state dynamics and observation
models}\label{uncertain-state-dynamics-and-observation-models}

While the presented framework is designed to be general, the focus will
be on geophysical monitoring of Geological Carbon Storage (GCS) aimed at
keeping track of the dynamics of CO\textsubscript{2} plumes via a
combination of multi-phase flow simulations and time-lapse multimodal
geophysical measurements, consisting of sparse localized direct
measurements of the saturation and pressure perturbations (shown in
figure~\ref{fig-ground-truth-multimodal}) collected within boreholes of
(monitoring) wells and indirect active-source seismic measurements,
collected at the surface. In this scenario, the time-varying
state-dynamics are measured at discrete time instants, \(k=1\cdots K\),
and consists of the time-varying spatial distribution for the
CO\textsubscript{2} saturation, \(\mathbf{x}_k[S_{\mathrm{CO_2}}]\), and
induced pressure perturbation, \(\mathbf{x}_k[\delta p]\), which can be
expressed as follows:

\begin{equation}\phantomsection\label{eq-dynamics}{
\begin{aligned}
\mathbf{x}_{k} & = \mathcal{M}_k\bigl(\mathbf{x}_{k-1}, \boldsymbol{\kappa}_k; t_{k-1}, t_k\bigr)\\
               & = \mathcal{M}_k\bigl(\mathbf{x}_{k-1}, \boldsymbol{\kappa}_k\bigr),\ \boldsymbol{\kappa}_k\sim p(\boldsymbol{\kappa}) \quad \text{for}\quad k=1\cdots K.
\end{aligned}
}\end{equation}

In this expression for the uncertain nonlinear dynamics, the possibly
time-dependent \emph{dynamics operator}, \(\mathcal{M}_k\), represents
multi-phase fluid-flow simulations that transition the state from the
previous timestep, \(t_{k-1}\), to the next timestep, \(t_k\), for all
\(K\) time-lapse timesteps. The dynamics operator, \(\mathcal{M}_k\),
which implements the underlying continuous dynamics between times
\(t_{k-1}\) and \(t_k\), may itself be time-dependent because of induced
chemical and/or rock mechanical changes. Because the spatial
distribution of the permeability field, collected in the vector
\(\boldsymbol{\kappa}\), is highly heterogeneous (Ringrose 2020, 2023),
unknown, and possibly time-dependent, it is treated as a stochastic
nuisance parameter (Radev et al. 2020). Without loss of generality, we
assume samples for the spatial gridded permeability field to be drawn
from the time-invariant probability distribution---i.e,
\(\boldsymbol{\kappa}\sim p(\boldsymbol{\kappa})\) where
\(p(\boldsymbol{\kappa})\) is assumed to be independent of the state and
time-lapse observations. The dynamics operator takes samples from the
spatial distribution for the subsurface permeability,
\(\boldsymbol{\kappa}\), and the CO\textsubscript{2} saturation and
reservoir pressure perturbations, collected at the previous timestep
\(k-1\) in the vector, \(\mathbf{x}_{k-1}\), as input and produces
samples for the CO\textsubscript{2} saturation and reservoir pressure
perturbations at the next timestep, \(\mathbf{x}_{k}\), as output. While
this procedure offers access to samples from the transition density, the
transition density itself can not be evaluated because of the presence
of the stochastic nuisance parameter, \(\boldsymbol{\kappa}\). To
leading order, the nonlinear dynamics consists of solutions of Darcy's
equation and mass conservation laws (Pruess and Nordbotten 2011; Li et
al. 2020). Despite the fact that the dynamics can be modeled accurately
with reservoir simulations (Krogstad et al. 2015; Settgast et al. 2018;
Rasmussen et al. 2021; Stacey and Williams 2017), relying on these
simulations alone is unfeasible because of a lack of knowledge on the
permeability.

To improve estimates for the state, CO\textsubscript{2} plumes are
monitored by collecting discrete time-lapse well (Freifeld et al. 2009)
and surface-seismic data (D. Lumley 2010). We model these multimodal
time-lapse datasets as

\begin{equation}\phantomsection\label{eq-obs}{
\mathbf{y}_k = \mathcal{H}_k(\mathbf{x}_k,\boldsymbol{\epsilon}_k, \mathbf{\bar{m}}_k),\ \boldsymbol{\epsilon}_k \sim p(\boldsymbol{\epsilon}) \quad\text{for}\quad k=1,2,\cdots,K
}\end{equation}

where the vector, \(\mathbf{y}_k\), represents the observed multimodal
data at timestep \(k\). To add realism, the seismic data is corrupted by
the noise term, \(\boldsymbol{\epsilon}_k\), sampled by filtering
realizations of standard zero-centered Gaussian noise with a
bandwidth-limited seismic source-time function. This filtered noise is
normalized to attain a preset signal-to-noise ratio before it is added
to the seismic shot records prior to reverse-time migration. The shot
records themselves are generated with 2D acoustic simulations with the
seismic properties (velocity and density) linked to the
CO\textsubscript{2} saturation through the patchy saturation
model\footnote{To arrive at this acquisition geometry, use was made of
  reciprocity---i.e., the computational experiments are calculated for
  \(16\) sources, reducing the computational costs.} (Avseth, Mukerji,
and Mavko 2010). The vector,
\(\mathbf{\bar{m}_k}=(\mathbf{\bar{v}}, \mathbf{\bar{\rho}})\) with
\(\mathbf{\bar{v}}\) and \(\mathbf{\bar{\rho}}\) the reference velocity
and density, represents a time-dependent seismic reference model from
which the migration-velocity model for imaging will be derived and with
respect to which the time-lapse differences in the seismic shot records
will be calculated. According to these definitions, the
\emph{observation operator}, \(\mathcal{H}_k\), models multimodal,
direct and indirect, time-lapse monitoring data, collected at the
(monitoring) wells and migrated from surface-seismic shot records. While
this observation operator may also depend on time, e.g.~because the
acquisition geometry changes, we assume it, like the dynamics operator,
to be time-invariant, and drop the time index \(k\) from the notation.

Due to sparsely sampled receiver and well locations, and the non-trivial
null-space of the modeling operators, the above measurement operator is
ill-posed, which renders monitoring the CO\textsubscript{2} dynamics
challenging, a situation that is worsened by the intrinsic stochasticity
of the dynamics. To overcome these challenges, we introduce a nonlinear
Bayesian filtering framework where the posterior distribution for the
state, conditioned on the collected multimodal time-lapse data, is
approximated by Variational Inference (Jordan et al. 1999; David M. Blei
and McAuliffe 2017) with machine learning (Rezende and Mohamed 2015;
Siahkoohi et al. 2023b; Orozco, Siahkoohi, et al. 2024), neural
posterior density estimation to be precise. With this approach, which
combines the latest techniques from sequential Bayesian inference and
amortized neural posterior density estimation (Papamakarios et al.
2021), we aim to quantify the uncertainty of the CO\textsubscript{2}
plumes in a principled manner.

\subsection{Nonlinear Bayesian
filtering}\label{nonlinear-bayesian-filtering}

Because both the dynamics and observation operators are nonlinear, we
begin by defining a general probabilistic model for the state and
observations in terms of the following conditional distributions:

\begin{equation}\phantomsection\label{eq-trans-density}{
\mathbf{x}_k \sim p(\mathbf{x}_k\mid\mathbf{x}_{k-1})
}\end{equation}

and

\begin{equation}\phantomsection\label{eq-likelihood}{
\mathbf{y}_k \sim p(\mathbf{y}_k\mid\mathbf{x}_{k}),
}\end{equation}

for \(k=0\cdots K\), representing, respectively, the \emph{transition
density}, defined by equation~\ref{eq-dynamics}, and the data
\emph{likelihood}, given by equation~\ref{eq-obs}. These conditional
distributions hold because the above monitoring problem (cf.~equations
\ref{eq-dynamics} and \ref{eq-obs}) is Markovian, which implies that the
likelihood does not depend on the time-lapse history---i.e.,
\(\left(\mathbf{y}_k\perp\mathbf{y}_{1:k-1}\right)\mid\mathbf{x}_k\)
with the shorthand,
\(\mathbf{y}_{1:k}= \left\{\mathbf{y}_1,\ldots,\mathbf{y}_k\right\}\),
so that
\(p(\mathbf{y}_k\mid \mathbf{x}_k, \mathbf{y}_{1:k-1})=p(\mathbf{y}_k\mid \mathbf{x}_k)\).
This orthogonality assumption holds as long as the nuisance parameters
are independent with respect to the state (Rubio, Marzouk, and Parno
2023). The aim of Bayesian filtering is to recursively compute the
posterior distribution of the state at time, \(k\), conditioned on the
complete history of collected time-lapse data---i.e., compute
\(p(\mathbf{x}_k\mid \mathbf{y}_{1:k})\). There exists a rich literature
on statistical inference for dynamical systems (McGoff, Mukherjee, and
Pillai 2015) aimed to accomplish this goal, which entails the following
three steps (Asch 2022):

\textbf{Initialization step:} Choose the prior distribution for the
state \(p(\mathbf{x}_0)\).

\textbf{Prediction step:} Compute the predictive distribution for the
state derived from the Chapman-Kolmogorov equation yielding (Tatsis,
Dertimanis, and Chatzi 2022)

\begin{equation}\phantomsection\label{eq-prediction}{
\begin{aligned}
p(\mathbf{x}_k\mid \mathbf{y}_{1:k-1})&=\int p(\mathbf{x}_k\mid \mathbf{x}_{k-1})p(\mathbf{x}_{k-1}\mid \mathbf{y}_{1:k-1})\mathrm{d}\mathbf{x}_{k-1}\\
& = \mathbb{E}_{\mathbf{x}_{k-1}\sim p(\mathbf{x}_{k-1}\mid\mathbf{y}_{1:k-1})}\bigl[p(\mathbf{x}_{k}\mid\mathbf{x}_{k-1})\bigr].
\end{aligned}
}\end{equation}

This step uses equation~\ref{eq-dynamics} to advance samples of the
distribution \(p(\mathbf{x}_{k-1}\mid \mathbf{y}_{1:k-1})\) to samples
of \(p(\mathbf{x}_k\mid \mathbf{y}_{1:k-1})\).

\textbf{Analysis step:} Apply Bayes' rule

\begin{equation}\phantomsection\label{eq-Bayes}{
p(\mathbf{x}_{k}|\mathbf{y}_{1:k})  \propto  p(\mathbf{y}_{k}|\mathbf{x}_{k})
p(\mathbf{x}_{k}|\mathbf{y}_{1:k-1})
}\end{equation}

with \(p(\mathbf{y}_{k}|\mathbf{x}_{k})\) and
\(p(\mathbf{x}_k\mid \mathbf{y}_{1:k-1})\) acting as the ``likelihood''
and ``prior''. During this step, the posterior distribution is
calculated by correcting the ``prior'' with the ``likelihood''. To this
end, equation~\ref{eq-likelihood} will be used to draw samples from the
likelihood.

These three steps are general and make no assumptions on the linearity
and Gaussianity of the distributions involved. The general solution of
the above Bayesian filtering problem is far from trivial for the
following reasons: \emph{(i)} equations \ref{eq-prediction} and
\ref{eq-Bayes} only permit analytical solutions in cases where the
state-dynamics and observation operators are linear and Gaussian
distributed; \emph{(ii)} the presence of the nuisance permeability
parameter in equation~\ref{eq-dynamics} requires access to the marginal
likelihood, which is computationally infeasible (Cranmer, Brehmer, and
Louppe 2020; Kruse et al. 2021a); \emph{(iii)} likewise, the
marginalization in equation~\ref{eq-prediction} is also not viable
computationally; and finally \emph{(iv)} computing the above conditional
probabilities becomes intractable in situations where the state is high
dimensional and the state-dynamics and observational operators are
computationally expensive to evaluate.

To address these challenges, numerous data-assimilation techniques have
been developed over the years (Asch, Bocquet, and Nodet 2016). The
ensemble Kalman filter (EnKF) (Evensen 1994; Carrassi et al. 2018)
features amongst the most advanced techniques and is widely used for
large nonlinear state-space and observation models. During the
\emph{prediction step} of the data-assimilation recursions, EnKF
propagates a set of \(M\) particles from the previous timestep,
\(\left\{\mathbf{x}^{(1)}_{k-1},\ldots,\mathbf{x}^{(M)}_{k-1}\right\}\),
to the next timestep by applying the state-space dynamics
(equation~\ref{eq-dynamics}) to each particle, followed by generating
simulated observation samples,
\(\left\{\mathbf{y}^{(1)}_{k},\ldots,\mathbf{y}^{(M)}_{k}\right\}\),
based on the observation model of equation~\ref{eq-obs}. During the
subsequent \emph{analysis step}, EnKF uses the pairs,
\(\{\mathbf{x}_k^{(m)}, \mathbf{y}_k^{(m)}\}_{m=1}^M\), to calculate
empirical Monte-Carlo based linear updates to the forecast ensemble.
While EnKF avoids the need for linearization, calculation of gradients,
and is relatively straightforward to scale, it relies on linear updates
that prevent it from solving the above nonlinear and non-Gaussian
Bayesian filtering problem accurately, even when increasing the ensemble
size (Spantini, Baptista, and Marzouk 2022).

Instead of relying on linear mappings, we consider the analysis step in
equation~\ref{eq-Bayes} as a trainable prior-to-posterior transformation
and employ techniques from amortized neural posterior density estimation
with CNFs (Kruse et al. 2021b; Gahlot et al. 2023; Herrmann 2023) to
carry out the analysis step, an approach also taken by Spantini,
Baptista, and Marzouk (2022), who proposes a closely related but
different coupling technique. Our proposed approach derives from
simulation-based inference (Cranmer, Brehmer, and Louppe 2020; Radev et
al. 2020) with invertible CNFs, which extends EnKF by allowing for
learned nonlinear prior-to-posterior density transformations. CNFs
represent a class of invertible neural networks that, dependent on a
condition term, convert samples from a complicated distribution, through
a series of nonlinear differentiable transformations, into samples from
a simpler distribution, e.g.~the standard Gaussian distribution. Instead
of explicit evaluations of the corresponding densities, and similar to
EnKF workflows, access is only needed to samples from the ``likelihood''
and ``prior'', available through simulations with equations
\ref{eq-dynamics} and \ref{eq-obs}.

\section{Methodology}\label{sec-method}

To free ourselves from limiting assumptions, which either rely on
simplifying linearity and Gaussianity assumptions or on computationally
unfeasible Markov chain Monte-Carlo (McMC) based approaches, extensive
use will be made of Variational Inference (VI) (Jordan et al. 1999) with
CNFs (Rezende and Mohamed 2015; Ardizzone et al. 2019a; Kruse et al.
2021a). In these approaches, scalable optimization techniques are used
to match parametric representations of distributions with CNFs to
unknown target distributions. During simulation-based inference (SBI)
(Cranmer, Brehmer, and Louppe 2020; Radev et al. 2020), the forward
Kullback-Leibler (KL) divergence (Kullback and Leibler 1951; Siahkoohi
et al. 2023a) is minimized on samples from the joint distribution for
the state and observations, which involves a relatively easy to evaluate
quadratic objective function. In this approach, the CNFs correspond to
generative neural networks that serve as conditional density estimators
(Radev et al. 2020) for the posterior distribution. After explaining the
training and inference phases of SBI, the technique of sequential
simulation-based inference will be introduced. This technique is
designed to serve as a nonlinear extension of EnKF. To this end,
algorithms for the recurrent \(\textsc{Forecast}\),
\(\textsc{Training}\), and \(\textsc{Analysis}\) stages of the proposed
neural Bayesian filtering approach will be presented.

\subsection{Amortized Bayesian Inference with CNFs}\label{sec-amortized}

Recent developments in generative Artificial Intelligence (genAI), where
neural networks are trained as surrogates for high-dimensional
(conditional) distributions, are key to the development and success of
our uncertainty-aware Digital Shadow, which relies on capturing the
posterior distribution from simulated samples for the state and
observations. While any amortized neural network implementation capable
of producing samples conditioned on samples for different observations
will work in our setting, and this includes conditional neural networks
based on conditional score-based diffusion (Batzolis et al. 2021), CNFs
have our preference because they allow us to quickly sample from the
posterior, \(\mathbf{x}\sim p(\mathbf{x}\mid\mathbf{y})\), and conduct
density calculations---i.e., evaluate the posterior density
\(p(\mathbf{x}\mid\mathbf{y})\) and its derivatives with respect to the
states, \(\mathbf{x}\), and observations, \(\mathbf{y}\). This latter
feature is necessary when designing Digital Twins capable of
uncertainty-informed decisions (Alsing, Edwards, and Wandelt 2023),
control (Gahlot et al. 2024), and Bayesian experimental design (Orozco,
Herrmann, and Chen 2024). While non-amortized approaches may be
feasible, amortization of the neural estimate of the posterior
distribution is essential because it allows us to marginalize over
stochastic nuisance parameters (the permeability field,
\(\boldsymbol{\kappa}\), in our case) while capturing its uncertainty
from the ensemble of dynamic states and associated time-lapse
observations.

Specifically, our goal is to capture the posterior distribution from
pairs, \(\{\mathbf{x}^{(m)}, \mathbf{y}^{(m)}\}_{m=1}^M\)
(time-dependence was dropped for simplicity). To achieve this goal, we
make use of Bayesian Variational Inference (Rezende and Mohamed 2015) in
combination with the ability to carry out simulations of the state
dynamics and observations (cf.~equations \ref{eq-dynamics} and
\ref{eq-obs}), which are of the type (Radev et al. 2020)

\begin{equation}\phantomsection\label{eq-simulations}{
\mathbf{y}=\mathcal{G}(\mathbf{x},\boldsymbol{\zeta})\quad\text{with}\quad \boldsymbol{\zeta}\sim p(\boldsymbol{\zeta}\mid\mathbf{x}),\, \mathbf{x}\sim p(\mathbf{x})
}\end{equation}

where \(\mathcal{G}\) is a generic simulator that depends on samples of
the prior, \(\mathbf{x}\sim p(\mathbf{x})\), and nuisance parameters,
\(\boldsymbol{\zeta}\sim p(\boldsymbol{\zeta}\mid\mathbf{x})\), possibly
conditioned on samples of the prior. These simulated samples give us
implicit access to the likelihood. Given \(M\) sampled pairs,
\(\{\mathbf{x}^{(m)}, \mathbf{y}^{(m)}\}_{m=1}^M\), simulation-based
inference with CNFs proceeds by training an invertible neural network to
approximate the posterior distribution, \(p(\mathbf{x}\mid\mathbf{y})\),
by minimizing the forward Kullback--Leibler (KL) divergence between the
true and the approximate posterior for all possible data,
\(\mathbf{y}\). The training objective reads (Radev et al. 2020;
Siahkoohi et al. 2023a)

\begin{equation}\phantomsection\label{eq-KL}{
\begin{aligned}
\widehat{\boldsymbol{\phi}}&= \mathop{\mathrm{argmin}\,}\limits_{\boldsymbol{\phi}}\mathbb{E}_{p(\mathbf{y})}\bigl[\mathbb{KL}\bigl(p(\mathbf{x}\mid \mathbf{y})\mid\mid p_{\boldsymbol{\phi}}(\mathbf{x}\mid \mathbf{y})\bigr)\bigr]\\
& =\mathop{\mathrm{argmin}\,}\limits_{\boldsymbol{\phi}}\mathbb{E}_{p(\mathbf{y})}\Bigl[\mathbb{E}_{p(\mathbf{x}\mid\mathbf{y})}\bigl[-\log p_{\boldsymbol{\phi}}(\mathbf{x}\mid\mathbf{y})+p(\mathbf{x}\mid\mathbf{y})\bigr]\Bigr] \\
& = \mathop{\mathrm{argmax}\,}\limits_{\boldsymbol{\phi}}\mathbb{E}_{p(\mathbf{x},\mathbf{y})}\bigl[\log p_{\boldsymbol{\phi}}(\mathbf{x}\mid\mathbf{y})\bigr].
\end{aligned}
}\end{equation}

To arrive at this objective involving samples from the joint
distribution for \(\mathbf{x}\) and \(\mathbf{y}\), use was made of the
law of total probability that allows us to rewrite the marginalization
over observations, \(\mathbf{y}\), in terms of the expectation over
samples from the joint distribution. The term
\(p(\mathbf{x}\mid\mathbf{y})\) can be dropped because it does not
depend on the parameters, \(\boldsymbol{\phi}\). For more details, see
Siahkoohi et al. (2023a). The above objective is used to approximate the
true posterior distribution as accurately as possible,
\(p_{\boldsymbol{\phi}}(\mathbf{x}\mid\mathbf{y})\approx p(\mathbf{x}\mid\mathbf{y})\)
for all possible \(\mathbf{x}\) and \(\mathbf{y}\), drawn from the joint
distribution, \(p(\mathbf{x},\mathbf{y})\). Following Radev et al.
(2020), the approximate posterior,
\(p_{\boldsymbol{\phi}}(\mathbf{x}\mid\mathbf{y})\) is parameterized
with a conditional Invertible Neural Network (cINN, Ardizzone et al.
(2019a)), \(f_{\boldsymbol{\phi}}\), which conditioned on,
\(\mathbf{y}\), defines an invertible mapping.

With the change-of-variable rule of probability

\[
p_{\boldsymbol{\phi}}(\mathbf{x}\mid\mathbf{y})=p\bigl(\mathbf{z}=f_{\boldsymbol{\phi}}(\mathbf{x};\mathbf{y})\bigr)\Bigl |\det\Bigl(\mathbf{J}_{f_{\boldsymbol{\phi}}}\Bigr)\Bigr |
\]

with \(\mathbf{J}_{f_{\boldsymbol{\phi}}}\) the Jacobian of
\(f_{\boldsymbol{\phi}}\) evaluated at \(\mathbf{x}\) and
\(\mathbf{y}\), the objective of maximizing the likelihood in
equation~\ref{eq-KL} becomes

\begin{equation}\phantomsection\label{eq-max-likelihood}{
\begin{aligned}
\widehat{\boldsymbol{\phi}}&=\mathop{\mathrm{argmax}\,}\limits_{\boldsymbol{\phi}}\mathbb{E}_{p(\mathbf{x},\mathbf{y})}\bigl[\log p_{\boldsymbol{\phi}}(\mathbf{x}\mid\mathbf{y})\bigr]\\
&= \mathop{\mathrm{argmax}\,}\limits_{\boldsymbol{\phi}}\mathbb{E}_{p(\mathbf{x},\mathbf{y})}\Bigl[\log p\bigl(\mathbf{z}=f_{\boldsymbol{\phi}}(\mathbf{x};\mathbf{y})\bigr)+\log\Bigl |\det\Bigl(\mathbf{J}_{f_{\boldsymbol{\phi}}}\Bigr)\Bigr |\Bigr].
\end{aligned}
}\end{equation}

Given the simulated training pairs,
\(\{\mathbf{x}^{(m)}, \mathbf{y}^{(m)}\}_{m=1}^M\), maximization of the
expectations in equation~\ref{eq-max-likelihood} can be approximated by
minimization of the sum:

\begin{equation}\phantomsection\label{eq-loss-CNF}{
\begin{aligned}
\widehat{\boldsymbol{\phi}}&= \mathop{\mathrm{argmin}\,}\limits_{\boldsymbol{\phi}}\frac{1}{M}\sum_{m=1}^M \Bigl(-\log p\bigl(f_{\boldsymbol{\phi}}(\mathbf{x}^{(m)};\mathbf{y}^{(m)})\bigr)-\log\Bigl |\det\Bigl(\mathbf{J}^{(m)}_{f_{\boldsymbol{\phi}}}\Bigr)\Bigr |\Bigr)\\
& =\mathop{\mathrm{argmin}\,}\limits_{\boldsymbol{\phi}} \frac{1}{M}\sum_{m=1}^M \Biggl(\frac{\Big\|f_{\boldsymbol{\phi}}(\mathbf{x}^{(m)};\mathbf{y}^{(m)})\Big\|_2^2}{2} - \log\Bigl |\det\Bigl(\mathbf{J}^{(m)}_{f_{\boldsymbol{\phi}}}\Bigr)\Bigr |\Biggr).
\end{aligned}
}\end{equation}

During derivation of this training loss for CNFs, use was made of the
property:
\(\mathcal{N}(0,\mathbf{I})\propto \exp(-(1/2)\|\mathbf{z}\|_2^2)\). In
the ideal situation where the estimated network weights,
\(\widehat{\boldsymbol{\phi}}\), attain a global minimum of the
objective in equation~\ref{eq-KL}, the latent distribution will be
statistically independent of the conditioning data (Radev et al. 2020).
In that case, samples from the posterior can be drawn based on the
following equivalence:

\begin{equation}\phantomsection\label{eq-ideal-sampling}{
\mathbf{x}\sim p_{\widehat{\boldsymbol{\phi}}}(\mathbf{x}\mid\mathbf{y}) \,\Longleftrightarrow\, f^{-1}_{\widehat{\boldsymbol{\phi}}}(\mathbf{z};\mathbf{y})\sim p(\mathbf{x}\mid\mathbf{y})\quad\text{with}\quad \mathbf{z}\sim \mathcal{N}(0,\mathbf{I}). 
}\end{equation}

According to this equivalence, sampling from the posterior corresponds
to generating samples from the multidimensional standard Gaussian
distribution, \(\mathbf{z}\sim \mathcal{N}(0,\mathbf{I})\), followed by
computing
\(f^{-1}_{\widehat{\boldsymbol{\phi}}}(\mathbf{z};\mathbf{y})\). In the
next section, we will employ and adapt these training and inference
phases for the less-than-ideal recursive training regime of
data-assimilation problems where the ensemble sizes are relatively small
for which the global minimum of equation~\ref{eq-KL} may not be
attained.

\begin{algorithm}[htb!]
\caption{Sequential Simulation-based Bayesian Inference}
\textbf{Require:} Datasets $\mathbf{y}_k^{\mathrm{obs}}$, collected sequentially in the field at times $k=1,\ldots,K$\\
\textbf{Output:} Estimate ensemble for the states $\{\widehat{\mathbf{x}}^{(m)}_k\}_{m=1}^M$, for timesteps $k=1,\ldots,K$
\label{alg-sSBI}
\begin{algorithmic}[1]
  \For{$m = 1$ \To $M$}
    \State Sample from initial state: $\widehat{\mathbf{x}}^{(m)}_0\sim p(\mathbf{x}_0)$ 
  \EndFor
  \For{$k = 1$ \To $K$}
    \State $\{(\mathbf{x}_k^{(m)},\mathbf{y}_k^{(m)}) \}_{m=1}^{M}= $\Call{Forcast}{$ p(\boldsymbol{\kappa}), p(\boldsymbol{\epsilon}_k),\{\widehat{\mathbf{x}}_{k-1}^{(m)}\}_{m=1}^M$}
    \State $\widehat{\boldsymbol{\phi}}_k$=\Call{Training}{$\{(\mathbf{x}_k^{(m)},\mathbf{y}_k^{(m)}) \}_{m=1}^{M}$}
    \State Collect data in the field $\mathbf{y}_k^{\mathrm{obs}}$
    \State $\{\widehat{\mathbf{x}}_k^{(m)}\}_{m=1}^M= $\Call{Analysis}{$\mathbf{y}^{\mathrm{obs}}_{k}, \widehat{\boldsymbol{\phi}}_k,\{(\mathbf{x}_k^{(m)},\mathbf{y}_k^{(m)}) \}_{m=1}^{M}$}
  \EndFor
\end{algorithmic}
\end{algorithm}

\subsection{Neural sequential simulation-based
inference}\label{sec-sSBI}

After randomized initialization of \(M\) particles of the
CO\textsubscript{2} plume's initial state,
\(\widehat{\mathbf{x}}^{(m)}_0\sim p(\mathbf{x}_0)\) (line 2 of
 Algorithm~\ref{alg-sSBI} ), training pairs for the state
(equation~\ref{eq-dynamics}) and observations (equation~\ref{eq-obs}),
\(\{\mathbf{x}_k^{(m)},\mathbf{y}_k^{(m)} \}_{m=1}^{M}\), for the next
time step, are predicted by simulations with the state dynamics and
detailed in the \(\textsc{Forecast}\) step of
 Algorithm~\ref{alg-training} ). For notational convenience, we already
introduced the symbol, \(\widehat{\quad}\), to distinguish between
predicted ``digital states'' and analyzed predictions (denoted by the
\(\widehat{\quad}\)) for the state conditioned on the field observed
data. During the \(\textsc{Forecast}\) step of
 Algorithm~\ref{alg-training} , random samples of the stochastic
permeability field and noise are drawn and used to draw samples from the
transition density (line 3 of  Algorithm~\ref{alg-training} ),
conditioned on analyzed samples,
\(\{\widehat{\mathbf{x}}_k^{(m)}\}_{m=1}^M\), from the previous time
step. These predictions for the advanced state are subsequently employed
to sample observations (line 4 of  Algorithm~\ref{alg-training} ),
producing the simulated training ensemble,
\(\{\mathbf{x}_k^{(m)},\mathbf{y}_k^{(m)} \}_{m=1}^{M}\). These
simulated pairs are utilized during the \(\textsc{Training}\) step (line
6 of  Algorithm~\ref{alg-sSBI} ) where equation~\ref{eq-loss-CNF} is
minimized with respect to the CNF's network weights, yielding
\(\widehat{\boldsymbol{\phi}}_k\). Given these network weights, \(M\)
latent-space representations for each pair are calculated via

\begin{equation}\phantomsection\label{eq-latent}{
\mathbf{z}_k^{(m)}= f_{\widehat{\boldsymbol{\phi}}_k}(\mathbf{x}_k^{(m)};\mathbf{y}_k^{(m)}),
}\end{equation}

for \(m=1\cdots M\) (Spantini, Baptista, and Marzouk 2022). Given
observed field data, \(\mathbf{y}_k^{\mathrm{obs}}\), collected at time
\(k\) (line 7 of  Algorithm~\ref{alg-sSBI} ), these latent variables are
used to correct the state for \(m=1\cdots M\) via

\begin{equation}\phantomsection\label{eq-analyses-samples}{
\widehat{\mathbf{x}}_k^{(m)}= f^{-1}_{\widehat{\boldsymbol{\phi}}_k}(\mathbf{z}_k^{(m)};\mathbf{y}_k^{\mathrm{obs}}).
}\end{equation}

In this expression, \(\{\widehat{\mathbf{x}}_k^{(m)}\}_{m=1}^M\),
corresponds for \(k\geq 1\) to analyzed states, produced by the
prior-to-posterior mapping (Algorithm 2 in Spantini, Baptista, and
Marzouk (2022)). To offset the potential effects of working with a
limited ensemble size, \(M\), we choose to adapt this approach of
producing samples from the latent space calculated from the forecast
ensemble. According to Spantini, Baptista, and Marzouk (2022), this
process mitigates the effects when a global minimum of
equation~\ref{eq-loss-CNF} is not attained, implying that
\(p_{\widehat{\boldsymbol{\phi}}}(\mathbf{z})\perp \mathbf{x}\) may no
longer hold. In that case, the latent output distribution is no longer
guaranteed to be the same for any fixed, \(\mathbf{y}\) (Radev et al.
2020). By producing latent samples that derive directly from the
predicted state-observation pairs, this issue is resolved. When
conditioned on the observed field data, \(\mathbf{y}_k^{\mathrm{obs}}\),
these latent samples improve the inference after running the CNF in the
reverse direction (cf. equation~\ref{eq-analyses-samples}).
 Algorithm~\ref{alg-training}  summarizes the three key phases, namely
the \(\textsc{Forecast}\), \(\textsc{Training}\), and
\(\textsc{Analysis}\) steps, which are used during the recursions in
 Algorithm~\ref{alg-sSBI}  that approximate equations
\ref{eq-prediction} and \ref{eq-Bayes}. During these recursions, for
timesteps \(k>1\), samples of the posterior distribution from the
previous timestep, \(\{\widehat{\mathbf{x}}^{(m)}_{k-1}\}_{m=1}^M\),
serve as ``priors'' for the next timestep, producing,
\(\{\widehat{\mathbf{x}}^{(m)}_k\}_{m=1}^M\), which are conditioned on
the observations, \(\mathbf{y}_k^{\mathrm{obs}}\). This procedure
corresponds to the nonlinear prior-to-posterior of Spantini, Baptista,
and Marzouk (2022).

\begin{algorithm}[htb!]
\caption{Forecast, Training, and Analysis}
\label{alg-training}
\begin{algorithmic}[1]
\Procedure{Forecast}{$ p(\boldsymbol{\kappa}), p(\boldsymbol{\epsilon}_k),\{\widehat{\mathbf{x}}_{k-1}^{(m)}\}_{m=1}^M$}
  \For{$m = 1$ \To $M$}
    \State Sample from transition density: $\mathbf{x}^{(m)}_k\sim p(\cdot\mid \widehat{\mathbf{x}}^{(m)}_{k-1})$ by running dynamics simulations (equation 1)
    \State Sample from the likelihood: $\mathbf{y}^{(m)}_k\sim p(\cdot\mid \mathbf{x}^{(m)}_{k})$ by simulating observations (equation 2)
  \EndFor 
  \Return $\{(\mathbf{x}_k^{(m)},\mathbf{y}_k^{(m)}) \}_{m=1}^{M}$
\EndProcedure 
\Procedure{Training}{$\{(\mathbf{x}_k^{(m)},\mathbf{y}_k^{(m)}) \}_{m=1}^{M}$}
  \State $\widehat{\boldsymbol{\boldsymbol{\phi}}}_k=\mathop{\mathrm{argmin}}_{\boldsymbol{\phi}} \frac{1}{M}\sum_{m=1}^M \Biggl(\frac{\Big\|f_{\boldsymbol{\phi}}(\mathbf{x}^{(m)};\mathbf{y}^{(m)})\Big\|_2^2}{2} - \log\Bigl |\det\Bigl(\mathbf{J}^{(m)}_{f_{\boldsymbol{\phi}}}\Bigr)\Bigr |\Biggr)$ 
  \Return $\widehat{\boldsymbol{\phi}}_k$
\EndProcedure
\Procedure{Analysis}{$\mathbf{y}^{\mathrm{obs}}_{k}, \widehat{\boldsymbol{\phi}}_k,\{(\mathbf{x}_k^{(m)},\mathbf{y}_k^{(m)}) \}_{m=1}^{M}$}
  \For{$m = 1$ \To $M$}
    \State Compute latent variables: $\mathbf{z}^{(m)}_k=f_{\widehat{\boldsymbol{\phi}}_k}(\mathbf{x}_k^{(m)};\mathbf{y}_k^{(m)})$
    \State Apply prior-to-posterior map: $\widehat{\mathbf{x}}_k^{(m)}=f^{-1}_{\widehat{\boldsymbol{\phi}}_k}(\mathbf{z}_k^{(m)};\mathbf{y}_k^{\mathrm{obs}})$
  \EndFor 
  \Return $\{\widehat{\mathbf{x}}^{(m)}_k\}_{m=1}^M$
\EndProcedure
\end{algorithmic}
\end{algorithm}

\begin{figure}

\centering{

\includegraphics[width=1\textwidth,height=\textheight]{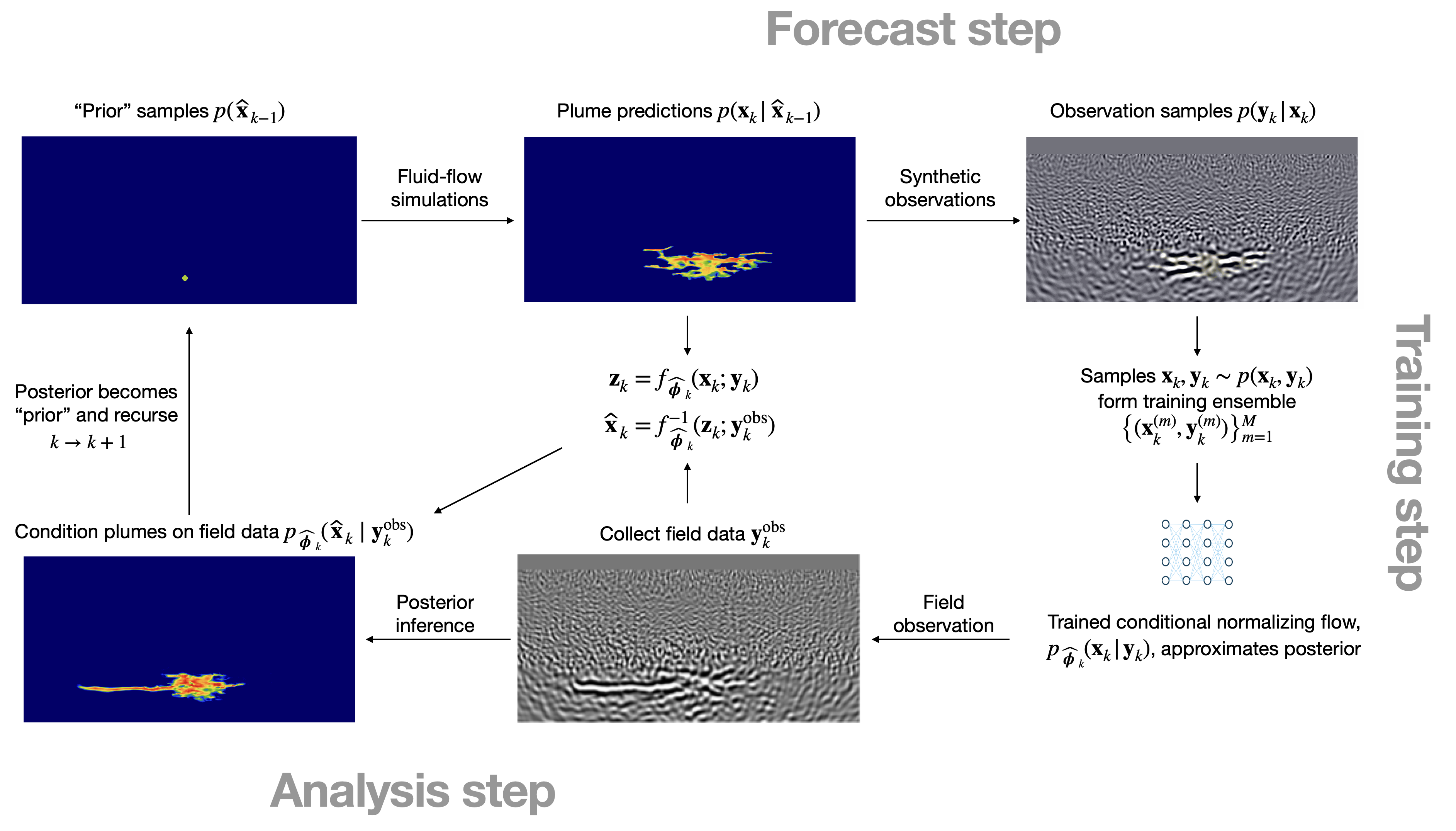}

}

\caption{\label{fig-life-cycle}The life cycle of the recurrent Digital
Shadow at timestep \(k=1\). Starting from the initial randomly chosen
CO\textsubscript{2} saturation,
\(\widehat{\mathbf{x}}_0\sim p(\mathbf{x}_0)\), samples from the joint
distribution,\(\mathbf{y}_{k}\sim p(\mathbf{x}_{k}, \mathbf{y}_{k})\),
are simulated during the \(\textsc{Forecast}\) step, consisting of
fluid-flow and observation simulations. Samples from the joint
distribution form a simulated training ensemble,
\(\{\mathbf{x}_k^{(m)},\mathbf{y}_k^{(m)} \}_{m=1}^{M}\), consisting of
\(M\) training pairs that are used to train the CNF,
\(p_{\widehat{\boldsymbol{\phi}}_k}(\mathbf{x}_{k} |\mathbf{y}_{k})\),
which approximates the posterior. After training, the predicted plume
is, during the \(\textsc{Analysis}\) step, conditioned on the observed
field data, \(\mathbf{y}^{\mathrm{obs}}_k\), and samples from the
CO\textsubscript{2} plume posterior distribution,
\(p_{\widehat{\boldsymbol{\phi}}_k}(\widehat{\mathbf{x}}_{k} \mid\mathbf{y}_{k}^{\mathrm{obs}})\),
are produced. These samples for the state are used as ``priors'' for the
next time step. The symbol \(\widehat{\quad}\) is used to distinguish
between predicted ``digital states'' and analyzed states, conditioned on
observed field data, \(\mathbf{y}^{\mathrm{obs}}_k\). During the
\(\textsc{Analysis}\) step, the predicted states are corrected by the
mapping \(\mathbf{x}_k\mapsto \widehat{\mathbf{x}}_k\).}

\end{figure}

To illustrate the different stages of the Digital Shadow's recursions,
figure~\ref{fig-life-cycle} is included to depict its ``lifecycle'' at
the first timestep, \(k=1\). The lifecycle entails the following stages.
Samples for the initial state of the CO\textsubscript{2} saturation,
denoted with a slight abuse of notation by
\(\widehat{\mathbf{x}}^{(m)}_0\sim p(\mathbf{x}_0)\), are drawn first
from the uniform distribution,
\(\widehat{\mathbf{x}}^{(m)}_0\sim \mathcal{U}(0.1,\, 0.6)\)
(figure~\ref{fig-life-cycle} (top-left)). By simulating the
state-dynamics equation~\ref{eq-dynamics} (figure~\ref{fig-life-cycle}
top-row middle), predictions for the state at the next timestep are
sampled,
\(\mathbf{x}_{k}\sim p(\mathbf{x}_{k}| \widehat{\mathbf{x}}_{k-1})\).
Given these predicted samples for the state, corresponding samples for
the observations are simulated with equation~\ref{eq-obs}
(figure~\ref{fig-life-cycle} top-row right). These two types of
simulations make up the \(\textsc{Forecast}\) step, which produces the
predicted training ensemble,
\(\bigl\{\mathbf{x}_{k}, \mathbf{y}_{k}\bigr\}_{m=1}^M\), consisting of
\(M\) samples drawn from the joint distribution,
\(\mathbf{y}_{k}\sim p(\mathbf{x}_{k}, \mathbf{y}_{k})\). This predicted
ensemble serves as input to the \(\textsc{Training}\) step
(figure~\ref{fig-life-cycle} bottom right) during which the weights of
the Digital Shadow's neural networks are trained by minimizing
equation~\ref{eq-max-likelihood}. This produces the optimized network
weights, \(\widehat{\boldsymbol{\phi}}_k\), at timestep, \(k\). After
collecting field observations, \(\mathbf{y}_k^{\mathrm{obs}}\),
posterior inference is conducted with the neural posterior density
estimator,
\(p_{\widehat{\boldsymbol{\phi}}_k}(\widehat{\mathbf{x}}_{k} \mid\mathbf{y}_{k}^{\mathrm{obs}})\),
conditioned on the observed field data, \(\mathbf{y}_k^{\mathrm{obs}}\).
To avoid finite sample size and training effects during the
\(\textsc{Analysis}\) step (figure~\ref{fig-life-cycle} lower left), a
sample from the posterior is drawn by evaluating the CNF in the forward
latent-space direction on the predicted pair,
\((\mathbf{x}_{k}, \mathbf{y}_{k})\), yielding
\(\mathbf{z}_k = f_{\widehat{\boldsymbol{\phi}}_k}(\mathbf{x}_k;\mathbf{y}_k)\).
This operation is followed by running the CNF, conditioned on the
observed field data, \(\mathbf{y}_{k}^{\mathrm{obs}}\), in the reverse
direction, yielding the sample
\(\widehat{\mathbf{x}}_k= f^{-1}_{\widehat{\boldsymbol{\phi}}_k}(\mathbf{z}_k;\mathbf{y}_k^{\mathrm{obs}})\).
By design, this sample approximates a sample from the true posterior,
\(\mathbf{x}_k\sim p(\mathbf{x}_k\mid \mathbf{y}_k^{\mathrm{obs}})\). At
the next timestep, this sample becomes a sample for the
``prior''---i.e., for \(k\rightarrow k+1\), and the recursions proceed
by repeating themselves for all, \(K\), timesteps. By comparing
figure~\ref{fig-life-cycle}'s sample for the plume prediction,
\(\mathbf{x}_{k}\sim p(\mathbf{x}_{k}| \widehat{\mathbf{x}}_{k-1})\),
with the corresponding analyzed sample conditioned on the observed field
data,
\(\widehat{\mathbf{x}}_{k} \sim p_{\widehat{\boldsymbol{\phi}}_k}(\widehat{\mathbf{x}}_{k} \mid\mathbf{y}_{k}^{\mathrm{obs}})\),
it becomes apparent that the predicted state produced by the
\(\textsc{Forecast}\) step undergoes a significant correction during the
\(\textsc{Analysis}\) step. This mapping during the
\(\textsc{Analysis}\) step between the predicted and inferred
states---i.e., \(\mathbf{x}_k\mapsto \widehat{\mathbf{x}}_k\), derives
from the invertibility of CNFs and the implicit conditioning of the
predictions on \(\mathbf{y}^{\mathrm{obs}}_{k-1}\) (via
\(\widehat{\mathbf{x}}_{k-1}\)) and explicit conditioning on
\(\mathbf{y}^{\mathrm{obs}}_k\). Despite the fact that the state at the
previous timestep is conditioned on \(\mathbf{y}^{\mathrm{obs}}_{k-1}\),
the prediction varies considerably. This variability is caused by the
unknown permeability field, which appears as a stochastic nuisance
parameter in the dynamics (cf. equation~\ref{eq-dynamics}), and explains
the significant time-lapse data-informed correction performed by the
\(\textsc{Analysis}\) step. The importance of the correction is nicely
illustrated by the example in figure~\ref{fig-life-cycle} where thanks
to the correction the plume is extended significantly in the leftward
direction, a feature completely missed by the predicted state.

By running through the Digital Shadow's lifecycle \(k\) times
recursively for an ensemble consisting of \(M\) particles, the
\(\textsc{Prediction}\) and \(\textsc{Analysis}\) steps approximate
samplings from \(p(\mathbf{x}_k\mid \mathbf{y}^{\mathrm{obs}}_{1:k-1})\)
and \(p(\mathbf{x}_k\mid \mathbf{y}^{\mathrm{obs}}_{1:k})\) of equations
\ref{eq-prediction} and \ref{eq-Bayes}. Because the proposed posterior
density estimation scheme recurrently trains networks, it can be
considered as an instance of neural sequential Bayesian inference, which
forms the basis of our Digital Shadow. Before going into further details
on how the simulations are performed, we introduce the important concept
of \emph{summary statistics}, which simplifies the observable-to-state
mapping, renders typically non-uniform seismic data space uniform, and
also learns to detect features that inform the posterior.

\subsection{Physics-based summary
statistics}\label{physics-based-summary-statistics}

Despite the adequate expressiveness of CNF's neural architecture based
on cINNs (Ardizzone et al. 2019b), the mapping of CO\textsubscript{2}
plume-induced acoustic changes to seismic shot records remains extremely
challenging because of the large degrees of freedom of Earth models and
seismic shot records (Cranmer, Brehmer, and Louppe 2020), which live in
three dimensions and five dimensions for 2- and 3D- Earth models,
respectively. This challenge is compounded by the relatively weak signal
strength of time-lapse changes, the complexity of the time-lapse changes
to seismic observations map, and the fact that raw seismic shot records
are collected non-uniformly---i.e.~different seismic surveys may have
different numbers of shot records or shot records may vary in the number
of traces (receivers), time samples, etc. To overcome these issues, we
introduce a hybrid approach, consisting of physics-based summary
statistics (Radev et al. 2020; Orozco et al. 2023), and a trainable
U-net (Ronneberger, Fischer, and Brox 2015) that forms the conditional
arm of the cINN (Ardizzone et al. 2019a). By mapping the shot records to
the image domain, uniformity is assured because the discretization can
be controlled in that domain. Moreover, the mapping from the image
domain to the velocity model is much simpler, so the conditional arm of
the cINN can dedicate itself to learning image-domain features that
maximally inform the posterior while the network's invertible arm can
concentrate on learning how to produce samples of the posterior from
realizations of standard Gaussian noise conditioned on these features.
More importantly, Alsing, Wandelt, and Feeney (2018) also showed that
under certain conditions the gradient of the score function (the
\(\log\)-likelihood) compresses physical data optimally. Motivated by
this work, Orozco et al. (2023) demonstrated that information in the
original shot records can be preserved during pre-stack reverse-time
migration (RTM) as long as the migration-velocity model is sufficiently
accurate. Yin, Orozco, et al. (2024) showed that this condition can be
relaxed in the WISE and WISER (Yin, Orozco, and Herrmann 2024)
frameworks. The network weights for the U-net in the conditional arm are
absorbed in the network weights, \(\boldsymbol{\phi}\), of the CNFs, and
trained during the minimization of the objective in
equation~\ref{eq-loss-CNF}.

To meet the challenges of imaging subtle time-lapse changes induced by
the replacement of brine by supercritical CO\textsubscript{2}, we
calculate the residual with respect to a reference model,
\(\mathbf{\bar{m}}_k\). To avoid tomographic artifacts in the gradient,
the residual is migrated with the inverse-scattering imaging condition
(Stolk, Hoop, and Op't Root 2009; Douma et al. 2010; Whitmore and
Crawley 2012; P. Wang et al. 2024). This corresponds to computing
sensitivities with respect to the acoustic impedance for a constant
density. Given this choice, the observation operator linking the state
of the CO\textsubscript{2} plume to migrated time-lapse seismic images
is given by

\begin{equation}\phantomsection\label{eq-image-observed}{
\begin{aligned}
\mathbf{y}_k&=\mathcal{H}_k(\mathbf{x}_k,\boldsymbol{\epsilon}_k, \mathbf{\bar{m}}_k)\\
&:=\Bigl({\mathbf{J}_{\mathcal{F}[\mathbf{v}]}}\Bigr |_{{\mathbf{\widetilde{v}}_k}}\Bigr)_{\mathsf{ic}}^{\top} \Bigl(\mathcal{F}[\mathbf{\bar{m}}_k]-\mathbf{d}_k\Bigr)
\end{aligned}
}\end{equation}

In this expression, the linear operator
\(\Bigl({\mathbf{J}_{\mathcal{F}[\mathbf{v}]}}\Bigr |_{{\mathbf{\widetilde{v}}_k}}\Bigr)_{\mathsf{ic}}\)
corresponds the Jacobian of the forward wave operator,
\(\mathcal{F}[\mathbf{m}]\), calculated at \(\mathbf{\widetilde{v}}_k\),
which is a smoothed version of the reference velocity model,
\(\mathbf{\bar{v}_k}\). The symbol \(^\top\) denotes the adjoint.
Together with the reference densities, the reference wavespeeds are
collected in the seismic reference model, \(\bar{\mathbf{m}}_k\). The
nonlinear forward model, \(\mathcal{F}\), represents the active-source
experiments part of the survey. Use of the inverse-scattering imaging
condition is denoted by the subscript, \(\mathsf{ic}\) in the definition
of the reverse-time migration operator,
\(\Bigl({\mathbf{J}_{\mathcal{F}[\mathbf{v}]}}\Bigr |_{{\mathbf{\widetilde{v}}_k}}\Bigr)^\top_{\mathsf{ic}}\).
Finally, the vector, \(\mathbf{d}_k\), collects the simulated shot
records at time-lapse timestep, \(k\), for all active-source experiments
and is simulated with

\begin{equation}\phantomsection\label{eq-forward-seismic}{
\mathbf{d}_k=\mathcal{F}[\mathbf{m}_k]+\boldsymbol{\epsilon}_k,\quad \text{where} \quad \mathbf{m}_k = \mathcal{R}({\mathbf{m}_0, \mathbf{x}_k}).
}\end{equation}

In this nonlinear forward model at timestep, \(k\), the
CO\textsubscript{2}saturation induced corrected time-lapse seismic
medium properties---i.e., the wavespeed, \(\mathbf{v}_k\), and the
density, \(\boldsymbol{\rho}_k\), are computed with respect to the
baseline, \(\mathbf{m}_0\), via
\(\mathbf{m}_k = \mathcal{R}({\mathbf{m}_0, \mathbf{x}_k})\). This
nonlinear operator, \(\mathcal{R}(\cdot,\cdot)\), captures the rock
physics, which models how changes in the state---i.e., the
CO\textsubscript{2} saturation, map to changes in the seismic medium
properties, \(\mathbf{m}_k\). Because ``observed field'' seismic
time-lapse data is assumed to come from the same forward model,
equation~\ref{eq-image-observed} will also be used to \emph{in silico}
simulate \(\mathbf{y}_k^{\mathrm{obs}}\), simply by replacing the
simulated time-lapse shot records, \(\mathbf{d}_k\), by the simulated
observed time-lapse seismic shot records,
\(\mathbf{d}_k^{\mathrm{obs}}\), that stand for data collected in the
field.

Given the exposition of the amortized Bayesian inference, the neural
sequential simulation-based Bayesian inference, and physics-based
summary statistics, we are now in a position to setup our Digital Shadow
for GCS prior to conducting our case study.

\section{\texorpdfstring{Digital Shadow for CO\textsubscript{2} storage
monitoring}{Digital Shadow for CO2 storage monitoring}}\label{sec-shadow}

To validate our Digital Shadow, an \emph{in silico} prototype is
developed and implemented for the off-shore setting of the South-Western
North Sea, an area actively considered for GCS (ETI 2016). For this
purpose, we consider the 2D cross-section of the Compass model (E. Jones
et al. 2012) and the experimental setup of figure~\ref{fig-acquisition},
which consists of sparse ocean-bottom nodes, dense sources, an
injection, and a monitor well. Based on the velocity model in
figure~\ref{fig-acquisition}, a probabilistic model for reservoir
properties (permeability) will be derived first, followed by the
establishment of the ground truth for the \emph{in sillico} simulated
states and associated observed time-lapse seismic and well data, which
serve as input to our Digital Shadow for CO\textsubscript{2} storage
monitoring.

\subsection{Probabilistic baseline model for the permeability
field}\label{sec-baseline}

\begin{figure}

\begin{minipage}{0.50\linewidth}

\includegraphics[width=0.9\textwidth,height=\textheight]{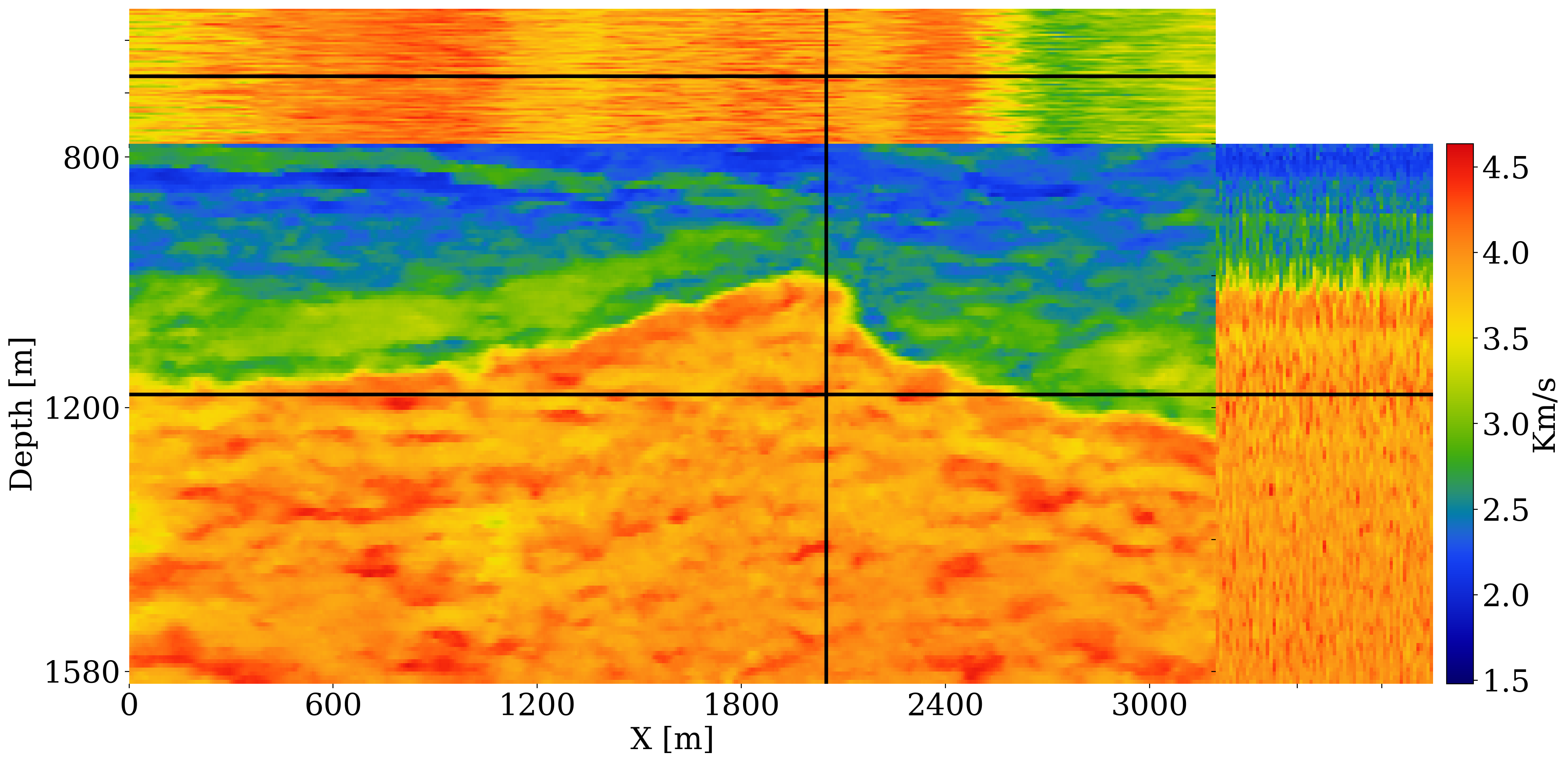}

\subcaption{\label{}a.}
\end{minipage}%
\begin{minipage}{0.50\linewidth}

\includegraphics[width=0.9\textwidth,height=\textheight]{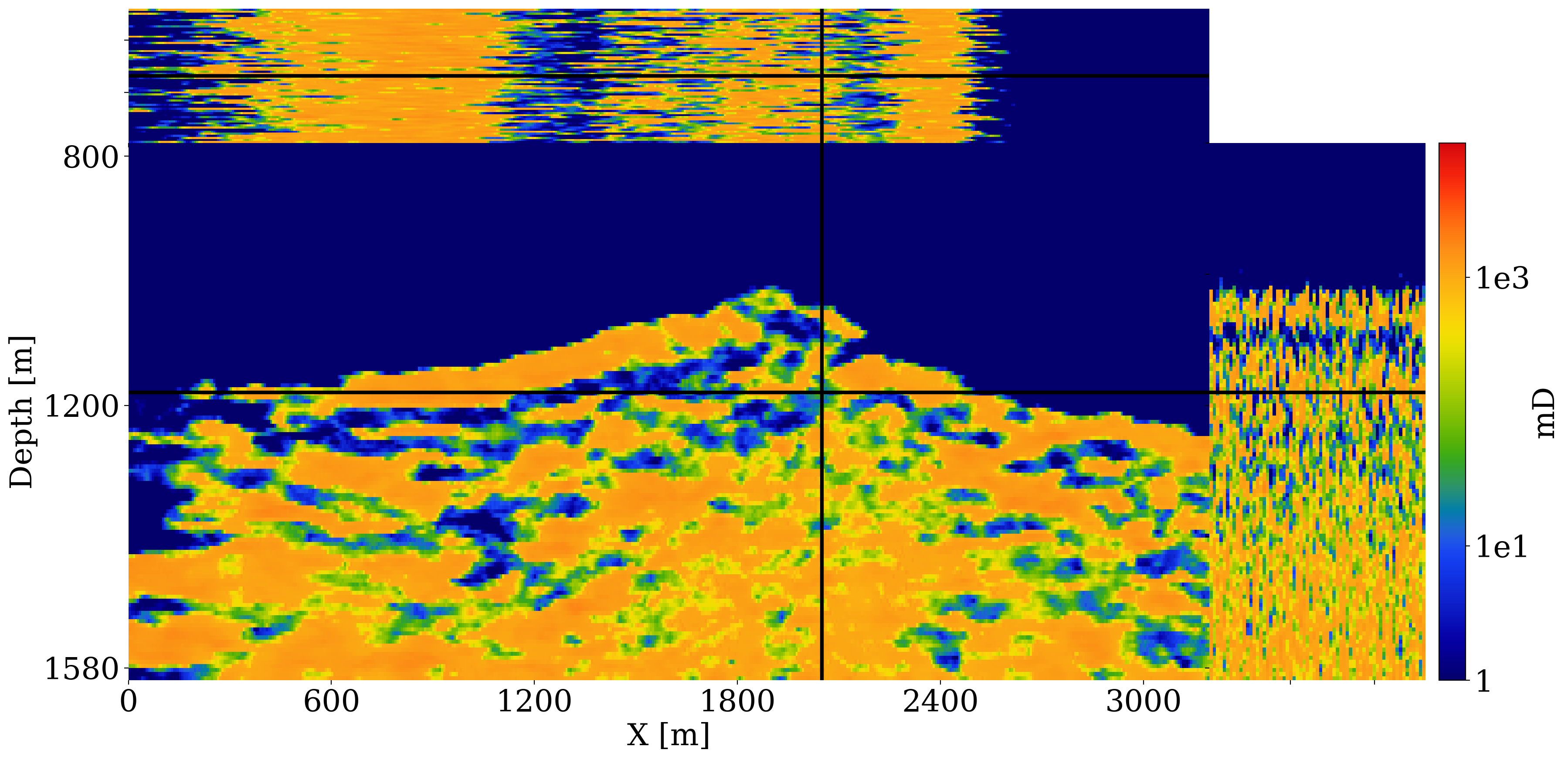}

\subcaption{\label{}b.}
\end{minipage}%

\caption{\label{fig-samples-wise}Samples of the probabilistic baseline
model for the permeabily field. \emph{(a)} samples from the posterior
for the wavespeeds,
\(\mathbf{v}_0\sim p(\mathbf{v}_0\mid \mathbf{y}_0^{\mathrm{obs}})\),
produced by WISE (Yin, Orozco, et al. 2024); \emph{(b)} samples of the
permeability field, \(\boldsymbol{\kappa}\sim p(\boldsymbol{\kappa})\),
derived from the velocities with equation~\ref{eq-conversion}.}

\end{figure}

To mimic a realistic CO\textsubscript{2} storage project where reservoir
properties are not known precisely, a stochastic baseline for the
permeability field is established. To this end, the velocity model
depicted in figure~\ref{fig-acquisition} is used to generate noisy
baseline seismic data (signal-to-noise ratio S/N \(12\mathrm{dB}\)).
This seismic baseline dataset serves as input to full-waveform
variational Inference via Subsurface Extensions (WISE (Yin, Orozco, et
al. 2024)). Given a poor 1D migration-velocity model and noisy shot
records, WISE draws samples from the posterior distribution for the
velocity. These samples for the wavespeed conditioned on shot data are
plotted in figure~\ref{fig-samples-wise} (a) and are converted to the
permeability fields plotted in figure~\ref{fig-samples-wise} (b). This
elementwise velocity-to-permeability conversion is done with the
following nonlinear elementwise mapping:

\begin{equation}\phantomsection\label{eq-conversion}{
\boldsymbol{\kappa}=\begin{cases}
\boldsymbol{\zeta}\odot\exp_\circ(\mathbf{v}-4.5)\oslash \exp_\circ(4.0-4.3) \quad &\mathrm{if}\quad \mathbf{v}\succcurlyeq 4.0\\
0.01 \exp_\circ\bigl((\log_\circ (\boldsymbol{\zeta}/0.01)/0.5)\odot(\mathbf{v}-3.5)\bigr) \quad &\mathrm{if}\quad \mathbf{v}\succcurlyeq 3.5 \\
0.01\exp_\circ(\mathbf{v}-1.5)\oslash\exp_\circ(3.5-1.5) \quad &\mathrm{else.}
\end{cases}
}\end{equation}

In this expression, the symbols \(\odot\),\(\oslash\), \(\exp_\circ\),
and \(\log_\circ\), refer to elementwise multiplication, division,
exponentiation, and logarithm. The conversion itself consists of
elementwise exponential scaling and randomization with
\(\boldsymbol{\zeta}\sim\mathrm{smooth}\bigl(\mathcal{N}(1000, 1200)\bigr)\),
where is \(\boldsymbol{\zeta}\) a realization of the standard Normal
distribution smoothed with Gaussian kernel of width \(5\mathrm{m}\). The
conversion is also controlled by elementwise (denoted by the elementwise
greater or equal symbol \(\succcurlyeq\)) operations, which involve the
wavespeeds. These inequalities are chosen such that the permeability
values are high (between \(1600 \mathrm{mD}\) and \(3000 \mathrm{mD}\))
for regions where the wavespeeds are above \(4.0\mathrm{km/s}\). These
regions correspond to clean sandstones. Areas with wavespeeds that vary
within the interval \(3.5-4.0\mathrm{km/s}\) stand for dirty sands with
intermediate permeability values between \(10 \mathrm{mD}\) and
\(1600 \mathrm{mD}\) while areas with lower wavespeeds correspond to
shales with the lowest permeability values of \(0.001 \mathrm{mD}\).

Because of the noise and poor migration-velocity model, the samples for
the baseline velocity models produced by WISE (see
figure~\ref{fig-samples-wise} (a) for a subset of these samples) exhibit
strong variability. This variability amongst the samples of the baseline
\(\mathbf{v}\sim p(\mathbf{v}\mid \mathbf{y}_0^{\mathrm{obs}})\)
produces, after conversion by equation~\ref{eq-conversion}, highly
variable samples for the permeability field,
\(\boldsymbol{\kappa}\sim p(\boldsymbol{\kappa})\), included in
figure~\ref{fig-samples-wise} (b). Because the permeability values are
small to extremely small in the primary and secondary seals, they are
denoted by solid blue-color overburden areas in
figure~\ref{fig-samples-wise} (b). These plots exhibit high variability,
especially in areas where the permeable sandstones are interspersed with
low-permeable shales and mudstones. This high degree of variability in
the permeability field reflects our lack of knowledge of this crucial
reservoir property, which complicates CO\textsubscript{2} plume
predictions based on reservoir simulations alone. Our Digital Shadow is
aimed at demonstrating how time-lapse data can be used to condition the
simulations, so the ground-truth CO\textsubscript{2} plume can be
estimated more accurately from monitoring data.

Geologically, the considered CO\textsubscript{2}-injection site
corresponds to a proxy for a storage complex in the South-Western corner
of the North Sea with a reservoir made of a thick stack of relatively
clean highly-permeable Bunter sandstone (Kolster et al. 2018). The study
area corresponds to a 2D slice selected from the synthetic Compass model
(E. Jones et al. 2012), which itself was created from imaged seismic and
well-log data collected in a region under active consideration for
Geological Carbon Storage (GCS) by the North-Sea Transition Authority
(ETI 2016). The seals of the storage complex comprise the Rot Halite
Member, as a primary seal about \(50 \mathrm{m}\) thick, situated
directly on top of the Blunt sandstone, and the Haisborough Group, as
the secondary seal over \(300\mathrm{m}\) thick consisting of
low-permeability mudstones.

The velocities of the 3D Compass model were also used to train the
networks of WISE (Yin, Orozco, et al. 2024). The reader is referred to
Yin, Orozco, et al. (2024) for further details on the training of these
neural networks. To create a ground-truth dataset, our simulation-based
workflow follows the numerical case study by Erdinc et al. (2022) and
Yin, Erdinc, et al. (2023), where realistic time-lapse seismic surveys
were generated in response to CO\textsubscript{2} injection in a saline
aquifer.

\subsection{\texorpdfstring{\emph{In silico} ground-truth
simulations}{In silico ground-truth simulations}}\label{sec-ground-truth}

The Digital Shadow will be validated against unseen ground-truth
multimodal simulations. At this early stage of development, we opt for
\emph{in silico} simulations for the ground-truth
CO\textsubscript{2}-plume dynamics and the seismic images of the induced
time-lapse changes. This choice, which is not an uncommon practice in
areas such as climate modeling where future data does not exist, also
offers more control and the ability to validate with respect to a ground
truth to which there is no access in practice. The fluid-flow
simulations themselves are carried out over \(24\) timesteps of \(80\)
days each. The time-lapse seismic images are derived from four sparsely
sampled noisy ocean-bottom node surveys, collected over a period of
\(1920=4\times 480\) days.

\subsubsection{Multi-phase flow
simulations}\label{multi-phase-flow-simulations}

\begin{figure}

\centering{

\includegraphics{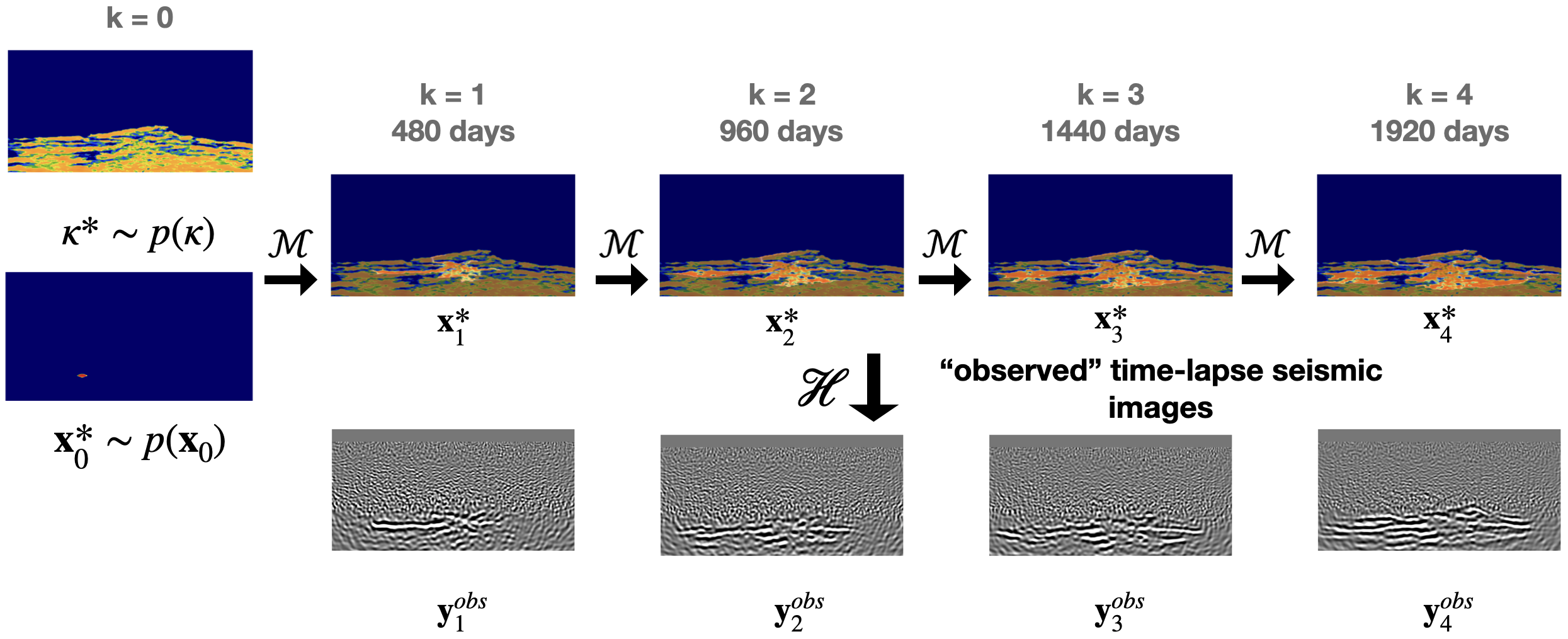}

}

\caption{\label{fig-Ground_truth}Schematic diagram showing ground-truth
data generation using a fixed realization for the permeability field
(CO\textsubscript{2} saturation shown in bright overlay on top of plots
for permeability field).}

\end{figure}

Figure \ref{fig-Ground_truth} contains a schematic overview on how the
ground-truth flow simulations are initialized by randomly sampling the
ground-truth (denoted by the symbol \(^\ast\)) permeability field,
\(\boldsymbol{\kappa}^\ast\), and the initial state for the
CO\textsubscript{2} plume at timestep, \(k=0\)---i.e.~the
CO\textsubscript{2} saturation is sampled,
\(\widehat{\mathbf{x}}^{\ast}_0\sim \mathcal{U}(0.2,\, 0.6)\). The
ground-truth anisotropic permeability field, \(\boldsymbol{\kappa}\),
consists of \(512\times 256\) gridpoints with a grid spacing of
\(6.25\mathrm{m}\) and has a vertical-to-horizontal permeability ratio
of \(\kappa_v/\kappa_h = 0.36\). The reservoir extends \(100\mathrm{m}\)
in the perpendicular \(y\)-direction. Supercritical CO\textsubscript{2}
is injected at a depth of \(1200\mathrm{m}\) over an injection interval
of \(37.5 \mathrm{m}\) for the \(1920\) day duration of the project with
an injection rate of \(0.0500\mathrm{m^3/s}\), pressurizing the
reservoir from a maximum pressure of \(9.7e6\) to \(1.1e7\mathrm{Pa}\).
The total amount of injected CO\textsubscript{2} is \(5.8 \mathrm{Mt}\)
with \(1.1. \mathrm{Mt}\) injected annually. The hydrostatic pressure at
the injection depth of \(1200\mathrm{m}\) is well below the depth at
which CO\textsubscript{2} becomes critical.

The resulting multi-phase fluid-flow simulations (Nordbotten and Celia
2011) by the reservoir simulator (Krogstad et al. 2015; Settgast et al.
2018; Rasmussen et al. 2021; Stacey and Williams 2017; Møyner et al.
2024) are collected in the vector, \(\mathbf{x}^\ast_{1:24}\), see
figure~\ref{fig-Ground_truth} top-row for the time-lapse states,
\(\mathbf{x}^\ast_{1:6:24}\) (\(1:6:24\) means taking each
\(6^{\mathrm{th}}\) sample). The simulations themselves are conducted
with an implicit method implemented with the open-source software
package
\href{https://github.com/sintefmath/JutulDarcy.jl}{JutulDarcy.jl}
(Møyner and Bruer 2023). This industry-scale reservoir modeling package
simulates how the injected CO\textsubscript{2} gradually displaces brine
within the connected pore space of rocks in the storage complex.
Initially, the reservoir rocks are filled with brine. Both brine and
CO\textsubscript{2} residual saturations are set to \(10\%\) each. As
can be seen from figure~\ref{fig-Ground_truth}, the CO\textsubscript{2}
plume follows regions of high permeability, forming extensive fingers in
the lateral direction. Because the density of critical
CO\textsubscript{2} is smaller than that of brine
(\(700\mathrm{kg/m^3}\) for CO\textsubscript{2} versus
\(1000\mathrm{kg/m^3}\) for water) buoyancy effects drive the
CO\textsubscript{2} plume upwards.

\subsubsection{Seismic simulations}\label{seismic-simulations}

To convert changes in the reservoir's state due to CO\textsubscript{2}
injection to modifications in the seismic properties, a nonlinear rock
physics model, \(\mathcal{R}\), is introduced as in Yin, Erdinc, et al.
(2023) and Louboutin, Yin, et al. (2023). At each of the \(K=4\)
time-lapse time instances, changes in the CO\textsubscript{2} saturation
are related to differences in the seismic properties by this model. When
ignoring velocity changes due to pressure changes and assuming a
heterogeneous distribution of the CO\textsubscript{2}, the patchy
saturation model (Avseth, Mukerji, and Mavko 2010) can be used to relate
increases in CO\textsubscript{2} saturation to reductions in the
acoustic wavespeed and density. Given baseline wavespeeds and densities
gathered in the seismic medium properties vector,
\(\mathbf{m}_0=(\mathbf{v}_0,\boldsymbol{\rho}_0)\), the rock physics at
each timestep, \(k\), reads

\begin{equation}\phantomsection\label{eq-rock-physics}{
\mathbf{m}_k=\mathcal{R}(\mathbf{m}_0,\mathbf{x}_k),\, k=1\cdots K.
}\end{equation}

\begin{figure}

\centering{

\includegraphics{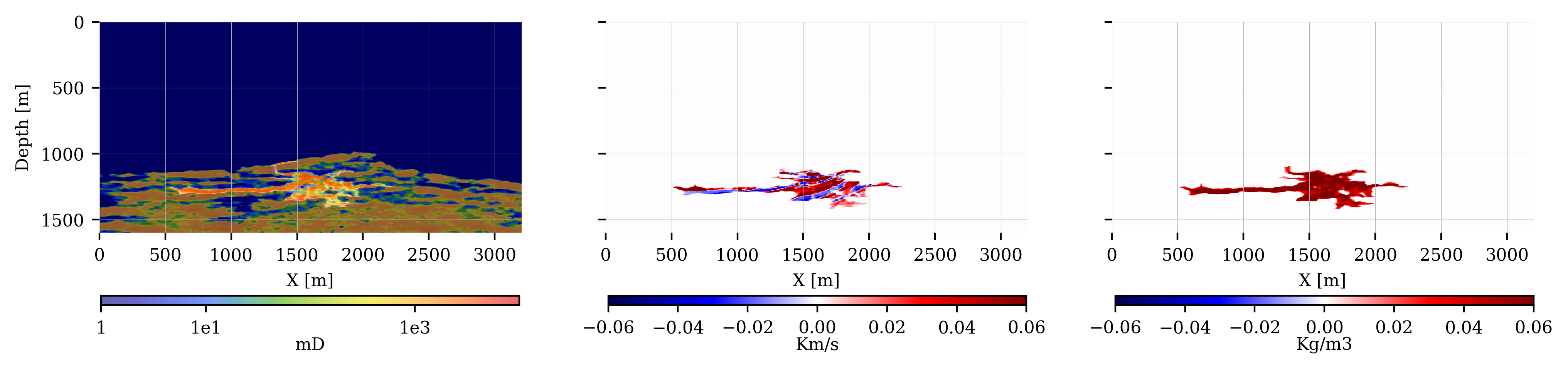}

}

\caption{\label{fig-velrho}Seismic property changes induced by
CO\textsubscript{2} injection where brine is replaced by super-critial
CO\textsubscript{2}. Left: CO\textsubscript{2} saturation shown as
bright overlay on top of the permeability field. Middle: time-lapse
changes in compressional wavespeed and Right: in density due to
CO\textsubscript{2} injection.}

\end{figure}

\begin{figure}

\begin{minipage}{0.25\linewidth}

\includegraphics[width=0.9\textwidth,height=\textheight]{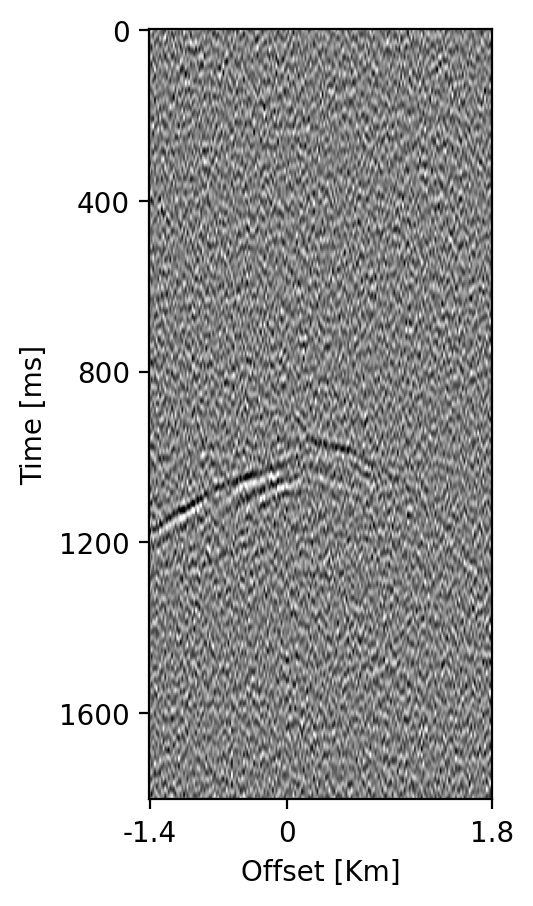}

\subcaption{\label{}\(k=1\) with \(\mathrm{NRMS}=2.56\)}
\end{minipage}%
\begin{minipage}{0.25\linewidth}

\includegraphics[width=0.9\textwidth,height=\textheight]{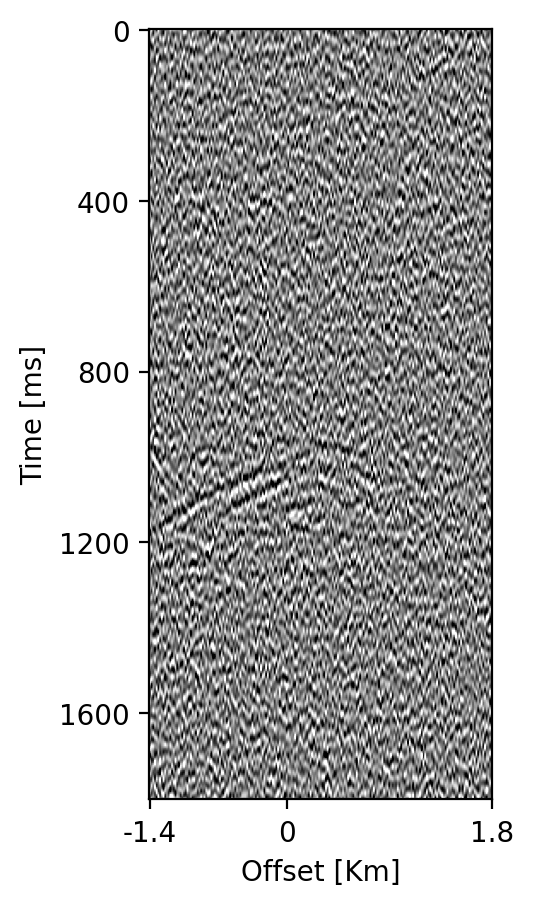}

\subcaption{\label{}\(k=2\) with \(\mathrm{NRMS}=4.60\)}
\end{minipage}%
\begin{minipage}{0.25\linewidth}

\includegraphics[width=0.9\textwidth,height=\textheight]{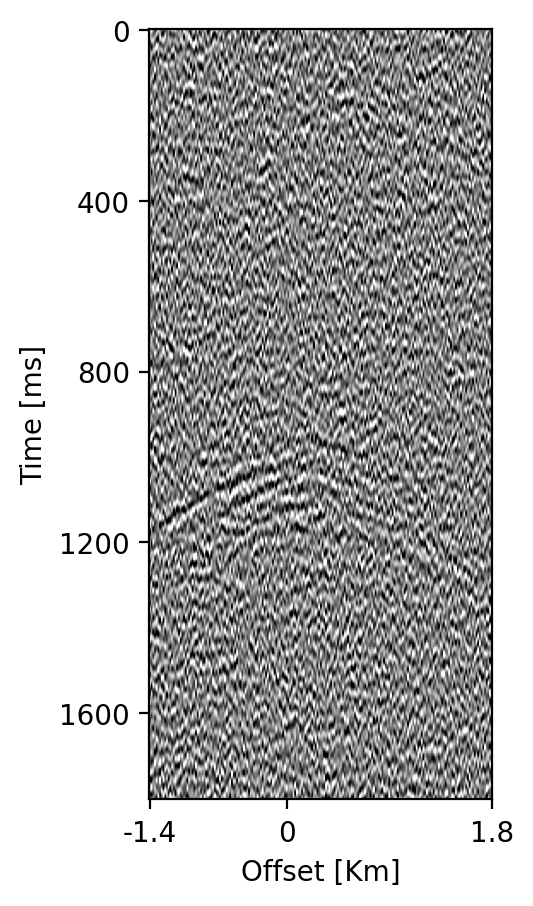}

\subcaption{\label{}\(k=3\) with \(\mathrm{NRMS}=4.60\)}
\end{minipage}%
\begin{minipage}{0.25\linewidth}

\includegraphics[width=0.9\textwidth,height=\textheight]{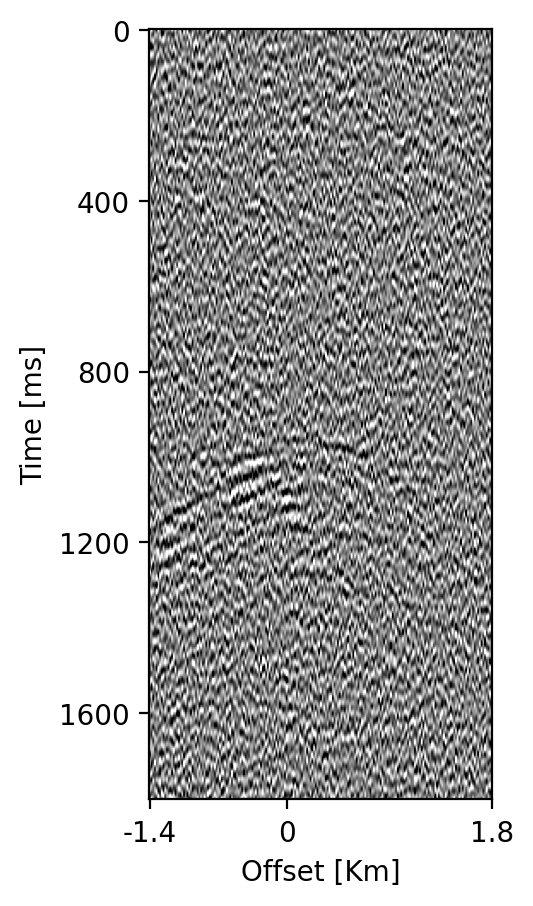}

\subcaption{\label{}\(k=4\) with \(\mathrm{NRMS}=4.61\)}
\end{minipage}%

\caption{\label{fig-shotrecord}Time-lapse difference seismic shot
records computed with respect to data generated with the reference
model. Except for the first timestep, the NRMS values remain relatively
constant.}

\end{figure}

Figure~\ref{fig-velrho} includes an example of how changes in the
CO\textsubscript{2} saturation at timestep, \(k=1\), are mapped to
changes in the acoustic wavespeed and density. In turn, these time-lapse
changes in the seismic properties lead to changes in the simulated shot
records as shown in figure~\ref{fig-shotrecord}, which depicts
time-lapse seismic differences with respect to data generated with the
reference model for \(k=1\cdots 4\). Shot records are simulated by
solving the acoustic wave equation for varying density and acoustic
wavespeeds using the open-source software package
\href{https://github.com/slimgroup/JUDI.jl}{JUDI.jl}, a Julia Devito
Inversion framework (P. A. Witte et al. 2019; Louboutin, Witte, et al.
2023), which uses highly optimized time-domain finite-difference
propagators of \href{https://www.devitoproject.org/}{Devito} (Louboutin
et al. 2019; Luporini et al. 2020).

The four ground-truth time-lapse seismic surveys are generated after
\(480, 960, 1440, \mathrm{and}\, 1920\) days since the start of the
CO\textsubscript{2} injection. There is no free surface. Throughout the
fixed acquisition setup, \(200\) shot records with airgun locations
sampled at \(16\mathrm{m}\) interval are collected sparsely by eight
Ocean Bottom Nodes (OBNs) with a receiver interval of \(400\mathrm{m}\),
yielding a maximum offset of \(3.2\mathrm{km}\). During the seismic
simulations, a Ricker wavelet with a dominant frequency of
\(15\mathrm{Hz}\) is used, and band-limited noise with a signal-to-noise
ratio (SNR) of \(28.0 \mathrm{dB}\) is added to the \(1800\mathrm{ms}\)
recordings sampled at \(4\mathrm{ms}\). Due to the complexity of the
model and time-lapse changes, the time-lapse differences with respect to
the reference model are noisy and have normalized root-mean-square
(NRMS) values of \(2.65,\, 4.60,\, 4.60\) and \(4.61\) after
\(480, 960, 1440, \mathrm{and}\, 1920\) days.

Ground-truth time-lapse shot records,
\(\mathbf{d}^{\mathrm{obs}}_{1:K}\) serve as input to the time-lapse
imaging scheme, detailed below, which produces the ground-truth
time-lapse images, \(\mathbf{y}^{\mathrm{obs}}_{1:K}\), depicted in the
bottom row of figure~\ref{fig-Ground_truth}. Despite the strong
artifacts, consisting of epistemic linearization errors and aleatoric
observation noise, these time-lapse seismic images display a distinct
imprint of the growing CO\textsubscript{2} plume. How these images are
formed and used to condition estimates for the dynamic state, will be
discussed next.

\subsection{\texorpdfstring{CO\textsubscript{2} monitoring with the
Digital
Shadow}{CO2 monitoring with the Digital Shadow}}\label{co2-monitoring-with-the-digital-shadow}

Amortized posterior inference stands at the basis of our Digital Shadow.
As described in Section~\ref{sec-amortized}, training CNFs requires the
creation of predicted training ensembles during the
\(\textsc{Forecast}\) step, which consists of input-output pairs made
out of samples from the ``prior'' and ``likelihood''. Forming these
ensembles is followed by the \(\textsc{Training}\) step during which the
network weights are calculated. Neural inference happens at the final
\(\textsc{Analysis}\) step during which members of the predicted
ensemble for the state are corrected by measurements from the field,
\(\mathbf{y}^{\mathrm{obs}}_{k}\). After exhaustive testing, the size of
the ensemble was chosen to be \(M=128\), a number that will be used
throughout our experiments. Aside from determining the ensemble size,
this number also determines how Monte-Carlo estimates the conditional
mean

\begin{equation}\phantomsection\label{eq-cmean}{
\mathbf{\bar{x}}= \mathbb{E}_{ p(\mathbf{x}\mid\mathbf{y})} \left[ \mathbf{x}\right]\approx \frac{1}{M}\sum_{m=1}^M\mathbf{x}^{(m)}\quad 
}\end{equation}

and conditional variance

\begin{equation}\phantomsection\label{eq-cstdev}{ 
\boldsymbol{\bar{\sigma}}^2 = \mathbb{E}_{ p(\mathbf{x}\mid\mathbf{y})} \Bigl[\bigl(\mathbf{x} - \mathbb{E}_{ p(\mathbf{x}\mid\mathbf{y})}   \bigr )^2\Bigr] \approx \frac{1}{M}\sum_{m=1}^M\bigl(\mathbf{x}^{(m)}-\mathbf{\bar{x}}\bigr)^2
}\end{equation}

are calculated.

\subsubsection{The Forecast step}\label{the-forecast-step}

The creation of realistic simulation-based ensembles forms a crucial
part of our Digital Shadow. Details on how these training ensembles for
the Digital Shadow are assembled are provided next.

\paragraph{Uncertain dynamics}\label{uncertain-dynamics}

\begin{figure}

\centering{

\includegraphics{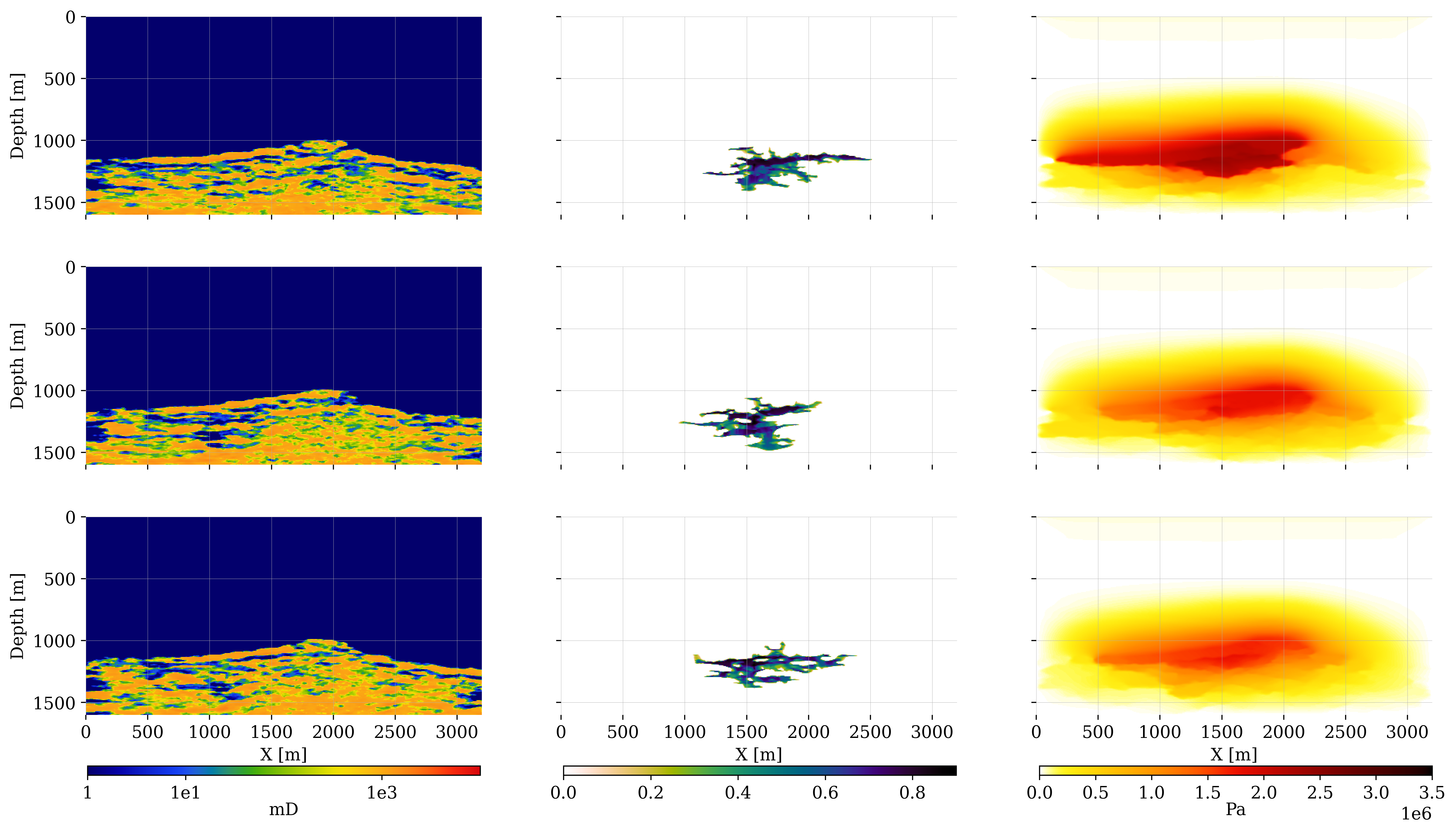}

}

\caption{\label{fig-k-1-K-x}Examples of the predicted state of the
CO\textsubscript{2} plume at timestep \(k=1\). Left column: random
realizations of the permeability field. Middle column: predictions for
the CO\textsubscript{2} concentration. Right column: predictions for the
pressure perturbations. The different realization for the permeability
field changes the shape of both the predicted saturations and pressure
perturbations.}

\end{figure}

The main purpose of the Digital Shadow for CO\textsubscript{2}
monitoring lies in its ability to handle uncertainties, in particular
epistemic uncertainty due to the unknown permeability field, which
throughout the simulations for the dynamics is considered as a random
vector (cf. equation~\ref{eq-dynamics}). This implies that during each
evaluation of the dynamics, \(\mathcal{M}\), a new sample for the
permeability field is drawn,
\(\boldsymbol{\kappa}_k\sim p(\boldsymbol{\kappa})\), according the
procedure describe in Section~\ref{sec-baseline}. Because the
high-permeability zones differ between samples, the injection depth
varies between \(1200-1250\mathrm{m}\). As can be seen from
figure~\ref{fig-k-1-K-x}, the variability amongst the permeabilities,
injection depth, and initial saturation, are all responsible for the
variable CO\textsubscript{2} plume predictions, even after a single
timestep at time \(k=1\). Both the CO\textsubscript{2} saturation and
pressure perturbations differ for each permeability realization. The
predicted CO\textsubscript{2} saturations have complex shapes that vary
significantly. The pressure perturbations vary less, are smooth, and as
expected cover a larger area.

\paragraph{Seismic reference model}\label{sec-reference}

While the different realizations for the permeability field give rise to
different CO\textsubscript{2} saturations, revealing their time-lapse
seismic signature calls for a dedicated RTM imaging procedure with
migration-velocity models, \(\widetilde{\mathbf{v}}_k\), that are
corrected at each time-lapse timestep, and with residuals that are
computed with respect to a carefully chosen reference model. To derive
these migration-velocity and reference models, the forecast ensemble for
the state, \(\{\mathbf{x}^{(m)}_k\}_{m=1}^M\), is used to update the
seismic properties with equation~\ref{eq-rock-physics}. Given this
predicted ensemble for the state, the seismic properties and simulated
shot records are given by

\begin{equation}\phantomsection\label{eq-shot-data-esemble}{
\mathbf{m}_k^{(m)} = \mathcal{R}({\mathbf{{m}}_0}, \mathbf{x}_k^{(m)})\quad \text{and}\quad \mathbf{d}_k^{(m)} = \mathcal{F}(\mathbf{m}_k^{(m)})+\boldsymbol{\epsilon}_k,\quad m=1\cdots M,
}\end{equation}

by virtue of equation~\ref{eq-forward-seismic}.

Since the time-lapse changes are small, and epistemic and aleatoric
uncertainties significant, the challenge is to define for each
time-lapse survey an appropriate reference velocity model. This
reference model serves two purposes. First, it is used to generate
reference shot records with respect to which the time-lapse differences
are calculated. Second, after smoothing, it yields migration-velocity
models, \(\widetilde{\mathbf{v}}_k\), for each timestep.

To compute this reference model, the seismic properties for the
baseline, \(\mathbf{m}_0\), are updated with the rock physics model,
\(\mathbf{\bar{m}}_k = \mathcal{R}({\mathbf{{m}}_0}, \mathbf{\bar{x}}_k)\)
where \(\mathbf{\bar{x}}_k=\frac{1}{M}\sum_{m=1}^M\mathbf{x}_k^{(m)}\)
is the average state at timestep \(k\). From these reference models,
\(\mathbf{\bar{m}}_k\), the corresponding migration-velocity models,
\(\widetilde{\mathbf{v}}_k\), are obtained via smoothing. Using
equation~\ref{eq-image-observed}, the migrated images for the ensemble
are now computed as follows:

\begin{equation}\phantomsection\label{eq-image-observed-final}{
\begin{aligned}
\mathbf{y}_k^{(m)}&=\mathcal{H}(\mathbf{x}_k^{(m)},\boldsymbol{\epsilon}_k, \mathbf{\bar{m}}_k)\\
&:=\bigl({\mathbf{J}_{\mathcal{F}[\mathbf{v}]}}\Bigr |_{{\mathbf{\widetilde{v}}_k}}\bigr)_{\mathsf{ic}}^{\top} \Bigl(\mathcal{F}[\mathbf{\bar{m}}_k]-\mathbf{d}_k^{(m)}\Bigr), \quad m=1\cdots M.
\end{aligned}
}\end{equation}

In this expression, seismic images for each ensemble member are
calculated by migrating differences between shot records computed in the
reference model and shot records of the ensemble given by
equation~\ref{eq-shot-data-esemble} with a migration-velocity model that
is given by \(\widetilde{\mathbf{v}}_k\). Because the surveys are
replicated, the time-lapse time-index subscript in the observation
operator is dropped.

With the uncertain dynamics and observation operators in place, we are
after initialization (lines 1---3 of  Algorithm~\ref{alg-sSBI} ) ready
to execute the \(\textsc{Forecast}\) step (cf.
 Algorithm~\ref{alg-training}  line 5). During this
\(\textsc{Forecast}\) step, tuples
\(\{\mathbf{x}_k^{(m)},\mathbf{y}_k^{(m)} \}_{m=1}^{M}\) are produced
that serve as input to the CNF training. By replacing the simulated
ensemble shot records with the ground-truth ``field'' shot records,
\(\mathbf{d}^{\mathrm{obs}}_k\), in
equation~\ref{eq-image-observed-final}, the observed images,
\(\mathbf{y}^{\mathrm{obs}}_k\), introduced in
Section~\ref{sec-ground-truth} are formed.

\subsubsection{The Training step}\label{the-training-step}

Given the \(\textsc{Forecast}\) ensemble,
\(\{\mathbf{x}_k^{(m)},\mathbf{y}_k^{(m)}\}_{m=1}^{M}\), the subsequent
step is to compute the optimized training weights,
\(\widehat{\boldsymbol{\phi}}_k\). After random initialization, the
network weights are trained by minimizing the objective on line 9 of
 Algorithm~\ref{alg-training}  with the optimizer \(\textsc{ADAM}\)
(Diederik P. Kingma and Ba 2014) and for a total of \(120\) epochs. As
we can see from figure~\ref{fig-learning-curve}, the plots for the
\(\ell_2\)-norm training objective and SSIM (see Appendix B) both behave
as expected during training and validation and there is no sign of
overfitting.

\begin{figure}

\begin{minipage}{0.50\linewidth}

\includegraphics{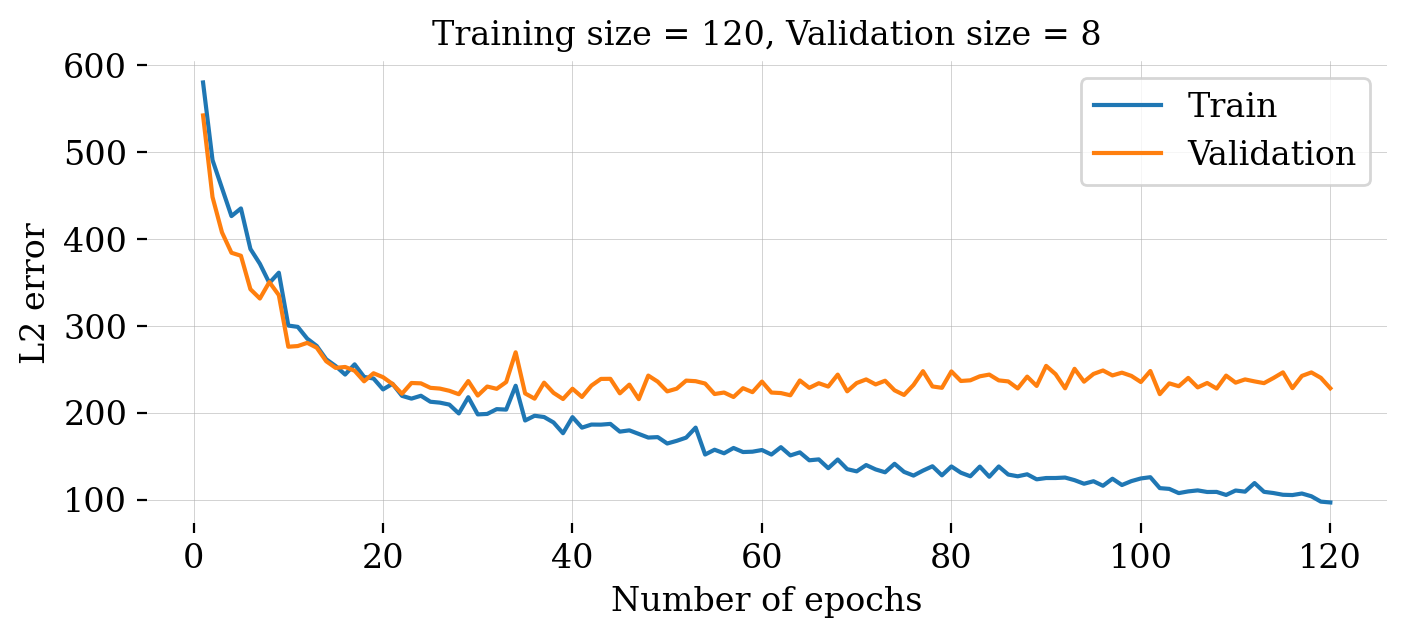}

\subcaption{\label{}a.}
\end{minipage}%
\begin{minipage}{0.50\linewidth}

\includegraphics{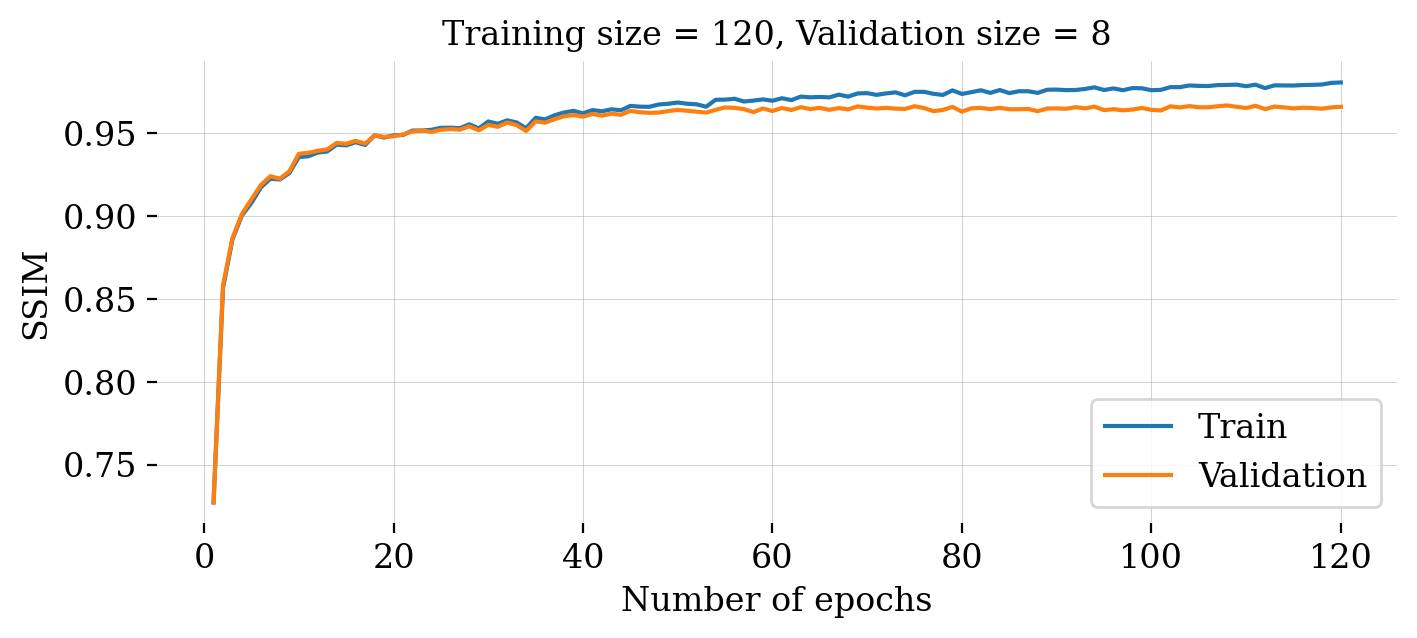}

\subcaption{\label{}b.}
\end{minipage}%

\caption{\label{fig-learning-curve}Training logs. \emph{(a)}
\(\ell_2\)-norm loss during training and validation. \emph{(b)} plot for
SSIM. Both plots behave as expected.}

\end{figure}%

To measure the performance of our network, we also plotted the
structural similarity index measure (SSIM, Z. Wang et al. (2004b)),
which is a metric designed to mimic human perception of image quality.
See Appendix B for further details.

\subsubsection{The Analysis step}\label{the-analysis-step}

\begin{figure}

\centering{

\includegraphics[width=0.95\textwidth,height=\textheight]{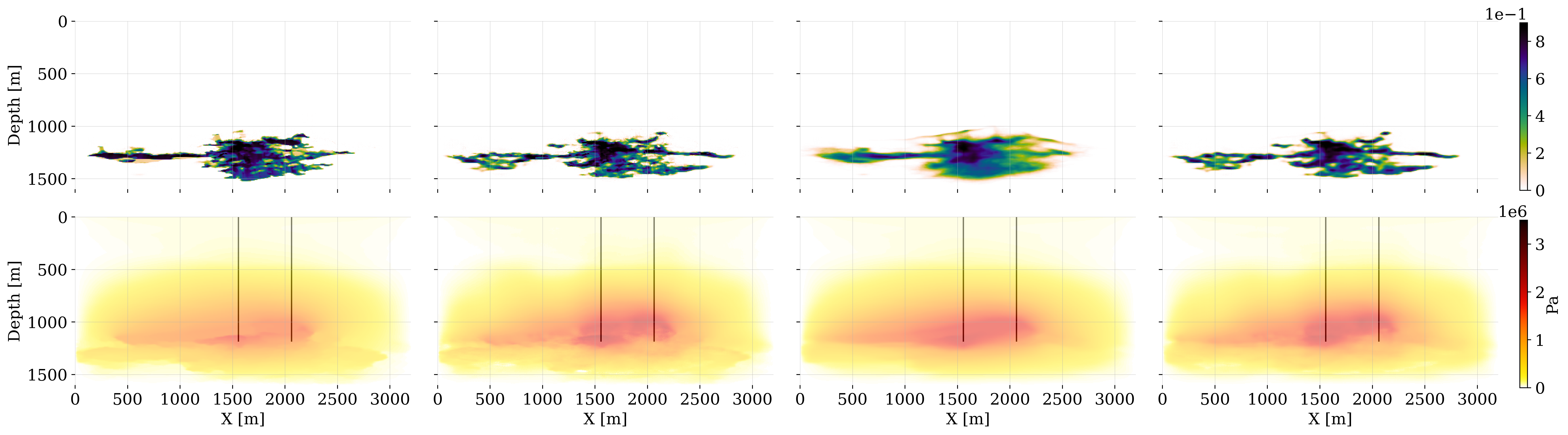}

}

\caption{\label{fig-analysis-samples-mean}Corrections by the
\(\textsc{Analysis}\) step at timestep, \(k=3\). Corrections on
predicted samples for the saturation,
\(\mathbf{x}_k[S_{\mathrm{CO_2}}]\) (top-row), and pressure
perturbation, \(\mathbf{x}_k[\delta p]\) (bottom-row). Left two columns
include individual predicted sample,
\(\mathbf{x}_3\sim p(\mathbf{x}_3\mid \widehat{\mathbf{x}}_2)\) with
\(\widehat{\mathbf{x}}_2\sim p_{\widehat{\phi}_2}(\cdot\mid \mathbf{y}_2^{\text{obs}})\)
for the tuple \((\mathbf{x}_k[S_{\mathrm{CO_2}}]\),
\(\mathbf{x}_k[\delta p])\) and corrected sample,
\(\mathbf{\widehat{x}}_3=\mathrm{Analysis}\bigl(\mathbf{y}^{\mathrm{obs}}_{3}, \widehat{\boldsymbol{\phi}}_3,(\mathbf{x}_3,\mathbf{y}_3)\bigr)\).
Conditional means before the \(\textsc{Analysis}\)\$ step,
\(\mathbf{\bar{x}}_3\) and after the \(\textsc{Analysis}\)\$ step,
\(\mathbf{\bar{\widehat{x}}}_3\), are included in the third and fourth
column. Notice the impact of conditioning the state on seismic
observations, \(\mathbf{y}^{\mathrm{obs}}_{3}\).}

\end{figure}

With the networks for the amortized posterior inference trained, the
predicted samples in the ensemble for the state, produced by the
\(\textsc{Forecast}\) step, are corrected by conditioning the posterior
distribution on the imaged ``field'' data. In our case, the field data
consists of the \emph{in silico} ground-truth simulations outlined in
Section~\ref{sec-ground-truth}. Given these ``time-lapse field
datasets'', \(\mathbf{y}_k^{\textrm{obs}}, \, k=1\cdots K\), the
predicted states of the ensemble, \(\{\mathbf{x}_k^{(m)}\}_{m=1}^{M}\),
are corrected during the \(\textsc{Analysis}\) step, which corresponds
to executing lines 14---15 of  Algorithm~\ref{alg-training}  for each
timestep, \(k=1\cdots K\). To illustrate the importance of the
\(\textsc{Analysis}\) step at timestep, \(k=3\), the first two columns
of figure~\ref{fig-analysis-samples-mean} include corrections on a
single instance of the predicted state,
\(\mathbf{x}_3\sim p(\mathbf{x}_3\mid \widehat{\mathbf{x}}_2)\) with
\(\widehat{\mathbf{x}}_2\sim p_{\widehat{\phi}_2}(\cdot\mid \mathbf{y}_2^{\text{obs}})\),
and the corrected plume,
\(\widehat{\mathbf{x}}_3\sim p_{\widehat{\phi}_3}(\cdot\mid \mathbf{y}_3^{\text{obs}})\).
The saturations are plotted in the top row and pressure differences in
the bottom row. The correction for the CO\textsubscript{2} saturation at
timestep, \(k=3\), is smaller than the correction at time step, \(k=1\),
as shown in figure~\ref{fig-life-cycle}. The reason for this is that the
early correction at the time step, \(k=1\), establishes the leftward
fingering feature that persists to the later timesteps. However, this
does not apply to the protruding rightward CO\textsubscript{2} plume
features that come in at later times and which are partially missed by
the saturation prediction. While less pronounced, the individual
pressure perturbation prediction is also updated considerably during the
\(\textsc{Analysis}\) Step.

\begin{figure}

\centering{

\includegraphics[width=0.95\textwidth,height=\textheight]{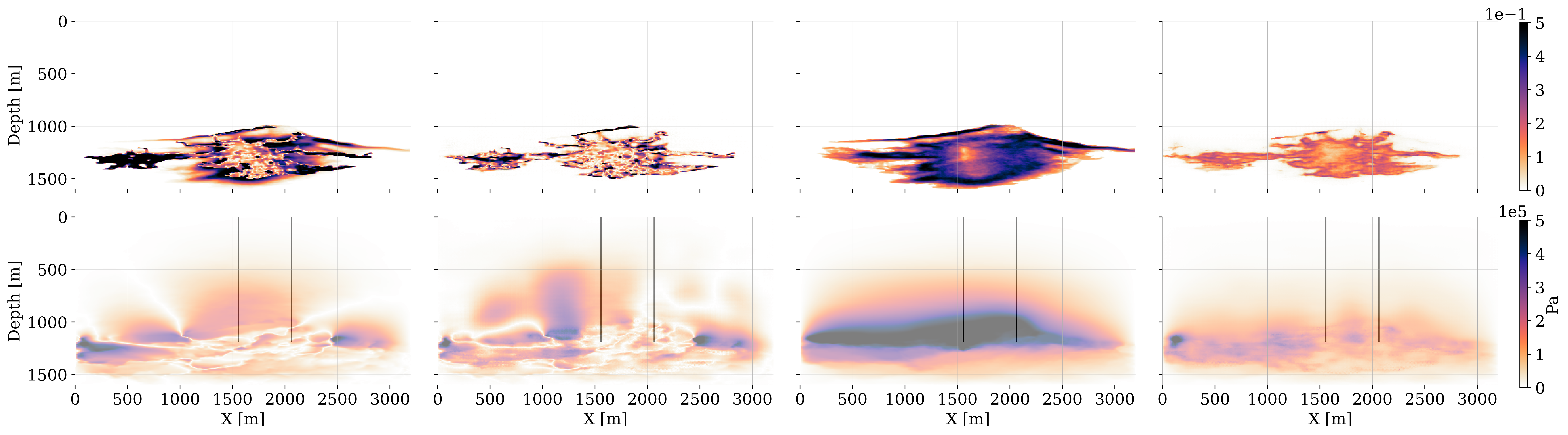}

}

\caption{\label{fig-analysis-CM-errors}Effect of the
\(\textsc{Analysis}\) step on predicted and corrected posterior errors
w.r.t. to the ground-truth \(\mathbf{x}^\ast_3\) and posterior standard
deviations. Two leftmost colums contain errors,
\(\text{Error}=|\mathbf{x}_3^\ast-\mathbf{\bar{x}}_3|\), before, and
after,
\(\text{Error}=|\mathbf{x}_3^\ast-\mathbf{\bar{\widehat{x}}}_3|\),
correction by the \(\textsc{Analysis}\) step. Two rightmost columns
depict the posterior standard deviations before and after correction.}

\end{figure}

The impact of the correction also becomes clear from the right-most two
columns of figure~\ref{fig-analysis-samples-mean} where the conditional
mean before, \(\bar{\mathbf{x}}_3\), and after the corrections,
\(\bar{\widehat{\mathbf{x}}}_3\), are juxtaposed. After conditioning on
the observed data, the corrected samples exhibit less variation, which
results in a significantly better resolved conditional mean for the
corrected CO\textsubscript{2} saturation. Similarly, the pressure
difference corrections yield a pressure field that is more consistent.
These improvements during the \(\textsc{Analysis}\) step are also
apparent when comparing the errors w.r.t. the ground truth,
\(\text{Error}=|\mathbf{x}_3^\ast-\mathbf{\bar{x}}_3|\)
vs.~\(\text{Error}=|\mathbf{x}_3^\ast-\mathbf{\bar{\widehat{x}}}_3|\),
plotted before and after corrections during the \(\textsc{Analysis}\)
step in figure~\ref{fig-analysis-CM-errors} for the CO\textsubscript{2}
saturations (top-row: first two columns). The effect of the correction
for the pressure differences (bottom-row: first two columns) is more
difficult to interpret. However, the uncertainty in terms of the
conditional deviations before, \(\boldsymbol{\bar{\sigma}}_3\), and
after the correction, \(\bar{\boldsymbol{\widehat{\sigma}}}_3\),
tightens significantly for both the CO\textsubscript{2} saturations
(top-row: rightmost columns) and pressure differences (bottom row:
rightmost columns). The plots for the pressure contain locations of the
wells where pressure differences can be measured.

While these results where the \(\textsc{Analysis}\) step reduces the
uncertainty are encouraging, we will introduce in the next section a set
of quantitative performance metrics before considering a case study
where the benefits of working with multimodal will be evaluated.

\subsection{Performance metrics}\label{sec-performance}

While the main goal of this paper is to present a principled workflow to
assess uncertainties in the state of the CO\textsubscript{2} plume,
another goal of establishing a Digital Shadow for Geological
CO\textsubscript{2} Storage is to accelerate innovations in monitoring
to ensure safe operations towards the regulators and the general public.
For this reason, we follow Donoho (2023)'s recipe to set the stage for
accelerated development in this field by

\begin{itemize}
\item
  releasing all our data under the Creative Commons License. This
  includes the ground-truth seismic and reservoir models, the creation
  of the samples for permeability, the generation of the ground-truth
  state and, ``observed'' time-lapse data;
\item
  making all our code available under the open-source software MIT
  license. While still work in progress, the aim is to make use of our
  software as frictionless as possible;
\item
  introducing benchmark metrics by which the success of the Digital
  Shadow is measured quantitatively. The aim is that new methods and
  proposed improvements to the monitoring life-cycle can quickly be
  validated in a trusted testbed of benchmarks.
\end{itemize}

For now, the proposed benchmarks include the methodology presented in
this paper. In a later companion paper, comparisons between results
obtained by alternative data-assimulation methods will be reviewed.

\subsubsection{Benchmark I: reconstruction
quality}\label{benchmark-i-reconstruction-quality}

Since the ground-truth state is available as part of our \emph{in
silico} simulations, quantitative assessments of the reconstruction
quality can be made by computing the Structural Similarity Index Metric
Error (SSIM-ERR) and Root-Mean-Square-Error (RMSE), both detailed in
Appendix B.

\subsubsection{Benchmark II: uncertainty
calibration}\label{benchmark-ii-uncertainty-calibration}

While the high-dimensionality and computational complexity of our
inference problem prevents rigorous validation of the presented
statistical analysis, quantitative establishment of how well the
predicted uncertainties, \(\widehat{\boldsymbol{\sigma}}_k\), and the
errors, \(\text{Error}_k=|\mathbf{x}_k^\ast-\mathbf{\bar{x}}_k|\),
correlate is computationally feasible. To this end, we follow Orozco,
Siahkoohi, et al. (2024) and introduce the calibration test developed by
Guo et al. (2017); Laves et al. (2020). During this test, Euclidean
distances are calculated between posterior mean estimates for the state
and the ground-truth state. According to this test, uncertainty is
well-calibrated when it is proportional to the error. To establish this
test, deviations for each gridpoint in \(\widehat{\boldsymbol{\sigma}}\)
are designated into \(L\) bins of equal width. The uncertainty for each
bin, \(B_l\), is then calculated with

\[
\mathrm{UQ}(B_l) = \frac{1}{|B_l|} \sum_{i\in B_l}\bar{\boldsymbol{\widehat{\sigma}}}[i]
\]

where \(\bar{\boldsymbol{\widehat{\sigma}}}[i]\) refers to the
\(i^{\text{th}}\) entry of the estimated deviation vector. For each bin,
the mean-square error w.r.t. the ground truth is calculated with

\[
\mathrm{MSE}(B_l) = \frac{1}{|B_l|} \sum_{i\in B_l}(\mathbf{x}^{\ast}[i] - \widehat{\mathbf{x}}[i])^2.
\]

\begin{figure}

\centering{

\includegraphics[width=0.5\textwidth,height=\textheight]{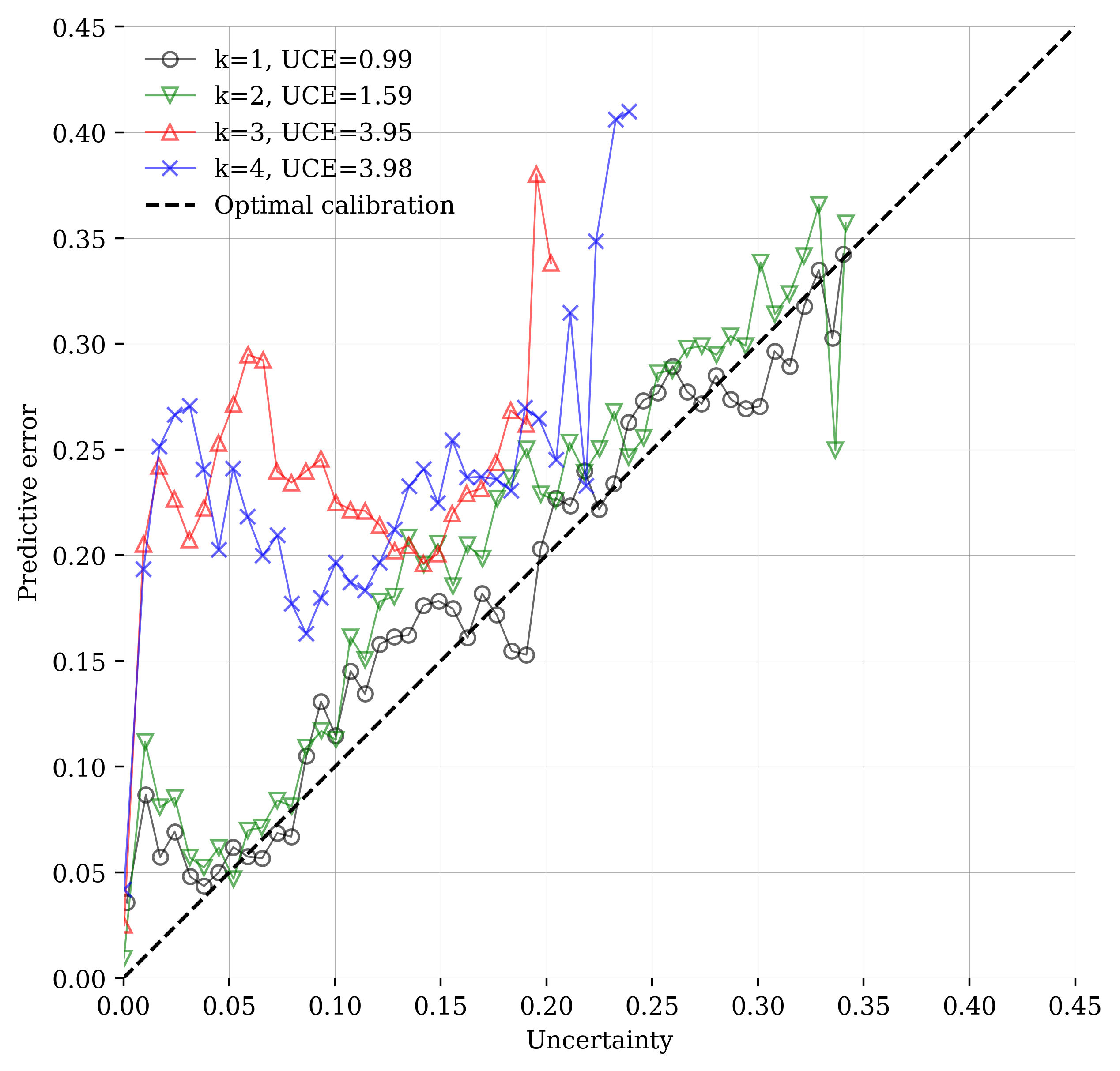}

}

\caption{\label{fig-Calibration}Calibration plot of binned mean-square
errors vs.~posterior deviation for the plumes conditioned on both
seismic and wells.}

\end{figure}

Qualitatively, good uncertainty calibration implies that points on the
crossplot between the predicted error and actual uncertainty fall along
the line \(y=x\). To quantify the quality of uncertainty calibration, we
follow Orozco, Siahkoohi, et al. (2024) and calculate the Uncertainty
Calibration Error (UCE), defined as the average absolute difference
between the predicted error and actual uncertainty across all bins. A
lower UCE indicates more precise calibration, which is approximately the
case in figure~\ref{fig-Calibration} for the posterior samples
conditioned on both seismic and wells. We observe a trend towards poorer
calibration as the monitoring time step increases. We hypothesize that
as the plume grows larger, there are more areas where errors can occur,
making the calibration test more difficult to pass. When comparing a new
method under this benchmark, we would qualitatively evaluate the
calibration plots on a per-monitoring time basis, and also in aggregate,
to summarize performance across all monitoring times. As a quantitative
metric, we would compare the UCE values between methods and select the
one with the lowest UCE as having the best calibration.

\section{Multimodal case study}\label{sec-case}

The task of the proposed Digital Shadow is to produce high-fidelity
estimates for the state of the CO\textsubscript{2} plume using a
combination of data-assimilation and machine learning techniques that
use multimodal time-lapse data as input. Aside from conducting
principled uncertainty quantification, the aim is also to maximally
benefit from having access to two complementary time-lapse datasets,
namely time-lapse pressure/saturation measurements at the two wells
(injection and monitoring) and 2D active-source time-lapse seismic
surveys. The aim of these \emph{in silico} experiments is to demonstrate
the possible benefits of multimodal data-assimilation over data
comprised of well or seismic data only. To this end, inferences will be
carried out on both time-lapse datasets separately and jointly. After
briefly describing the relative ease by which the presented Digital
Shadow can be extended to include data collected at (monitoring)
well(s), systematic quantitative comparisons will be made between the
different approaches. This includes quantitative analyses based on the
metrics introduced in Section~\ref{sec-performance}.

\subsection{Data collection, simulation, and
training}\label{data-collection-simulation-and-training}

While inverting multimodal data in classical (read without machine
learning) inversion/inference problems has proven challenging,
especially when the different datasets are scaled differently, or
collected differently, extending the proposed neural data-assimilation
framework is relatively straightforward as long as the time-lapse data
are collected on the same grid and at the same discrete observation
times. Under that assumption, observations of the saturation/pressure
(\(\mathbf{x}_k[S_{\mathrm{CO_2}}]\) and \(\mathbf{x}_k[\delta p]\))
data can be modeled by a restriction operator, collecting these
quantities at gridpoints that correspond to the well locations. During
training and inference, time-lapse saturation, pressure, and seismic
images each serve as input to separate input channels of the conditional
branch of the cINN. The saturation and pressure are both converted to
all-zero images except for values of the saturation and pressure at the
well locations. Based on conditioning, by seismic only, well only, or
both, the cINN is trained to produce samples of the state, consisting of
gridded images for saturation and pressure perturbation. These gridded
quantities are produced by separate output channels of the network. The
experiment setup for the ground truth and is the same as outlined in
Section~\ref{sec-shadow}.

\subsection{Experimental protocol}\label{experimental-protocol}

Experiments to investigate the possible benefits of multimodal
acquisition proceed as follows:

\textbf{Initialization step:} Draw \(M=128\) samples from the initial
distribution for the state,
\(\widehat{\mathbf{x}}_0\sim p(\mathbf{x}_0)\). The initial saturation
is selected from the uniform distribution on the interval
\([0.1,\, 0.6]\) and chosen to occupy a cylinder with a diameter of four
grid cells (\(4\times 6.25\mathrm{m}\)). The location of the cylinder
coincides with the injection interval at the well, which extends
\(37.5\mathrm{m}\), vertically. The vertical location of the well's
injection interval is positioned adaptively within the closest
high-permeability layer in the randomly drawn permeability fields,
\(\boldsymbol{\kappa}\sim p(\boldsymbol{\kappa})\). These permeability
fields are drawn with replacement from a set of \(1000\) converted
velocity models produced by WISE. The resulting injection depth varies
between \(1200-1250\mathrm{m}\). The initial pressure is set to the
hydraulic pressure, which increases with depth. This initialization
produces the initial ensemble for the state,
\(\{\widehat{\mathbf{x}}_0^{(m)}\}_{m=1}^{128}\).

\textbf{Forecast step:} For each timestep, the state and multimodal
observations are predicted by time advancing the state ensemble,
\(\{\widehat{\mathbf{x}}_{k-1}^{(m)}\}_{m=1}^{128}\), to
\(\{\mathbf{x}_k^{(m)}\}_{m=1}^{128}\) by running the dynamics (cf.
equation~\ref{eq-dynamics}), followed by sampling from the
``likelihood'' (cf.~equations \ref{eq-obs}, \ref{eq-image-observed},
\ref{eq-forward-seismic}). This produces the predicted ensemble,
\(\{(\mathbf{x}_k^{(m)}, \mathbf{y}_k^{(m)})\}_{m=1}^{128}\). These
predicted states consist of the saturation, \(S_{CO_2}\), and pressure
perturbation, \(\delta\mathbf{p}\), with respect to the initial
hydraulic pressure. Depending on the time-lapse data modality, the
simulated observations, \(\{\mathbf{y}_k^{(m)}\}_{m=1}^{128}\), consist
of imaged seismic, saturation/pressure, or both.

\textbf{Training step:} The predicted ensembles,
\(\{\mathbf{x}_k^{(m)}\}_{m=1}^{128}\), are split into \(120\) training
samples and eight validation samples. After the network training is
completed, the inference is carried out on one sample of the test set,
which consists of the ground-truth seismic images,
\(\mathbf{y}_k^{\mathrm{obs}}\).

\textbf{Analyses step:} Conditioned on observed time-lapse ``field''
data, \(\mathbf{y}^{\mathrm{obs}}_k\), the predicted states are
corrected by computing the latent representation for each ensemble pair,
followed by sampling from the posterior conditioned on the field
observations by running the cINN in reverse on the latent variables
(cf.~equations \ref{eq-latent} and \ref{eq-analyses-samples}). This
produces the corrected ensemble for the states,
\(\{\widehat{\mathbf{x}}_{k}^{(m)}\}_{m=1}^{128}\).

By applying the \(\textsc{Forecast}\) and \(\textsc{Analysis}\) steps
recursively, the full history of the estimated state for each member of
the ensemble is collected in the vectors
\(\{\widehat{\mathbf{x}}_{1:24}^{(m)}\}_{m=1}^{128}\). These complete
state histories are formed by appending six dynamic simulation timesteps
after each member of the ensemble is corrected during the
\(\textsc{Analysis}\) step. Given these complete histories for the
state, we proceed by calculating point estimates, in terms of the
conditional mean, \(\bar{\widehat{\mathbf{x}}}_{1:24}\) and their
uncertainty in terms of the conditional deviations,
\(\bar{\widehat{\boldsymbol{\sigma}}}_{1:24}\), for all \(24\)
timesteps. To quantify the data-assimilation accuracy of the inferred
states, the errors with respect to the ground truth,
\(\mathbf{e}_{1:24}=|\mathbf{x}^\ast_{1:24}-\mathbf{\bar{\widehat{x}}}_{1:24}|\),
will be calculated as well.

\subsection{Qualitative results}\label{qualitative-results}

\begin{figure}

\centering{

\includegraphics{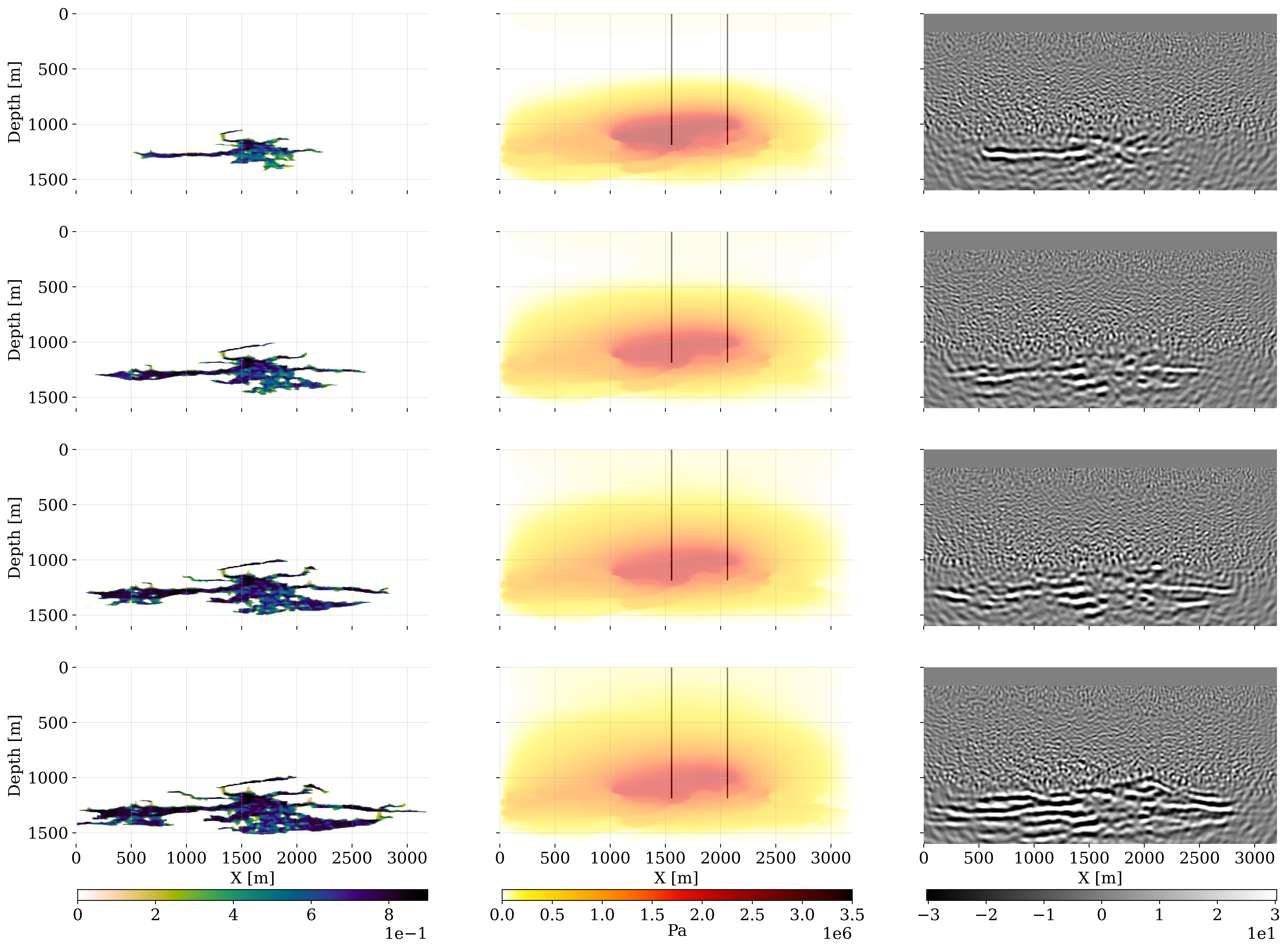}

}

\caption{\label{fig-ground-truth-multimodal}Ground-truth
\(\mathbf{x}^\ast_k[S_{\mathrm{CO_2}}]\) (first column) and
\(\mathbf{x}^\ast_k[\delta p]\) (second column) for \(k=1\cdots 4\),
including locations for the wells. The corresponding seismic
observations, \(\mathbf{y}^{\mathrm{obs}}_k\), are included in the third
column. Notice that the pressure difference field precedes the
CO\textsubscript{2} saturation.}

\end{figure}

To establish the performance of the proposed Digital Shadow,
juxtapositions are made between data-assimilations for three different
data modalities, namely well only, seismic only, and well + seismic. The
importance of data-assimilation compared to relying on reservoir
simulations alone will also be demonstrated by comparing plots for the
conditional mean (\(\bar{\widehat{\mathbf{x}}}_{1:6:24}\)), the errors
with respect to the ground truth,
\(\mathbf{e}_{1:6:24}=|\mathbf{x}^\ast_{1:6:24}-\mathbf{\bar{\widehat{x}}}_{1:6:24}|\),
and the conditional standard deviation,
\(\bar{\widehat{\boldsymbol{\sigma}}}_{1:6:24}\) for these three
different modalities. These quantities are calculated from posterior
samples, conditioned on the simulated ground-truth time-lapse data
included in figure~\ref{fig-ground-truth-multimodal} for
\(k=1\cdots 4\). This figure contains plots for the ground-truth
saturation, \(\mathbf{x}^\ast_k[S_{\mathrm{CO_2}}]\); pressure
perturbations, \(\mathbf{x}^\ast_k[\delta p]\), including locations of
the wells, and plots for the corresponding noisy imaged seismic
observations in the third column. Observations at wells and/or imaged
surface seismic are collected in the vectors,
\(\mathbf{y}^{\mathrm{obs}}_k\), for \(k=1\cdots 4\), and serve as input
to our Digital Shadow whose task is to infer the state of the
CO\textsubscript{2} plume for each timestep.

\begin{figure}

\centering{

\includegraphics{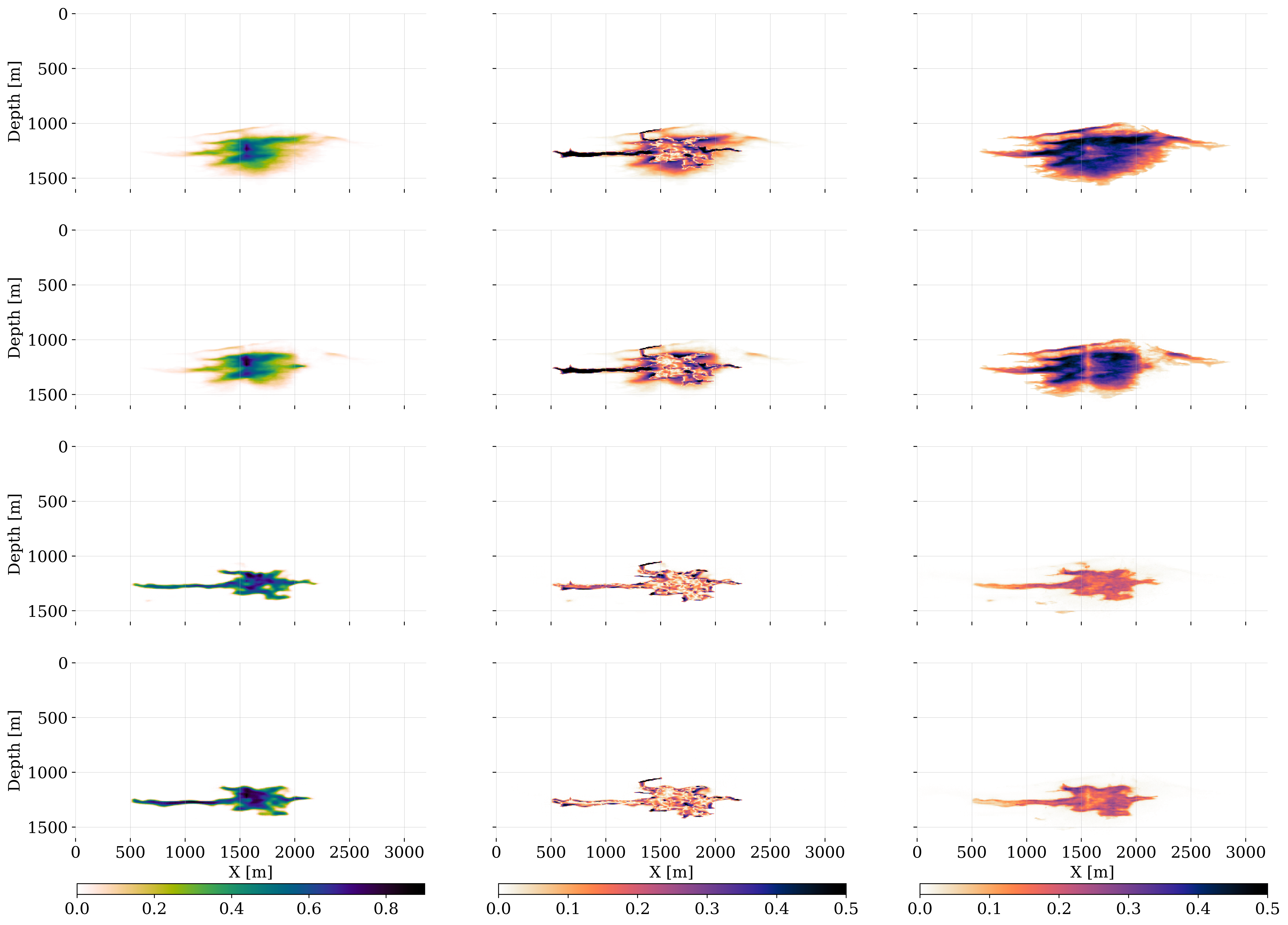}

}

\caption{\label{fig-inference-t1}At k=1, (from left to right column)
plots for the posterior mean, the error, and standard deviations of the
inferred CO\textsubscript{2} saturations. Unconditioned
CO\textsubscript{2} saturation (first row); CO\textsubscript{2}
saturation conditioned on pressure and saturation wells (second row); on
seismic data (third row); and on both seismic and wells (fourth row).}

\end{figure}

\begin{figure}

\centering{

\includegraphics{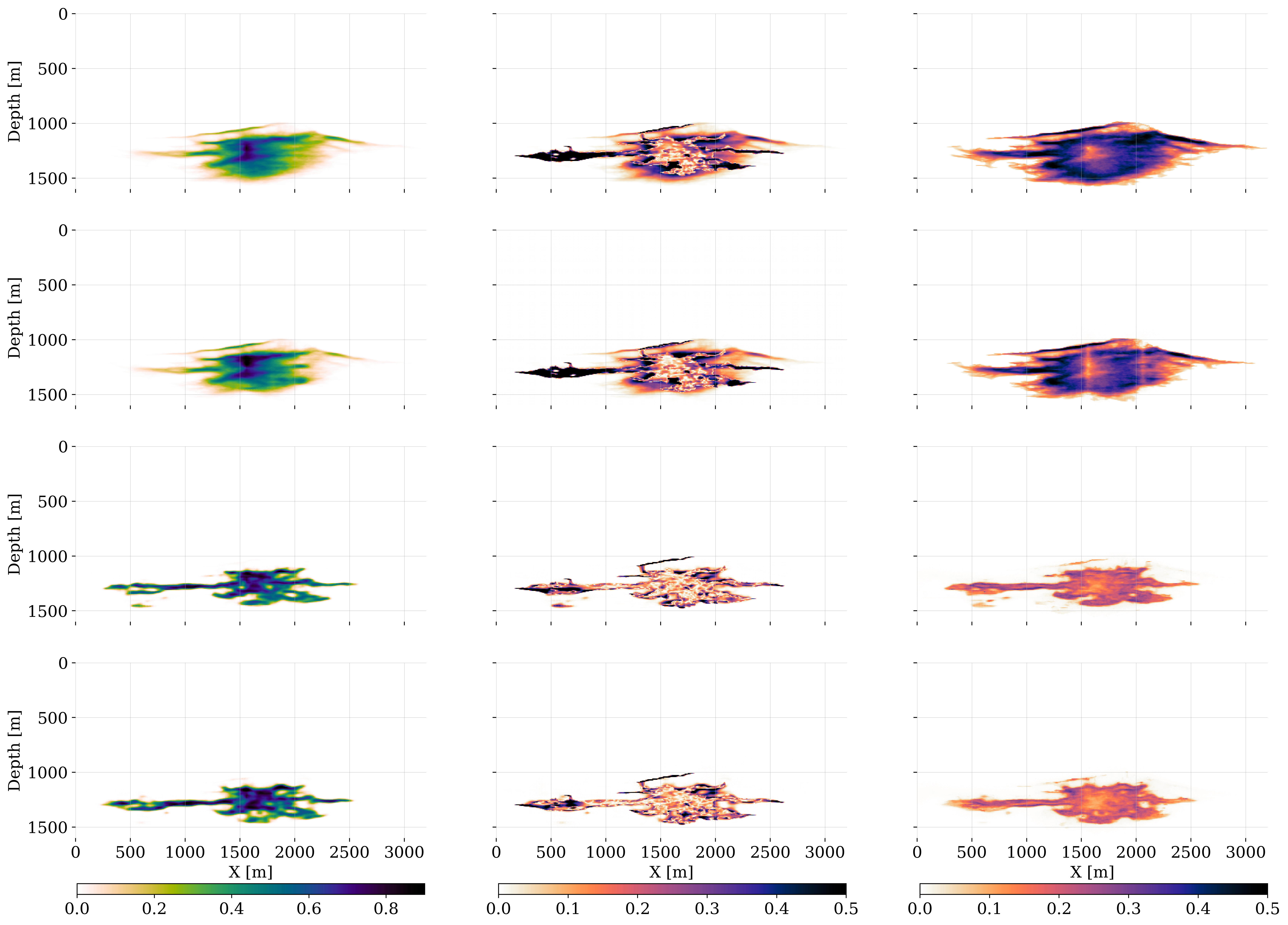}

}

\caption{\label{fig-inference-t2}At k=2, (from left to right column)
plots for the posterior mean, the error, and standard deviations of the
inferred CO\textsubscript{2} saturations. Unconditioned
CO\textsubscript{2} saturation (first row); CO\textsubscript{2}
saturation conditioned on pressure and saturation wells (second row); on
seismic data (third row); and on both seismic and wells (fourth row).}

\end{figure}

\begin{figure}

\centering{

\includegraphics{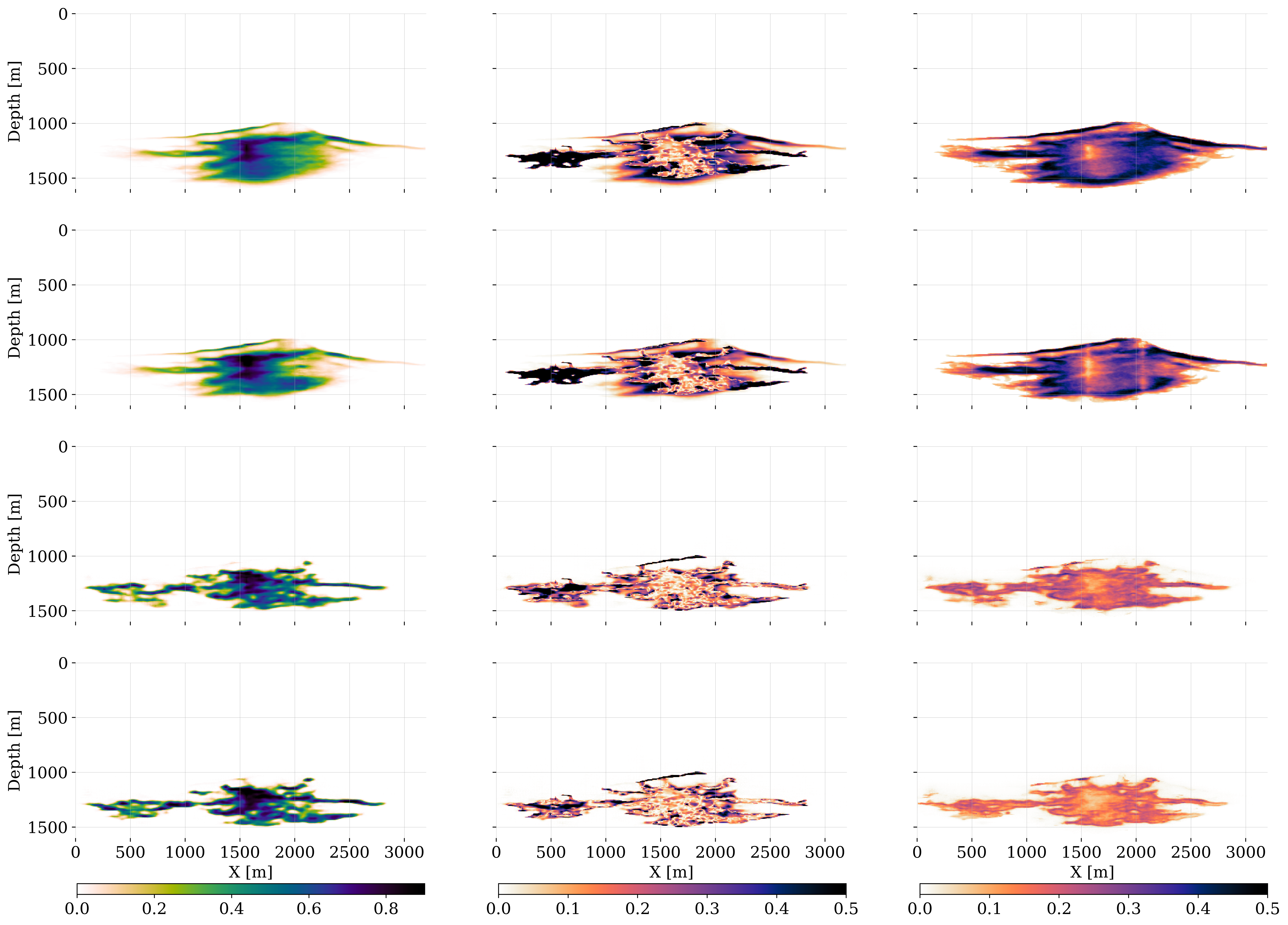}

}

\caption{\label{fig-inference-t3}At k=3, (from left to right column)
plots for the posterior mean, the error, and standard deviations of the
inferred CO\textsubscript{2} saturations. Unconditioned
CO\textsubscript{2} saturation (first row); CO\textsubscript{2}
saturation conditioned on pressure and saturation wells (second row); on
seismic data (third row); and on both seismic and wells (fourth row).}

\end{figure}

\begin{figure}

\centering{

\includegraphics{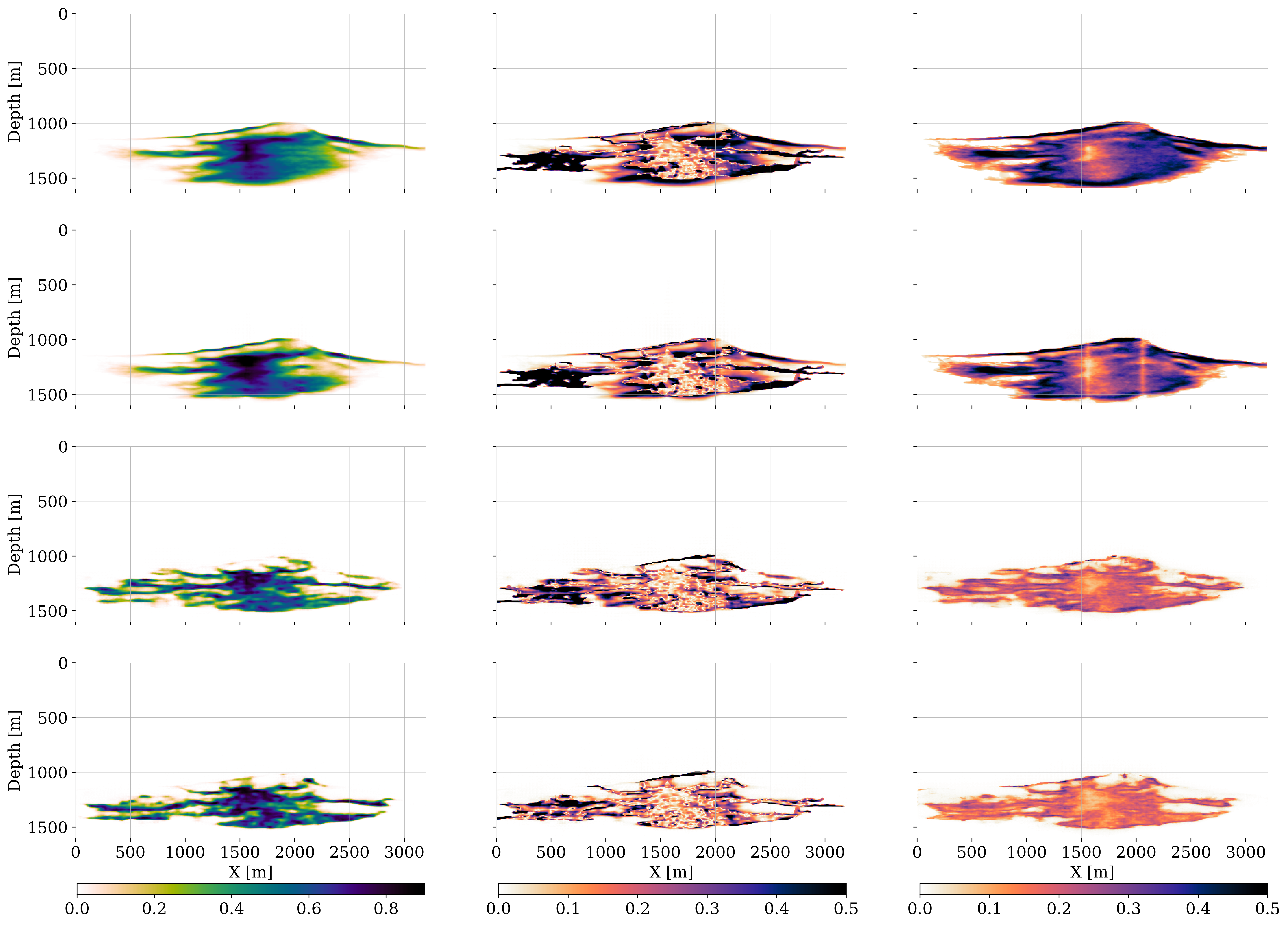}

}

\caption{\label{fig-inference-t4}At k=4, (from left to right column)
plots for the posterior mean, the error, and standard deviations of the
inferred CO\textsubscript{2} saturations. Unconditioned
CO\textsubscript{2} saturation (first row); CO\textsubscript{2}
saturation conditioned on pressure and saturation wells (second row); on
seismic data (third row); and on both seismic and wells (fourth row).}

\end{figure}

Estimates for the CO\textsubscript{2} saturation states
\(\bar{\widehat{\mathbf{x}}}_{1:6:24}\),
\(\mathbf{e}_{1:6:24}=|\mathbf{x}^\ast_{1:6:24}-\mathbf{\bar{\widehat{x}}}_{1:6:24}|\),
and \(\bar{\widehat{\boldsymbol{\sigma}}}_{1:6:24}\) are, for each of
the four scenarios (unconditioned, conditioned on well only, seismic
only, well + seismic) included in the rows of figures
\ref{fig-inference-t1}---\ref{fig-inference-t4} for \(k=1\cdots 4\). The
corresponding estimates for the pressure perturbations
\(\mathbf{x}_{1:6:24}[\delta p]\) are shown in Appendix B. The columns
of these figures contain estimates for the conditional mean, errors, and
uncertainty in terms of the standard deviation. From these plots, the
following qualitative observations can be made.

First, and foremost, a comparison between the unconditioned and
conditioned CO\textsubscript{2} plume estimates shows that conditioning
on any type of time-lapse data nudges, for each timestep, the corrected
fluid-flow predictions closer to the ground-truth CO\textsubscript{2}
saturations. As expected, the unconditioned CO\textsubscript{2} plume
diverges due to epistemic uncertainties induced by the unknown
permeability field at each time-lapse timestep (\(k=1\cdots 4\)).
Remember, every time the dynamic state equation (cf.
equation~\ref{eq-dynamics}) is run, a new sample for the permeability
field is drawn, introducing new uncertainties into the system. This
effect is also present for the CO\textsubscript{2} plume estimates
conditioned on time-lapse data but to a much lesser degree because the
insertion of time-lapse data cancels some of the systematic uncertainty.
However, including time-lapse data at the wells by itself does not move
the overall shape of the inferred CO\textsubscript{2} plume much closer
to the ground truth. Nevertheless, when the conditioning involves
seismic data, the corrections capture the overall shape of the
CO\textsubscript{2} saturations very well.

Second, the conditional mean estimates for the CO\textsubscript{2}
saturation appear smooth compared to the ground truth. This smoothing
effect is due to the remaining uncertainty of the estimated plumes (see
column three of figures \ref{fig-inference-t1}---\ref{fig-inference-t4})
and persists, and even increases somewhat, as time progresses.

Third, the CO\textsubscript{2} plume estimates at \(k=1\) obtained for
the wells only completely miss the important leftward protruding
CO\textsubscript{2}-plume feature, which eventually leads to a violation
of Containment. Conversely, this feature is well recovered by inferences
based on time-lapse data that includes imaged seismic. Compared to
seismic only, joint inference from well + seismic data shows further
improvements near the wells and away from the wells, reducing the error
with respect to the ground-truth CO\textsubscript{2} saturation (see
column two of figures \ref{fig-inference-t1}---\ref{fig-inference-t4}).
As expected, including pressure and saturation data at the wells reduces
uncertainty in the direct vicinity of the wells but there are also
indications that it may have an effect away from the wells when combined
with time-lapse seismic. Finally, the errors and uncertainty in terms of
the conditional standard deviations correlate reasonably well when the
CO\textsubscript{2} plume is conditioned on both time-lapse seismic and
well data.

Fourth, the overall trend continues at timestep, \(k=2\), leading to
reduced errors and uncertainty in the main body of the plume when the
CO\textsubscript{2}-plume inference is conditioned on well + seismic
data. However, because of a lack of lateral information the results
based on well data alone more or less completely miss the important
leftward-protruding streamline feature induced by the unknown
high-permeability streak.

Fifth, the complexity of the CO\textsubscript{2} plume increases further
at later timesteps. This challenges the inference. However, we note
again that the estimated conditional mean improves considerably when the
inference is conditioned on time-lapse seismic data. As can be observed
in figure~\ref{fig-inference-t3} and \ref{fig-inference-t3} bottom first
and third column, the inference improves due to the inclusion of
information at the wells. As a result, uncertainty is reduced in regions
away from the wells both horizontally, and vertically below the
wellhead. Interestingly, while the well-only inference result lacks
detail, its uncertainty at the horizontal extremities seems to suggest
protruding fingers at the periphery of the CO\textsubscript{2} plume.
These features are indeed present in the ground-truth
CO\textsubscript{2} plume and in the inferences that include time-lapse
seismic data.

Sixth, the conditional mean estimates for the CO\textsubscript{2} plume
are for most features best recovered when the inference is conditioned
on time-lapse seismic data with visible improvements when pressure and
saturation data collected at the wells are included. When comparing the
uncertainty (column three of figures
\ref{fig-inference-t1}---\ref{fig-inference-t4}), it appears that adding
pressure/saturation data from the wells also reduces the uncertainty in
areas relatively far away from the well-bore. Finally, the well-only
inference seems to include increased CO\textsubscript{2} saturation at
the seal, a feature that is only partly present in the ground truth. The
presence of this feature can attribute to the average behavior of the
permeability, an observation shared by the unconditioned results. It is
encouraging to see that this feature at the seal has large uncertainty
judged by the large \(\bar{\widehat{\boldsymbol{\sigma}}}_{24}\) in this
region. Except for an area near the well-bore, CO\textsubscript{2}-plume
saturations are not visible on the conditional mean estimates
conditioned on the seismic. However, there is a small imprint on the
plots for the uncertainty, which means that the posterior samples do
include sporadic examples of detectable CO\textsubscript{2} saturation
in this narrow area near the seal.

While these inference results by the Digital Shadow are encouraging,
they only provide snapshots at the four timesteps time-lapse data is
collected. They also do not provide quantitative information on the
performance of the proposed Digital Shadow.

\subsection{Quantitative results}\label{quantitative-results}

\begin{figure}

\begin{minipage}{0.33\linewidth}

\includegraphics{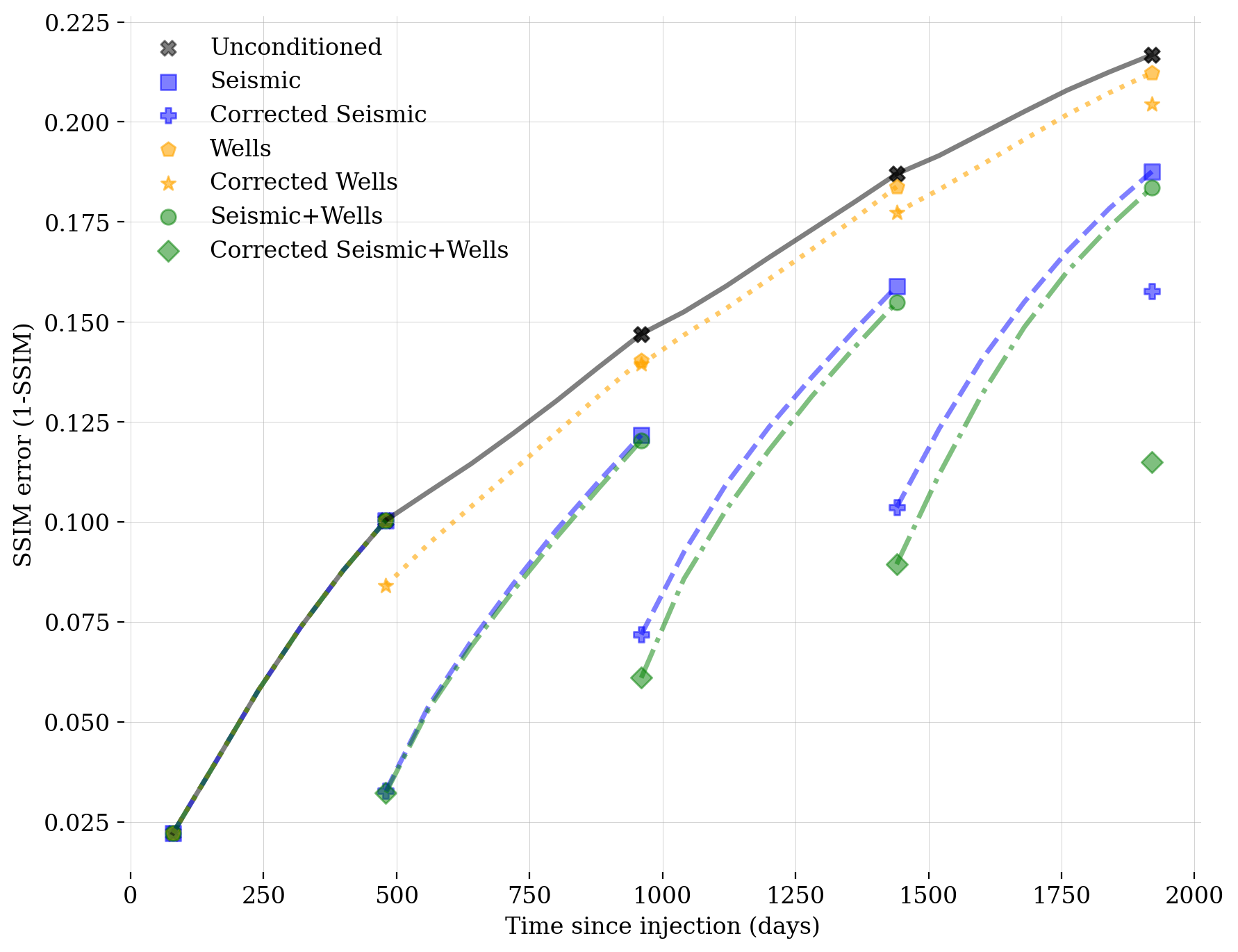}

\subcaption{\label{}a.}
\end{minipage}%
\begin{minipage}{0.33\linewidth}

\includegraphics{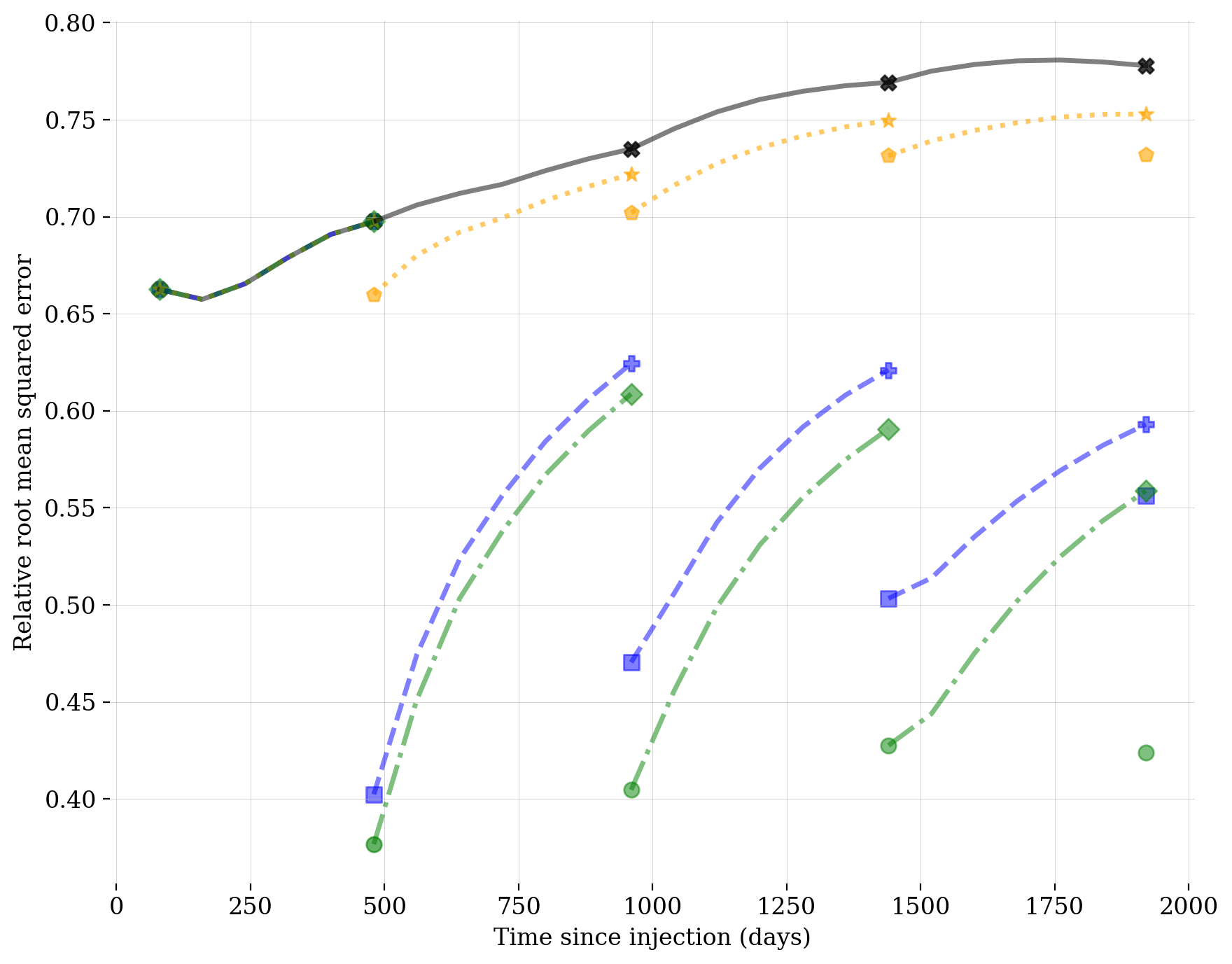}

\subcaption{\label{}b.}
\end{minipage}%
\begin{minipage}{0.33\linewidth}

\includegraphics{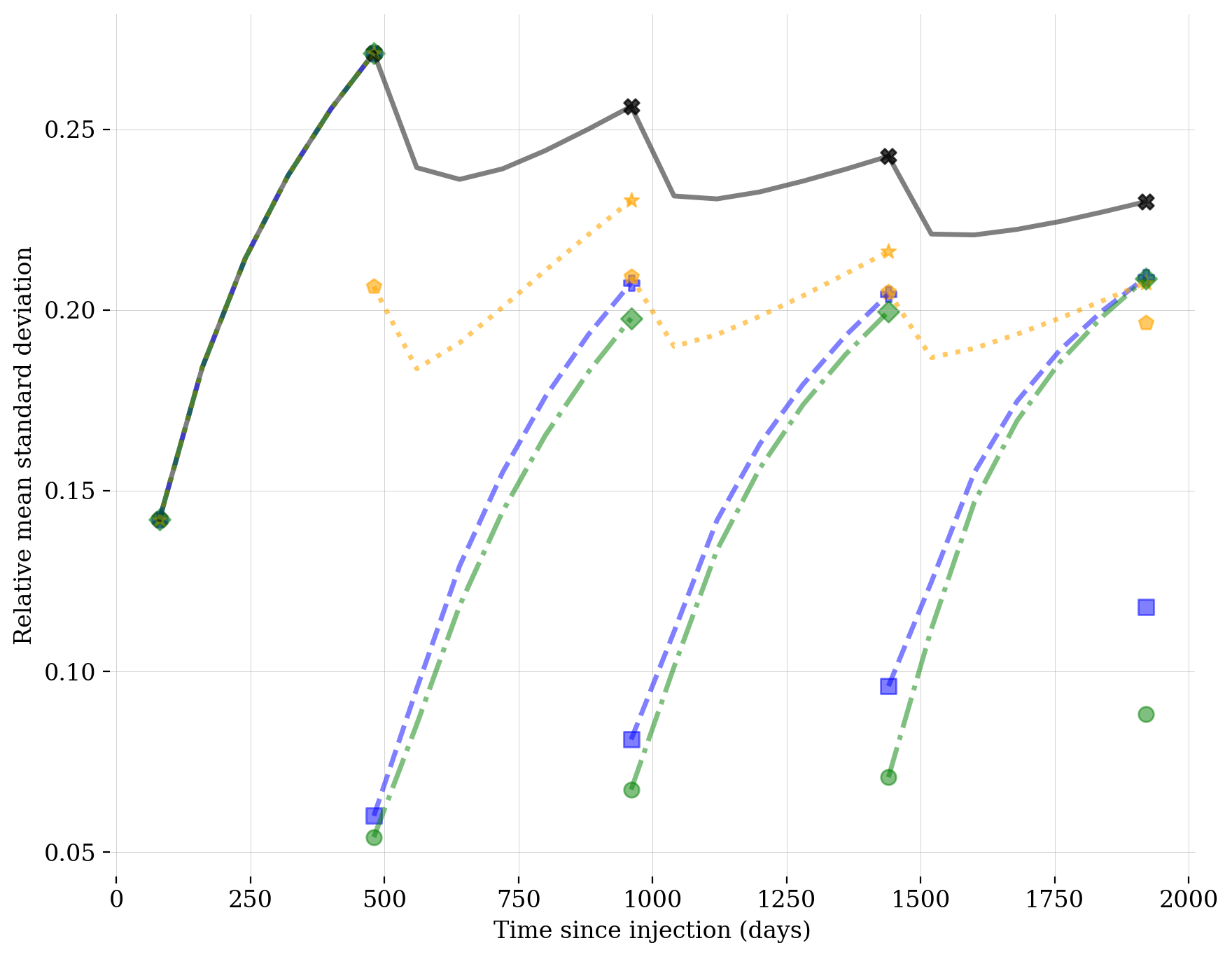}

\subcaption{\label{}c.}
\end{minipage}%

\caption{\label{fig-DT-Statistics}Quantitative performance metric of
CO\textsubscript{2} plume recovery with the proposed Digital Shadow.
Before and after corrections by the \(\textsc{Analysis}\) step are
denoted by blue square/cross symbols for seismic/corrected seismic; the
orange pentagon/star symbols for wells/corrected wells, and the green
circles/diamodns for seismic + wells/corrected seismic + wells.
\emph{(a)} \(\mathrm{SIMMD-Err}\) as a function of time since injection;
\emph{(b)} \(\Delta\mathrm{RMSE}_k\), and \emph{(c)}
\(\Delta\mathrm{STD}_k\).}

\end{figure}

Since the inference results produced by the Digital Shadow will be used
during decision making by a Digital Twin, a better understanding needs
to be gained on the performance of the Digital Shadow by means of a
quantitative analysis of the different inference strategies. To this
end, we include in figure~\ref{fig-DT-Statistics} calculations of the
SSIM-ERR defined as \(1-\mathrm{SSIM}(\cdot)\), the mean absolute error
(MAE), and the mean standard deviation (\(\bar{\sigma}_k\)), for all
\(24\) timesteps spanning a total of \(1920\) days. To compensate for
biases induced by the growth of the CO\textsubscript{2} plumes
themselves, the following relative measures will be used for the errors
and standard deviations:

\begin{equation}\phantomsection\label{eq-relative-metrics}{
\Delta\mathrm{RMSE}_k(\widehat{\mathbf{x}}_k) = \frac{\mathrm{RMS}(\mathbf{e}_k)}{\mathrm{RMS}(\mathbf{x}^\ast_k)}\quad \text{and}\quad \Delta\mathrm{STD}_k=\frac{\mathrm{mean}(\bar{\mathbf{\widehat{\boldsymbol{\sigma}}}}_k)}{\mathrm{RMS}(\widehat{\mathbf{\bar{x}}}_k)}
}\end{equation}

where \(\mathrm{RMS}\) stands for Root-Mean Square. The normalization of
the \(\mathrm{RMS}\) for the error and the mean standard deviation by
the \(\mathrm{RMS}\) of the ground truth and estimated conditional mean
is meant to correct for increased variability exhibited by growing
CO\textsubscript{2} plumes.

Each subplot in figure~\ref{fig-DT-Statistics} contains estimates for
these quantities based on inferences conditioned on seismic only (dashed
blue lines), on wells only (dotted orange lines), and seismic + wells
(dashed-dotted green lines). To exemplify the importance of conditioning
fluid-flow simulations on time-lapse observations, we also plot
estimates obtained without conditioning (solid black lines) obtained by
running equation~\ref{eq-dynamics} recursively for \(k=1\cdots 4\). From
these three plots, the following observations can be made.

First, the curves for the SSIM-Error, \(\Delta\mathrm{RMSE}_k\), and
\(\Delta\mathrm{STD}_k\) overall show improved quality in all three
metrics when constrained by time-lapse observations (all other than
black curves). Moreover, significant enhancements as a result of the
\(\textsc{Analysis}\) step can be observed thanks to time-lapse
data-informed corrections on the predictions of the
\(\textsc{Forecast}\) step. These corrections start by being relatively
small when conditioning on the well measurements alone (cf.~the orange
star (corrected) and pentagon symbols) but become more significant when
conditioning on the seismic (cf.~the blue cross and square symbols) and
even more pronounced when conditioning on well + seismic (cf.~green
diamond and round symbols). As time evolves, the magnitude of the
data-informed corrections increases for all metrics with the largest
improvement in \(\mathrm{SSIM-Err}\) at the last timestep, which can be
explained because the bulk of the CO\textsubscript{2} plume will have
reached the monitor well at that time. Since the state dynamics and
observation model are both nonlinear, this significant improvement can
not be explained by the corrections from the well data alone.

Second, for each timestep in between the time-lapse data collection,
there is, as expected, a general downward trend in plume recovery
quality from the time the corrections are made until the next timestep
during which the next time-lapse data is collected and the subsequent
corrections are made. This deterioration in recovery quality is observed
for the \(\mathrm{SSIM-Err}\) for all modalities and is to leading order
the result of differences between the randomly drawn permeability field
and the ground truth that lead during the intermediate timesteps to
increasing deviations between the actual CO\textsubscript{2} plume and
the predicted ensemble produced by the \(\textsc{Forecast}\) step. The
normalization in the definition of \(\Delta\mathrm{RMSE}_k\) in
equation~\ref{eq-relative-metrics} compensates partially for this
effect, yielding more prominent improvements, especially between
conditioning on seismic data alone or on wells + seismic. In addition,
relative errors after correction remain reasonably flat and even
decrease at the final timestep. This suggests that relative errors
remain more or less constant thanks to the conditioning on multimodal
time-lapse observations.

Third, the relative quantity, \(\Delta\mathrm{STD}_k\), aims to answer
the question of whether including time-lapse data, and multimodal
time-lapse data in particular, reduces the uncertainty. Thanks to the
normalization by the \(\mathrm{RMS}\) of the (conditional) mean, the
uncertainty plotted in figure~\ref{fig-DT-Statistics} (c) remains
relatively constant with slightly larger corrections produced by
conditioning the CO\textsubscript{2} plumes on multimodal time-lapse
data. While this behavior is consistent with the behavior displayed by
the relative errors in figure~\ref{fig-DT-Statistics} (b), the
uncertainty measured in terms of the variability amongst the
intermediate plume predictions seems to dip further during the immediate
timesteps after the corrections are made. This behavior is observed for
the unconditioned and conditioned by well-only modalities and can be
explained by the fact that after each correction new samples for the
permeability distribution field are drawn at which point the flow will
first explore regions of high permeability before growing. This explains
the observed reduction in variability for the unconditioned and
conditioned on well-data-only modalities that produce
CO\textsubscript{2} plumes that are more widely spread and therefore can
explore more regions to fill in. After these regions are filled, the
plumes grow again increasing the variability and therefore the
uncertainty.

In summary, the overall trends observed in
figure~\ref{fig-DT-Statistics}, are consistent with a highly nonlinear
data-assimilation problem where the unknown permeability field that
controls the flow is modeled as a random nuisance parameter. Even though
there is a slight deterioration in the quality of the inferred
CO\textsubscript{2} plumes, the observed relative errors and uncertainty
remain relatively stable over time when the predicted simulations are
conditioned on time-lapse data that includes seismic surveys. While
including time-lapse well data alone does not lead to significant
improvements in CO\textsubscript{2} plume reconstruction quality and
uncertainty, combining this data modality with time-lapse seismic data
leads to significantly improved results.

\section{Discussion}\label{sec-discussion}

The main goal of this work is the design, implementation, and \emph{in
sillico} validation of a Digital Shadow prototype for Geological Carbon
Storage. IBM's definition of a Digital Twin reads:

\begin{quote}
A digital twin is a virtual representation of an object or system
designed to \emph{reflect a physical object accurately}. It spans the
object's \emph{lifecycle}, is \emph{updated from real-time data} and
uses \emph{simulation, machine learning} and reasoning to help make
decisions.
\end{quote}

When considering this definition, it is clear that the presented neural
data-assimilation framework does indeed capture several key aspects
(denoted by italics) of a Digital Twin. Because our formulation does not
yet support control, reasoning, and decision-making, we follow Asch
(2022) and use the term Digital Shadow for our prototype monitoring
framework instead. To help future decision-making and optimize
underground CO\textsubscript{2} storage operations, we leverage recent
breakthroughs in neural posterior-density estimation and generative AI
to make our Digital Shadow uncertainty aware, which will help the
Digital Twin with future tasks including underground monitoring and
control, vital for cost-effective risk mitigation and management.

By taking a neural approach, where the simulated state-observation
predictions of the well-established ensemble Kalman filter (EnKF, Asch
(2022)) are considered as training samples, we arrive at a formulation
where the important \(\textsc{Analysis}\) step no longer relies on
empirical estimates for the first- and second-order moments (mean and
covariance). Instead, our approach hinges on generative conditional
neural networks' ability to characterize non-Gaussian posterior
distributions with high degrees of freedom. These distributions result
from nonlinear models for the state dynamics (multi-phase flow) and
observations (e.g.~seismic images), which contain probabilistic
components that give rise to complex uncertainties. Compared to other
ensemble-based data-assimilation methods (Frei and Künsch 2013), which
are typically contingent on Monte-Carlo approximations of the classical
Kalman recursions together with limiting Gaussian assumptions, our
nonlinear extension of the EnKF enjoys certain important advantages.
Backed by the recent work of Spantini, Baptista, and Marzouk (2022),
there are strong indications that our approach does not suffer from an
unresolvable lack of accuracy of EnKF whose accuracy can not be improved
by increasing the ensemble size. Similar to Spantini, Baptista, and
Marzouk (2022), we identify corrections during the stochastic
prior-to-posterior map as a Kalman-like \(\textsc{Analysis}\) step,
implemented in the Gaussianized latent space of a conditional Invertible
Neural Network (cINN, Ardizzone et al. (2019b)). Compared to existing
ensemble-based methods, the use of conditional neural networks has the
advantage that their accuracy, and hence the latent-space correction,
increases when adding computational training effort, e.g.~by growing the
ensemble size or by training with more epochs. When the network is
adequately expressive, these accuracy improvements are likely when the
set of simulated training pairs, which correspond to the ensemble, are
sufficiently large, diverse, and close in distribution to the actual
state and observations.

Throughout each lifecycle, the Digital Shadow trains its networks with
Simulation-Based Inference (SBI, Cranmer, Brehmer, and Louppe (2020)) by
using nonlinear simulations of the time-advanced state (calculated with
equation~\ref{eq-dynamics}) and associated, possibly multimodal,
observations (calculated with equation~\ref{eq-obs}). Because our
approach is simulation-based, uncertainties associated with the previous
state, the random permeability field, and observational noise are
marginalized implicitly (Cranmer, Brehmer, and Louppe 2020). As a
result, conditional neural networks act as amortized surrogates for the
posterior of the state given (multimodal) observations. Thanks to the
marginalization, contributions due to the stochasticity of the
permeability distribution and other noise terms are factored into the
resulting posterior, which is no longer dependent on this nuisance
parameter and on specifics of the noise. Moreover, theoretical
connections exist between the EnKF's \(\textsc{Analysis}\) step and the
proposed stochastic latent space prior-to-posterior mapping. Similar to
Spantini, Baptista, and Marzouk (2022), this stochastic mapping reads in
the latent space of the CNF as follows:

\begin{equation}\phantomsection\label{eq-analysis-equivalence}{
\mathbf{x}_k^{(m)}\mapsto \widehat{\mathbf{x}}_k^{(m)}= f^{-1}_{\widehat{\boldsymbol{\phi}}_k}(\mathbf{z}_k^{(m)};\mathbf{y}_k^{\mathrm{obs}})\quad\text{with}\quad \mathbf{z}_k^{(m)}= f_{\widehat{\boldsymbol{\phi}}_k}(\mathbf{x}_k^{(m)};\mathbf{y}_k^{(m)}). 
}\end{equation}

Unlike the EnKF, this correction is nonlinear and corresponds to a
transport map where samples from the joint distribution for the state
and observations are mapped to the latent space, followed by
conditioning on the observed time-lapse data during inversion from the
latent space back to the physical state. As reported by Ramgraber et al.
(2023), conditioning on the observed data in the latent space can have
certain advantages thanks to the Gaussinization of the latent space, a
topic which we will further explore in future work. Finally, the use of
physics-based summary statistics in our approach to our knowledge also
differs from current data-assimilation approaches where dimensionality
reduction approaches are more common for the state and not for the
observations.

Even though the current implementation of our Digital Shadow is
specialized towards multimodal monitoring of Geological Carbon Storage,
the presented neural data-assimilation approach is general and ready to
be applied to other fields as long as implementations for the offline
simulation-based \(\textsc{Forecast}\) step, conditional neural
networks, and network training procedure are available to approximate
the prior-to-posterior mapping. However, specific choices for the
implementation of our Digital Shadow prototype were made contingent on
scalability, computational cost, memory footprint, and accuracy of the
multi-phase flow and wave simulations. For this reason, our
implementation is based on the low-memory footprint Julia implementation
for CNFs, \texttt{InvertibleNetworks.jl} (Orozco, Witte, et al. 2024a),
in combination with the industry-strength parallel multi-phase
fluid-flow simulator,
\href{https://github.com/sintefmath/JutulDarcy.jl}{JutulDarcy.jl}
(Møyner and Bruer 2023), and parallel wave-based Julia Devito Inversion
framework inversion package
\href{https://github.com/slimgroup/JUDI.jl}{JUDI.jl},(P. A. Witte et al.
2019; Louboutin, Witte, et al. 2023). While our framework allows for
other choices, our prototype for the Digital Shadow benefits from the
intrinsic parallel capabilities of the fluid-flow and wave simulators
and from the fact that the ensemble calculations are embarrasingly
parallel. In that respect, the presented scalable approach differs from
recent score-based work by (Rozet and Louppe 2023; Bao et al. 2024; X.
Huang, Wang, and Alkhalifah 2024), which either rely on many, and
therefore fast evaluations of the dynamics and observation operators, or
on fast access to the complete history of the CO\textsubscript{2}
plume's dynamics, neither of which are conducive to realistic scale
data-assimilation for GCS projects.

Because of the relatively long lead time between the collection of
time-lapse seismic surveys, we so far did not opt for the use of our
parallel implementation of Fourier Neural Operators (FNO, Grady et al.
(2023)), which in principle can serve as a suitable substitute to the
expensive multi-phase flow simulations (Louboutin, Yin, et al. 2023;
Yin, Orozco, et al. 2023). We also reserve inclusion of geomechanical,
chemical, and (poro-) elastic effects, to future research even though
these effects can in principle be included in the Digital Shadow's
simulations. These extensions would allow for the use of our Digital
Shadow in more complex situations, including GCS on land, where
geomechanics could play a more important role, or within reservoirs that
undergo geochemical changes.

Compared to contemporary approaches to CO\textsubscript{2} monitoring
(C. Huang and Zhu 2020; Grana, Liu, and Ayani 2020), the proposed
Digital Shadow's conditional neural networks are capable of capturing
the nonlinear state dynamics, given by the multi-phase flow, and
nonlinear seismic observation model, involving nonlinear solutions of
the wave equation. Our current implementation achieves this at a scale
where extended Kalman filters struggle to store explicit matrix
representations for the Kalman filter that includes explicit inverses of
large matrices. Like recent work by Bruer et al. (2024), the proposed
approach reaps information from an ensemble of states, which is an
approach that scales well. By treating the permeability field as a
stochastic nuisance parameter, which is drawn independently at each
time-lapse timestep for each ensemble member, our Digital Shadow
distinguishes itself from this latest work on an EnKF for GCS where the
randomly drawn permeability field remains fixed within each member of
the ensemble. Aside from incorporating stochasticity induced by the
unknown permeable field, which can be interpreted as a modeling error
connecting this work to Si and Chen (2024), our approach is also based
on a nonlinear seismic observation model.

Finally, the presented approach made assumptions on the distribution of
the permeability field as well as on the underlying rock and wave
physics. In future work, we will relax these assumptions by \emph{(i)}
extending recent work by Yin, Louboutin, et al. (2024) to carry out
permeability inference from time-lapse seismic and/or well data, which
would allow for narrowing of the baseline probability distribution for
the permeability field, as time-lapse data becomes available;
\emph{(ii)} marginalizing over uncertainties in the rock-physics model,
e.g.~by augmenting the Forecast ensemble with samples generated from
models other than the patchy saturation model (Avseth, Mukerji, and
Mavko 2010); \emph{(iii)} calibrating rock-physical relations at
(monitoring) wells (Feng et al. 2022); \emph{(iv)} robustifying the
inference with respect to distribution shifts either through
physics-based latent-space corrections, as proposed by Yin, Orozco, and
Herrmann (2024) and Orozco, Siahkoohi, et al. (2024), or by adapting
score-based methods to solve the weak formulation proposed by Siahkoohi,
Rizzuti, and Herrmann (2020) and Orozco, Siahkoohi, et al. (2024);
\emph{(iv)} no longer insisting on costly seismic survey replication to
achieve repeatability of time-lapse seismic surveys (Oghenekohwo et al.
2017; Wason, Oghenekohwo, and Herrmann 2017; López et al. 2023; Y. Zhang
et al. 2023) \emph{(v)} increasing detectibility of time-lapse changes
in the density by using SH-polarized shear waves (Clochard et al. 2018).

\section{Conclusions}\label{sec-conclusion}

The inherent high degrees of freedom and complications arising from the
intricate interplay between the nonlinear multi-phase flow, nonlinear
rock physics, and possibly nonlinear multimodal time-lapse observations,
challenge the development of monitoring systems for subsurface storage
of supercritical CO\textsubscript{2}. In this work, we meet these
challenges by introducing an uncertainty-aware Digital Shadow, which
combines state-of-the-art high-fidelity simulation capabilities with
data-assimilation and machine-learning techniques to arrive at a
scalable formulation that through the deployment of a stochastic
nuisance parameter incorporates nonlinearities and inherent
uncertainties in the permeability field. Thanks to the Digital Shadow's
use of simulation-based inference, its neural networks marginalize over
this stochastic permeability field, factoring lack of knowledge on this
important reservoir property into Digital Shadows' uncertainty
quantification. As a result, dependence on the permeability field is
eliminated and its uncertainty is absorbed into the recursively inferred
posterior distributions for the CO\textsubscript{2} plume's state
(pressure/saturation), conditioned at each timestep on multimodal
time-lapse measurements. The training itself takes place recurrently on
\(\textsc{Forecast}\) ensembles, consisting of predictions for the state
and simulations of induced time-lapse data collected at the surface
and/or in wells.

To validate the proposed Digital Shadow, a realistic simulation-based
case study was conducted, starting from a probabilistic model for the
permeability field derived from a baseline surface seismic survey. Next,
the Digital Shadow's network was trained on \(\textsc{Forecast}\)
ensembles, consisting of simulations for the time-advanced
CO\textsubscript{2} plume's state and corresponding multimodal
time-lapse data. After training was completed, the Digital Shadow's
networks were applied to unseen simulated ground-truth time-lapse data.
From this computational case study, the following observations can be
made. First, CO\textsubscript{2}plume estimates based on averages over
the \(\textsc{Forecast}\) ensembles behave poorly compared to estimates
corrected by observed time-lapse observations. These improvements are
achieved irrespective of the permeability field. Second, the crucial
imprint of an unknown high-permeability streak was missed when the
CO\textsubscript{2} plume's state was conditioned on pressure/saturation
measurements at wells alone. Plume inferences conditioned on seismic or
seismic and well measurements succeeded in detecting this event that
ultimately leads to Containment failure. Third, the inferred pointwise
standard deviation amongst the samples of the posterior, which serves as
a measure of their uncertainty, correlates well with errors with respect
to the ground-truth state. This property persists over the different
timesteps and is an indication that the uncertainty quantification is
reasonably well calibrated. Fourth, visual as well as quantitative
measures of the reconstruction quality show across the board the best
results when the CO\textsubscript{2} plume's state is conditioned on
both time-lapse seismic and well measurements. While conditioning the
CO\textsubscript{2} plume on wells only leads to marginal improvements
compared to uncorrected forecasts, combination with seismic leads to
disproportional improvements even in areas away from the wells. We
attribute these global improvements in the inferred state of the
CO\textsubscript{2} plume to the strong nonlinearity of multi-phase
flow. All in all, we argue from this case study that the proposed
Digital Shadow tracks the unseen ground-truth physical state accurately
for the duration of a realistic CO\textsubscript{2}injection project. To
our knowledge, the proposed methodology constitutes the first
proof-of-concept of an uncertainty-aware Digital Shadow that principally
encapsulates uncertainty due to unknown reservoir properties and noise
present in the observed data. This Digital Shadow will form the basis
for a Digital Twin designed to mitigate risks and optimize operations of
underground storage projects.

\section*{Data and Software
Availability}\label{data-and-software-availability}
\addcontentsline{toc}{section}{Data and Software Availability}

The data underlying the findings of this study and the software
developed for it will be made publicly available upon acceptance of the
manuscript for publication. Also, full movies of the snapshots shown in
the manuscripts will be included during final submission in the online
supplementary material.

\section*{Acknowledgement}\label{acknowledgement}
\addcontentsline{toc}{section}{Acknowledgement}

This research was carried out with the support of Georgia Research
Alliance and partners of the ML4Seismic Center. This research was also
supported in part by the US National Science Foundation grant OAC
2203821.

\begin{center}\rule{0.5\linewidth}{0.5pt}\end{center}

\section*{References}\label{references}
\addcontentsline{toc}{section}{References}

\phantomsection\label{refs}

\setcounter{section}{0}
\renewcommand{\thesection}{\Alph{section}}

\setcounter{table}{0}
\renewcommand{\thetable}{A\arabic{table}}

\setcounter{figure}{0}
\renewcommand{\thefigure}{A\arabic{figure}}

\section*{Appendices}\label{appendices}
\addcontentsline{toc}{section}{Appendices}

\subsection*{Appendix A}\label{appendix-a}
\addcontentsline{toc}{subsection}{Appendix A}

\subsection*{Network architecture cINNs}\label{sec-network}
\addcontentsline{toc}{subsection}{Network architecture cINNs}

For CNFs, we followed conditional GLOW network architecture (Durk P.
Kingma and Dhariwal 2018) implemented in the open-source package
\href{https://github.com/slimgroup/InvertibleNetworks.jl}{InvertibleNetworks.jl}
(P. Witte et al. 2023). The architecture of Conditional GLOW typically
consists of invertible neural network layers, coupled with affine
coupling layers, which enable efficient computation of the determinant
of the Jacobian matrix. These layers are organized in a hierarchical
manner, which offers better expressiveness when compared to the
conventional invertible architectures (Dinh, Sohl-Dickstein, and Bengio
2016). This expressivity comes from applying a series of conventional
invertible layers (Dinh, Sohl-Dickstein, and Bengio 2016) to the input
in a hierarchical manner which results in an invertible architecture
exhibiting the potential to depict intricate bijective transformations
(Siahkoohi et al. 2023a; Orozco, Witte, et al. 2024b) and hence the
ability to learn complex distributions.

\subsubsection*{Training Setting}\label{training-setting}
\addcontentsline{toc}{subsubsection}{Training Setting}

\begin{longtable}[]{@{}ll@{}}
\caption{Training
Hyperparameters}\label{tbl-hyperparameters}\tabularnewline
\toprule\noalign{}
Hyperparameters & Values \\
\midrule\noalign{}
\endfirsthead
\toprule\noalign{}
Hyperparameters & Values \\
\midrule\noalign{}
\endhead
\bottomrule\noalign{}
\endlastfoot
Batch Size & \(8\) \\
Optimizer & \(\mathrm{Adam}\) (Diederik P. Kingma and Ba 2014) \\
Learning rate (LR) & \(5^{-4}\) \\
No.~of training epochs & \(120\) \\
Fixed Noise Magnitude & \(0.005\) \\
No.~of training samples & \(120\) \\
No.~of validation samples & \(8\) \\
No.~of testing samples & \(1\) \\
\end{longtable}

\subsection*{Appendix B}\label{sec-appendix-B}
\addcontentsline{toc}{subsection}{Appendix B}

\subsubsection*{Useful Definitions}\label{useful-definitions}
\addcontentsline{toc}{subsubsection}{Useful Definitions}

After the completion of training, we use the following procedure to
calculate the posterior mean:

\begin{equation}\phantomsection\label{eq-posteriormean}{
\mathbf{x}_{\text{pm}} = \mathbb{E}_{\mathbf{x} \sim p(\mathbf{x}|\mathbf{y})} [ \mathbf{x} ]
\approx 
\frac{1}{M}  \sum^{M}_{n=1}\mathbf{x}_{\text{gen}}^{i} \text{ where } \mathbf{x}_{\text{gen}}^{i} = f_{\hat{\boldsymbol{\phi}}}^{-1}(\mathbf{z_i};\mathbf{y}) \text{ with } \mathbf{z_i} \sim \mathcal{N}(0,I),
}\end{equation}

where \(\hat{\boldsymbol{\phi}}\) is the minimizer of Equation
equation~\ref{eq-loss-CNF}.

\(\textbf{SSIM}\) - Structural Similarity Index Metric quantifies the
similarity between two images and is commonly used to assess how closely
a generated image resembles a ground truth or reference image. It
considers image quality aspects such as luminance, contrast, and
structure. For the mathematical formulation of SSIM, please refer to the
study by (Z. Wang et al. 2004a).

\(\textbf{RMSE}\) - Root mean squared error is used to represent the
measure of the difference between ground truth CO\(_2\) saturation image
and the posterior mean of the samples generated by the trained network.

\subsection*{Appendix C}\label{sec-appendix-C}
\addcontentsline{toc}{subsection}{Appendix C}

\(\textbf{Pressure perturbation}\) - Total reservoir pressure, excluding
hydrostatic pressure.

\begin{figure}

\centering{

\includegraphics{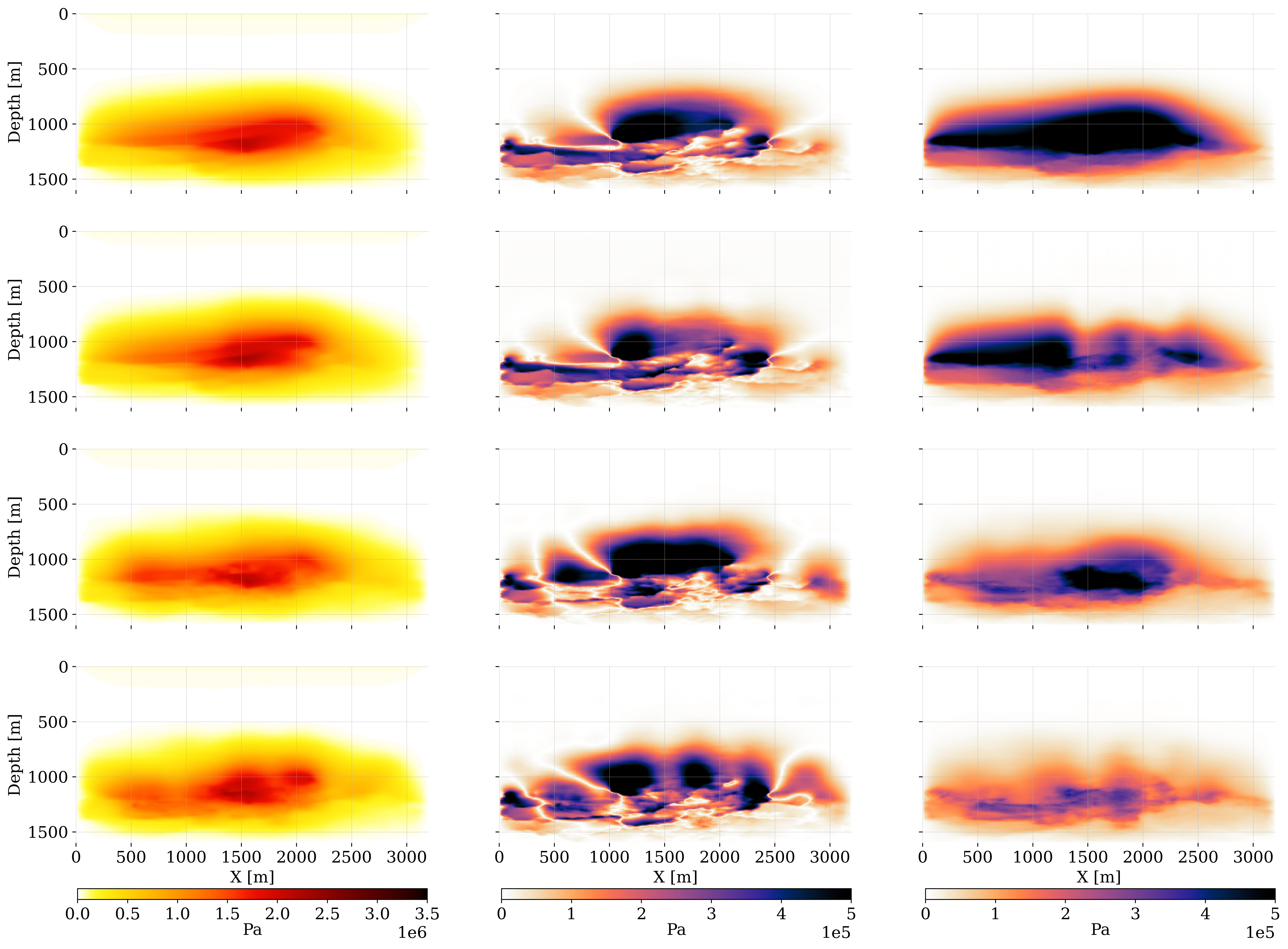}

}

\caption{\label{fig-inference-t1P}At k=1, (from left to right column)
plots for the posterior mean, the error, and standard deviations of the
inferred CO\textsubscript{2} pressure perturbation. Unconditioned
pressure perturbation (first row); pressure perturbation conditioned on
pressure and saturation wells (second row); on seismic data (third row);
and on both seismic and wells (fourth row).}

\end{figure}

\begin{figure}

\centering{

\includegraphics{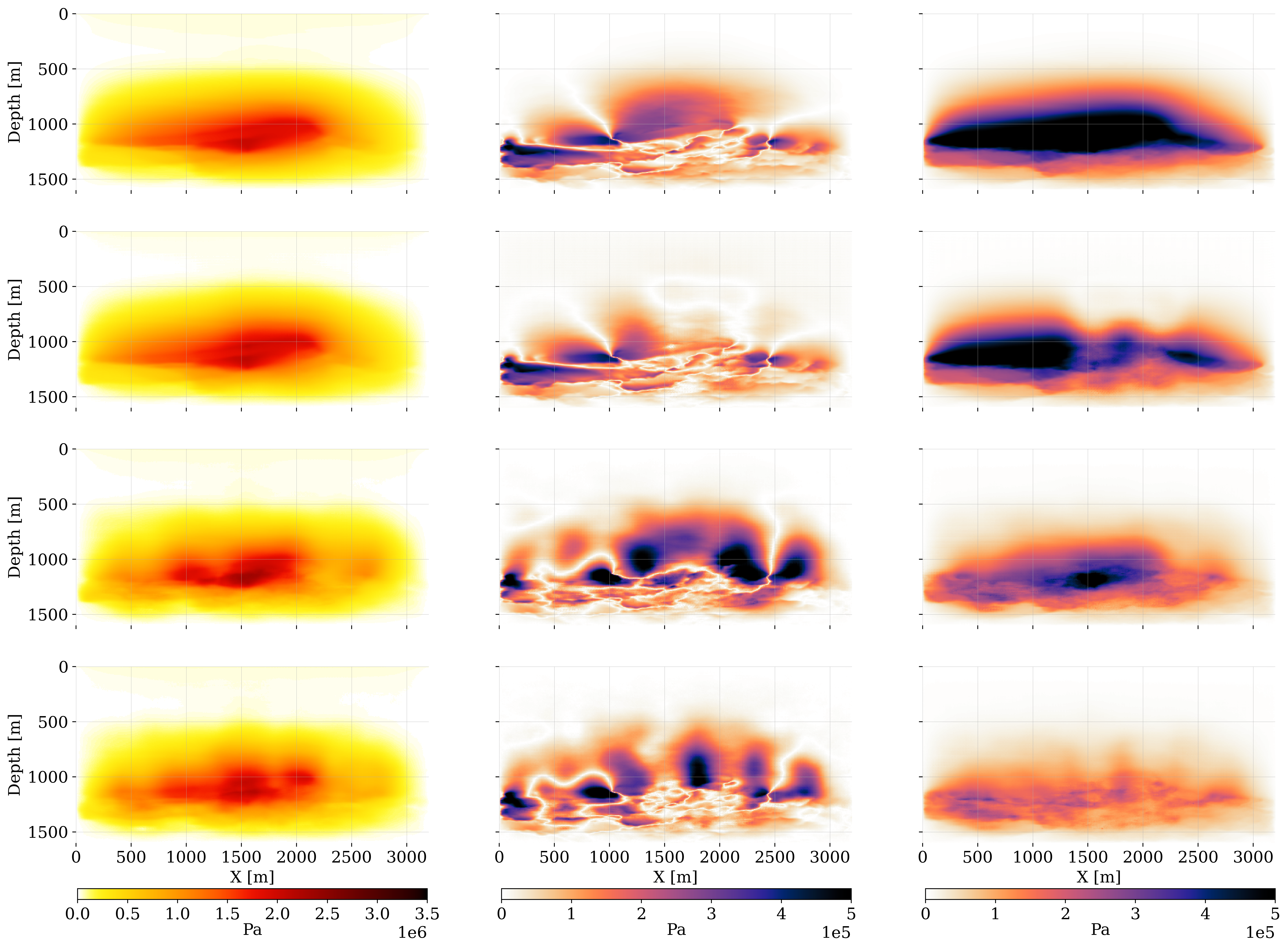}

}

\caption{\label{fig-inference-t2P}At k=2, (from left to right column)
plots for the posterior mean, the error, and standard deviations of the
inferred CO\textsubscript{2} pressure perturbation. Unconditioned
pressure perturbation (first row); pressure perturbation conditioned on
pressure and saturation wells (second row); on seismic data (third row);
and on both seismic and wells (fourth row).}

\end{figure}

\begin{figure}

\centering{

\includegraphics{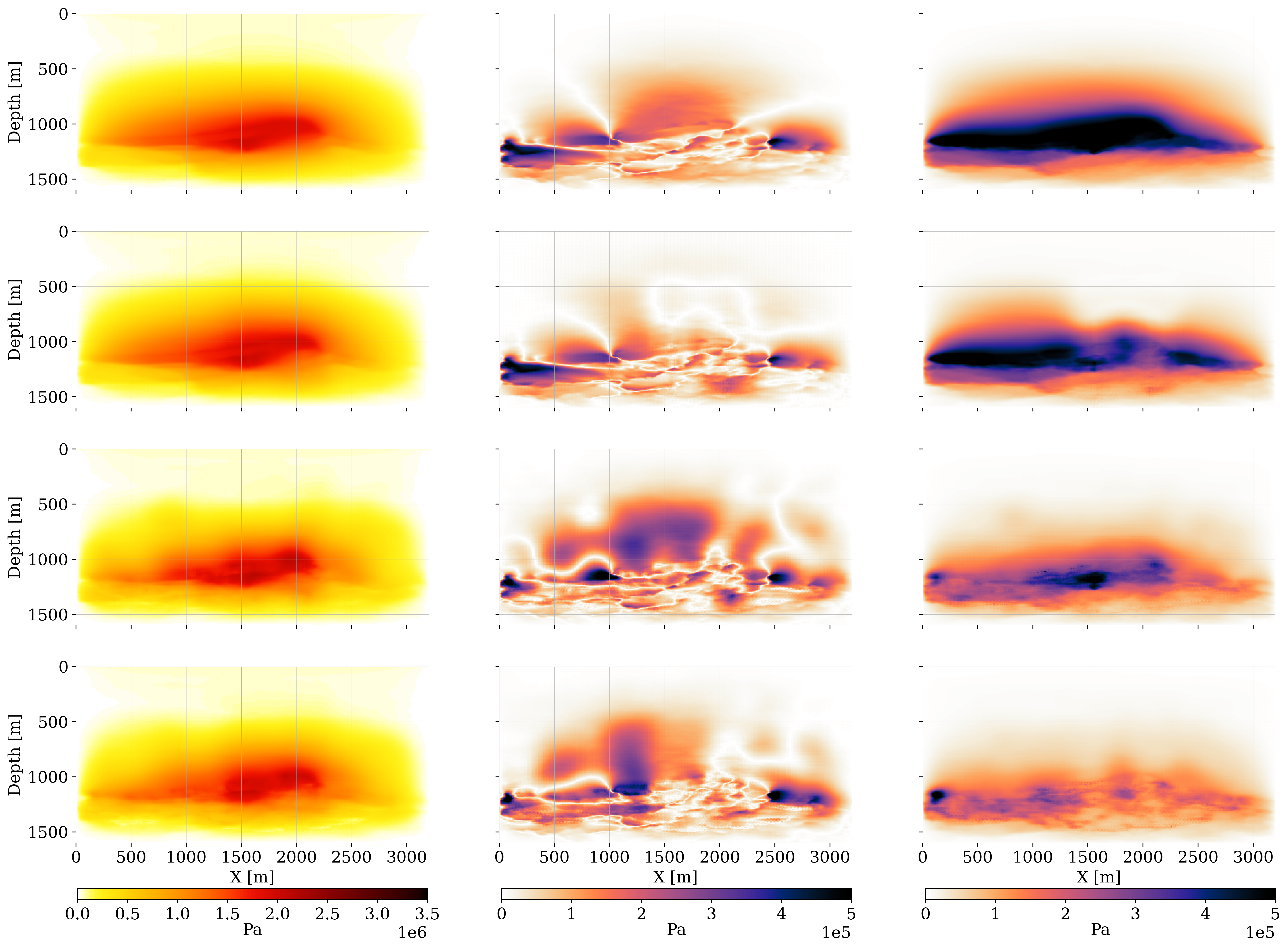}

}

\caption{\label{fig-inference-t3P}At k=3, (from left to right column)
plots for the posterior mean, the error, and standard deviations of the
inferred CO\textsubscript{2} pressure perturbation. Unconditioned
pressure perturbation (first row); pressure perturbation conditioned on
pressure and saturation wells (second row); on seismic data (third row);
and on both seismic and wells (fourth row).}

\end{figure}

\begin{figure}

\centering{

\includegraphics{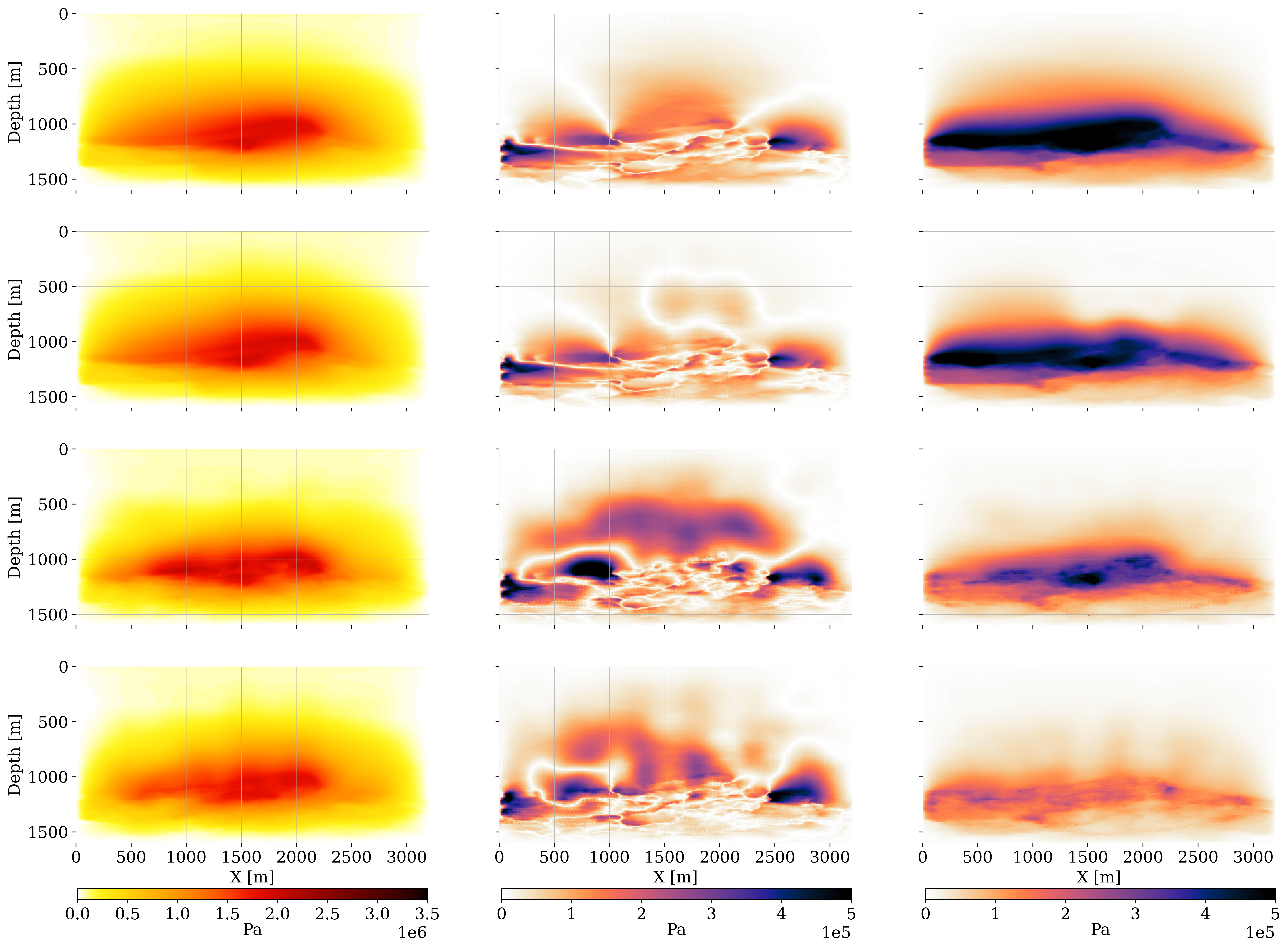}

}

\caption{\label{fig-inference-t4P}At k=4, (from left to right column)
plots for the posterior mean, the error, and standard deviations of the
inferred CO\textsubscript{2} pressure perturbation. Unconditioned
pressure perturbation (first row); pressure perturbation conditioned on
pressure and saturation wells (second row); on seismic data (third row);
and on both seismic and wells (fourth row).}

\end{figure}

Similar to CO\textsubscript{2} saturation states, estimates for the
CO\textsubscript{2} pressure perturbation,
\(\bar{\widehat{\mathbf{x}}}_{1:6:24}[\delta p]\),
\(\mathbf{e}_{1:6:24}=|\mathbf{x}^\ast_{1:6:24}[\delta p]-\mathbf{\bar{\widehat{x}}}_{1:6:24}[\delta p]|\),
and \(\bar{\widehat{\boldsymbol{\sigma}}}_{1:6:24}\) are, for each of
the four scenarios (unconditioned, conditioned on well only, seismic
only, well + seismic) included in the rows of figures
\ref{fig-inference-t1P}---\ref{fig-inference-t4P} for \(k=1\cdots 4\).
The columns of these figures contain estimates for the conditional mean,
errors, and uncertainty in terms of the standard deviation. From these
plots, the following qualitative observations can be made.

Unlike the CO\textsubscript{2} saturation states, the pressure
perturbations estimates are relatively smooth similar to the ground
truth (cf. figure~\ref{fig-ground-truth-multimodal}, second column), so
the conditional mean estimates (first column of figures
\ref{fig-inference-t1P}---\ref{fig-inference-t4P}) do not appear to
differ substantially amongst the four different scenarios during the
four time-lapse time steps. As observed for infered saturations,
variations between the posterior samples conditioned on the wells or
wells + sesimic (third columns) show reduced uncertainty near the wells.
Errors (second colums) with respect to the ground truth do not change
much between the different modalities.

\begin{CSLReferences}{1}{0}
\bibitem[\citeproctext]{ref-alsing2023optimal}
Alsing, Justin, Thomas DP Edwards, and Benjamin Wandelt. 2023.
{``Optimal Simulation-Based Bayesian Decisions.''} \emph{arXiv Preprint
arXiv:2311.05742}.

\bibitem[\citeproctext]{ref-jalsing2018massive}
Alsing, Justin, Benjamin Wandelt, and Stephen Feeney. 2018. {``Massive
Optimal Data Compression and Density Estimation for Scalable,
Likelihood-Free Inference in Cosmology.''} \emph{Monthly Notices of the
Royal Astronomical Society} 477 (3): 2874--85.

\bibitem[\citeproctext]{ref-doi:https:ux2fux2fdoi.orgux2f10.1002ux2f9781119156871.ch14}
Appriou, Delphine, and Alain Bonneville. 2022. {``Monitoring Carbon
Storage Sites with Time-Lapse Gravity Surveys.''} In \emph{Geophysical
Monitoring for Geologic Carbon Storage}, 211--32. American Geophysical
Union (AGU).
https://doi.org/\url{https://doi.org/10.1002/9781119156871.ch14}.

\bibitem[\citeproctext]{ref-cnf}
Ardizzone, Lynton, Carsten Lüth, Jakob Kruse, Carsten Rother, and
Ullrich Köthe. 2019b. {``Guided Image Generation with Conditional
Invertible Neural Networks.''} \emph{CoRR} abs/1907.02392.
\url{http://arxiv.org/abs/1907.02392}.

\bibitem[\citeproctext]{ref-ardizzone2019guided}
---------. 2019a. {``Guided Image Generation with Conditional Invertible
Neural Networks.''} \emph{arXiv Preprint arXiv:1907.02392}.

\bibitem[\citeproctext]{ref-asch2022toolbox}
Asch, Mark. 2022. \emph{A Toolbox for Digital Twins: From Model-Based to
Data-Driven}. SIAM.

\bibitem[\citeproctext]{ref-asch2016data}
Asch, Mark, Marc Bocquet, and Maëlle Nodet. 2016. \emph{Data
Assimilation: Methods, Algorithms, and Applications}. SIAM.

\bibitem[\citeproctext]{ref-avseth2010quantitative}
Avseth, Per, Tapan Mukerji, and Gary Mavko. 2010. \emph{Quantitative
Seismic Interpretation: Applying Rock Physics Tools to Reduce
Interpretation Risk}. Cambridge university press.

\bibitem[\citeproctext]{ref-AYANI2020103098}
Ayani, Mohit, Dario Grana, and Mingliang Liu. 2020. {``Stochastic
Inversion Method of Time-Lapse Controlled Source Electromagnetic Data
for CO\(_{2}\) Plume Monitoring.''} \emph{International Journal of
Greenhouse Gas Control} 100: 103098.
https://doi.org/\url{https://doi.org/10.1016/j.ijggc.2020.103098}.

\bibitem[\citeproctext]{ref-bao2024nonlinear}
Bao, Feng, Hristo G Chipilski, Siming Liang, Guannan Zhang, and Jeffrey
S Whitaker. 2024. {``Nonlinear Ensemble Filtering with Diffusion Models:
Application to the Surface Quasi-Geostrophic Dynamics.''} \emph{arXiv
Preprint arXiv:2404.00844}.

\bibitem[\citeproctext]{ref-batzolis2021conditional}
Batzolis, Georgios, Jan Stanczuk, Carola-Bibiane Schönlieb, and
Christian Etmann. 2021. {``Conditional Image Generation with Score-Based
Diffusion Models.''} \emph{arXiv Preprint arXiv:2111.13606}.

\bibitem[\citeproctext]{ref-baysal1983reverse}
Baysal, Edip, Dan D Kosloff, and John WC Sherwood. 1983. {``Reverse Time
Migration.''} \emph{Geophysics} 48 (11): 1514--24.

\bibitem[\citeproctext]{ref-bruer2024seismicmonitoringco2plume}
Bruer, Grant, Abhinav Prakash Gahlot, Edmond Chow, and Felix Herrmann.
2024. {``Seismic Monitoring of CO\(_{2}\) Plume Dynamics Using Ensemble
Kalman Filtering.''} \url{https://arxiv.org/abs/2409.05193}.

\bibitem[\citeproctext]{ref-bui2018carbon}
Bui, Mai, Claire S Adjiman, André Bardow, Edward J Anthony, Andy Boston,
Solomon Brown, Paul S Fennell, et al. 2018. {``Carbon Capture and
Storage (CCS): The Way Forward.''} \emph{Energy \& Environmental
Science} 11 (5): 1062--1176.

\bibitem[\citeproctext]{ref-carrassi2018data}
Carrassi, Alberto, Marc Bocquet, Laurent Bertino, and Geir Evensen.
2018. {``Data Assimilation in the Geosciences: An Overview of Methods,
Issues, and Perspectives.''} \emph{Wiley Interdisciplinary Reviews:
Climate Change} 9 (5): e535.

\bibitem[\citeproctext]{ref-clochard2018quadri}
Clochard, Vincent, Bryan C DeVault, David Bowen, Nicolas Delépine, and
Kanokkarn Wangkawong. 2018. {``Quadri-Joint Inversion: Method and
Application to the Big Sky 9C 3D Data Set in Northern Montana.''}
\emph{Interpretation} 6 (4): SN101--18.

\bibitem[\citeproctext]{ref-cranmer2020frontier}
Cranmer, Kyle, Johann Brehmer, and Gilles Louppe. 2020. {``The Frontier
of Simulation-Based Inference.''} \emph{Proceedings of the National
Academy of Sciences} 117 (48): 30055--62.
\url{https://doi.org/10.1073/pnas.1912789117}.

\bibitem[\citeproctext]{ref-doi:10.1080ux2f01621459.2017.1285773}
David M. Blei, Alp Kucukelbir, and Jon D. McAuliffe. 2017.
{``Variational Inference: A Review for Statisticians.''} \emph{Journal
of the American Statistical Association} 112 (518): 859--77.
\url{https://doi.org/10.1080/01621459.2017.1285773}.

\bibitem[\citeproctext]{ref-dinh2016density}
Dinh, Laurent, Jascha Sohl-Dickstein, and Samy Bengio. 2016. {``Density
Estimation Using Real Nvp.''} \emph{arXiv Preprint arXiv:1605.08803}.

\bibitem[\citeproctext]{ref-donoho2023data}
Donoho, David. 2023. {``Data Science at the Singularity.''} \emph{arXiv
Preprint arXiv:2310.00865}.

\bibitem[\citeproctext]{ref-douma2010connection}
Douma, Huub, David Yingst, Ivan Vasconcelos, and Jeroen Tromp. 2010.
{``On the Connection Between Artifact Filtering in Reverse-Time
Migration and Adjoint Tomography.''} \emph{Geophysics} 75 (6): S219--23.

\bibitem[\citeproctext]{ref-BG}
E. Jones, C., J. A. Edgar, J. I. Selvage, and H. Crook. 2012.
{``Building Complex Synthetic Models to Evaluate Acquisition Geometries
and Velocity Inversion Technologies.''} \emph{In 74th EAGE Conference
and Exhibition Incorporating EUROPEC 2012}, cp--293.
https://doi.org/\url{https://doi.org/10.3997/2214-4609.20148575}.

\bibitem[\citeproctext]{ref-erdinc2022AAAIdcc}
Erdinc, Huseyin Tuna, Abhinav Prakash Gahlot, Ziyi Yin, Mathias
Louboutin, and Felix J. Herrmann. 2022. {``De-Risking Carbon Capture and
Sequestration with Explainable CO\(_2\) Leakage Detection in Time-Lapse
Seismic Monitoring Images.''}
\url{https://slim.gatech.edu/Publications/Public/Conferences/AAAI/2022/erdinc2022AAAIdcc/erdinc2022AAAIdcc.pdf}.

\bibitem[\citeproctext]{ref-eti395ccs}
ETI, Strategic UK. 2016. {``CCS Storage Appraisal Project Report: D14
WP5E--Viking a Storage Development Plan, 10113ETIS-Rep-21-03. 2016.''}
\emph{Energy Technologies Institute. P} 395.

\bibitem[\citeproctext]{ref-evensen1994sequential}
Evensen, Geir. 1994. {``Sequential Data Assimilation with a Nonlinear
Quasi-Geostrophic Model Using Monte Carlo Methods to Forecast Error
Statistics.''} \emph{Journal of Geophysical Research: Oceans} 99 (C5):
10143--62.

\bibitem[\citeproctext]{ref-feng2022extremelyweaksupervisioninversion}
Feng, Shihang, Peng Jin, Xitong Zhang, Yinpeng Chen, David Alumbaugh,
Michael Commer, and Youzuo Lin. 2022. {``Extremely Weak Supervision
Inversion of Multi-Physical Properties.''}
\url{https://arxiv.org/abs/2202.01770}.

\bibitem[\citeproctext]{ref-folgoc2021mc}
Folgoc, Loic Le, Vasileios Baltatzis, Sujal Desai, Anand Devaraj, Sam
Ellis, Octavio E Martinez Manzanera, Arjun Nair, Huaqi Qiu, Julia
Schnabel, and Ben Glocker. 2021. {``Is MC Dropout Bayesian?''}
\emph{arXiv Preprint arXiv:2110.04286}.

\bibitem[\citeproctext]{ref-frei2013bridging}
Frei, Marco, and Hans R Künsch. 2013. {``Bridging the Ensemble Kalman
and Particle Filters.''} \emph{Biometrika} 100 (4): 781--800.

\bibitem[\citeproctext]{ref-freifeld2009recent}
Freifeld, Barry M, Thomas M Daley, Susan D Hovorka, Jan Henninges, Jim
Underschultz, and Sandeep Sharma. 2009. {``Recent Advances in Well-Based
Monitoring of CO\(_{2}\) Sequestration.''} \emph{Energy Procedia} 1 (1):
2277--84.

\bibitem[\citeproctext]{ref-gahlot2023NIPSWSifp}
Gahlot, Abhinav Prakash, Huseyin Tuna Erdinc, Rafael Orozco, Ziyi Yin,
and Felix J. Herrmann. 2023. {``Inference of CO\(_{2}\) Flow Patterns
{\textendash} a Feasibility Study.''}
\url{https://doi.org/10.48550/arXiv.2311.00290}.

\bibitem[\citeproctext]{ref-gahlot2024digital}
Gahlot, Abhinav Prakash, Haoyun Li, Ziyi Yin, Rafael Orozco, and Felix J
Herrmann. 2024. {``A Digital Twin for Geological Carbon Storage with
Controlled Injectivity.''} \emph{arXiv Preprint arXiv:2403.19819}.

\bibitem[\citeproctext]{ref-gluyas2019passive}
Gluyas, Jon, Lee Thompson, Dave Allen, Charlotte Benton, Paula Chadwick,
Sam Clark, Joel Klinger, et al. 2019. {``Passive, Continuous Monitoring
of Carbon Dioxide Geostorage Using Muon Tomography.''}
\emph{Philosophical Transactions of the Royal Society A} 377 (2137):
20180059.

\bibitem[\citeproctext]{ref-grady2023model}
Grady, Thomas J, Rishi Khan, Mathias Louboutin, Ziyi Yin, Philipp A
Witte, Ranveer Chandra, Russell J Hewett, and Felix J Herrmann. 2023.
{``Model-Parallel Fourier Neural Operators as Learned Surrogates for
Large-Scale Parametric PDEs.''} \emph{Computers \& Geosciences} 178:
105402.

\bibitem[\citeproctext]{ref-grana2020prediction}
Grana, Dario, Mingliang Liu, and Mohit Ayani. 2020. {``Prediction of
CO\(_{2}\) Saturation Spatial Distribution Using Geostatistical
Inversion of Time-Lapse Geophysical Data.''} \emph{IEEE Transactions on
Geoscience and Remote Sensing} 59 (5): 3846--56.

\bibitem[\citeproctext]{ref-9186368}
---------. 2021. {``Prediction of CO\(_{2}\) Saturation Spatial
Distribution Using Geostatistical Inversion of Time-Lapse Geophysical
Data.''} \emph{IEEE Transactions on Geoscience and Remote Sensing} 59
(5): 3846--56. \url{https://doi.org/10.1109/TGRS.2020.3018910}.

\bibitem[\citeproctext]{ref-guo2017calibration}
Guo, Chuan, Geoff Pleiss, Yu Sun, and Kilian Q Weinberger. 2017. {``On
Calibration of Modern Neural Networks.''} In \emph{International
Conference on Machine Learning}, 1321--30. PMLR.

\bibitem[\citeproctext]{ref-he2011risk}
He, Manchao, Sousa Luis, Sousa Rita, Gomes Ana, Vargas Euripedes Jr, and
Na Zhang. 2011. {``Risk Assessment of CO\(_{2}\) Injection Processes and
Storage in Carboniferous Formations: A Review.''} \emph{Journal of Rock
Mechanics and Geotechnical Engineering} 3 (1): 39--56.

\bibitem[\citeproctext]{ref-herrmann2023president}
Herrmann, Felix J. 2023. {``President's Page: Digital Twins in the Era
of Generative AI.''} \emph{The Leading Edge} 42 (11): 730--32.

\bibitem[\citeproctext]{ref-huang2020towards}
Huang, Chao, and Tieyuan Zhu. 2020. {``Towards Real-Time Monitoring:
Data Assimilated Time-Lapse Full Waveform Inversion for Seismic Velocity
and Uncertainty Estimation.''} \emph{Geophysical Journal International}
223 (2): 811--24.

\bibitem[\citeproctext]{ref-huang2024diffusionbasedsubsurfacemultiphysicsmonitoring}
Huang, Xinquan, Fu Wang, and Tariq Alkhalifah. 2024. {``Diffusion-Based
Subsurface Multiphysics Monitoring and Forecasting.''}
\url{https://arxiv.org/abs/2407.18426}.

\bibitem[\citeproctext]{ref-metz2005ipcc}
IPCC special report. 2005. \emph{IPCC Special Report on Carbon Dioxide
Capture and Storage}. Cambridge: Cambridge University Press.

\bibitem[\citeproctext]{ref-masson2018global}
---------. 2018. {``Global Warming of 1.5 c.''} \emph{An IPCC Special
Report on the Impacts of Global Warming of} 1 (5): 43--50.

\bibitem[\citeproctext]{ref-jordan1998introduction}
Jordan, Michael I, Zoubin Ghahramani, Tommi S Jaakkola, and Lawrence K
Saul. 1998. {``An Introduction to Variational Methods for Graphical
Models.''} \emph{Learning in Graphical Models}, 105--61.

\bibitem[\citeproctext]{ref-jordan1999introduction}
---------. 1999. {``An Introduction to Variational Methods for Graphical
Models.''} \emph{Machine Learning} 37 (2): 183--233.

\bibitem[\citeproctext]{ref-Ketzer2012}
Ketzer, J. Marcelo, Rodrigo S. Iglesias, and Sandra Einloft. 2012.
{``Reducing Greenhouse Gas Emissions with CO\(_{2}\) Capture and
Geological Storage.''} In \emph{Handbook of Climate Change Mitigation},
1405--40. New York, NY: Springer US.
\url{https://doi.org/10.1007/978-1-4419-7991-9_37}.

\bibitem[\citeproctext]{ref-Kingma2014AdamAM}
Kingma, Diederik P., and Jimmy Ba. 2014. {``Adam: A Method for
Stochastic Optimization.''} \emph{CoRR} abs/1412.6980.
\url{https://api.semanticscholar.org/CorpusID:6628106}.

\bibitem[\citeproctext]{ref-kingma2018glow}
Kingma, Durk P, and Prafulla Dhariwal. 2018. {``Glow: Generative Flow
with Invertible 1x1 Convolutions.''} In \emph{Advances in Neural
Information Processing Systems}, edited by S. Bengio, H. Wallach, H.
Larochelle, K. Grauman, N. Cesa-Bianchi, and R. Garnett. Vol. 31. Curran
Associates, Inc.
\url{https://proceedings.neurips.cc/paper_files/paper/2018/file/d139db6a236200b21cc7f752979132d0-Paper.pdf}.

\bibitem[\citeproctext]{ref-kolster2018impact}
Kolster, Clea, Simeon Agada, Niall Mac Dowell, and Samuel Krevor. 2018.
{``The Impact of Time-Varying {CO\(_2\)} Injection Rate on Large Scale
Storage in the {UK} Bunter Sandstone.''} \emph{International Journal of
Greenhouse Gas Control} 68: 77--85.

\bibitem[\citeproctext]{ref-krogstad2015mrst}
Krogstad, Stein, Knut--Andreas Lie, Olav Møyner, Halvor Møll Nilsen,
Xavier Raynaud, and Bård Skaflestad. 2015. {``MRST-AD--an Open-Source
Framework for Rapid Prototyping and Evaluation of Reservoir Simulation
Problems.''} In \emph{SPE Reservoir Simulation Conference?},
D022S002R004. SPE.

\bibitem[\citeproctext]{ref-kruse2021hint}
Kruse, Jakob, Gianluca Detommaso, Ullrich Köthe, and Robert Scheichl.
2021a. {``HINT: Hierarchical Invertible Neural Transport for Density
Estimation and Bayesian Inference.''} In \emph{Proceedings of the AAAI
Conference on Artificial Intelligence}, 35:8191--99. 9.

\bibitem[\citeproctext]{ref-kruse2021hintb}
---------. 2021b. {``HINT: Hierarchical Invertible Neural Transport for
Density Estimation and Bayesian Inference.''}
\url{https://arxiv.org/abs/1905.10687}.

\bibitem[\citeproctext]{ref-kullback1951information}
Kullback, Solomon, and Richard A Leibler. 1951. {``On Information and
Sufficiency.''} \emph{The Annals of Mathematical Statistics} 22 (1):
79--86.

\bibitem[\citeproctext]{ref-laves2020well}
Laves, Max-Heinrich, Sontje Ihler, Jacob F Fast, Lüder A Kahrs, and
Tobias Ortmaier. 2020. {``Well-Calibrated Regression Uncertainty in
Medical Imaging with Deep Learning.''} In \emph{Medical Imaging with
Deep Learning}, 393--412. PMLR.

\bibitem[\citeproctext]{ref-https:ux2fux2fdoi.orgux2f10.1029ux2f2019WR027032}
Li, Dongzhuo, Kailai Xu, Jerry M. Harris, and Eric Darve. 2020.
{``Coupled Time-Lapse Full-Waveform Inversion for Subsurface Flow
Problems Using Intrusive Automatic Differentiation.''} \emph{Water
Resources Research} 56 (8): e2019WR027032.
https://doi.org/\url{https://doi.org/10.1029/2019WR027032}.

\bibitem[\citeproctext]{ref-lopez2023spectral}
López, Oscar, Rajiv Kumar, Nick Moldoveanu, and Felix J Herrmann. 2023.
{``Spectral Gap-Based Seismic Survey Design.''} \emph{IEEE Transactions
on Geoscience and Remote Sensing} 61: 1--9.

\bibitem[\citeproctext]{ref-louboutin2018dae}
Louboutin, Mathias, Fabio Luporini, Michael Lange, Navjot Kukreja,
Philipp A. Witte, Felix J. Herrmann, Paulius Velesko, and Gerard J.
Gorman. 2019. {``Devito (V3.1.0): An Embedded Domain-Specific Language
for Finite Differences and Geophysical Exploration.''}
\emph{Geoscientific Model Development}.
\url{https://doi.org/10.5194/gmd-12-1165-2019}.

\bibitem[\citeproctext]{ref-JUDI}
Louboutin, Mathias, Philipp Witte, Ziyi Yin, Henryk Modzelewski, Kerim,
Carlos da Costa, and Peterson Nogueira. 2023. {``Slimgroup/JUDI.jl:
V3.2.3.''} Zenodo. \url{https://doi.org/10.5281/zenodo.7785440}.

\bibitem[\citeproctext]{ref-louboutin2023lmi}
Louboutin, Mathias, Ziyi Yin, Rafael Orozco, Thomas J. Grady II, Ali
Siahkoohi, Gabrio Rizzuti, Philipp A. Witte, Olav Møyner, Gerard J.
Gorman, and Felix J. Herrmann. 2023. {``Learned Multiphysics Inversion
with Differentiable Programming and Machine Learning.''} \emph{The
Leading Edge} 42 (July): 452--516.
\url{https://library.seg.org/doi/10.1190/tle42070474.1}.

\bibitem[\citeproctext]{ref-lueckmann2017flexible}
Lueckmann, Jan-Matthis, Pedro J Goncalves, Giacomo Bassetto, Kaan Öcal,
Marcel Nonnenmacher, and Jakob H Macke. 2017. {``Flexible Statistical
Inference for Mechanistic Models of Neural Dynamics.''} \emph{Advances
in Neural Information Processing Systems} 30.

\bibitem[\citeproctext]{ref-lumley20104d}
Lumley, David. 2010. {``4D Seismic Monitoring of CO 2 Sequestration.''}
\emph{The Leading Edge} 29 (2): 150--55.

\bibitem[\citeproctext]{ref-doi:10.1190ux2f1.1444921}
Lumley, David E. 2001. {``Time-Lapse Seismic Reservoir Monitoring.''}
\emph{GEOPHYSICS} 66 (1): 50--53.
\url{https://doi.org/10.1190/1.1444921}.

\bibitem[\citeproctext]{ref-luporini2020architecture}
Luporini, Fabio, Mathias Louboutin, Michael Lange, Navjot Kukreja,
Philipp Witte, Jan Hückelheim, Charles Yount, Paul HJ Kelly, Felix J
Herrmann, and Gerard J Gorman. 2020. {``Architecture and Performance of
Devito, a System for Automated Stencil Computation.''} \emph{ACM
Transactions on Mathematical Software (TOMS)} 46 (1): 1--28.

\bibitem[\citeproctext]{ref-mcgoff2015statistical}
McGoff, Kevin, Sayan Mukherjee, and Natesh Pillai. 2015. {``Statistical
Inference for Dynamical Systems: A Review.''}

\bibitem[\citeproctext]{ref-metz2005carbon}
Metz, Bert et al. 2005. \emph{Carbon Dioxide Capture and Storage:
Special Report of the Intergovernmental Panel on Climate Change}.
Cambridge University Press.

\bibitem[\citeproctext]{ref-muxf8yner2023}
Møyner, Olav, and Grant Bruer. 2023. \emph{Sintefmath/JutulDarcy.jl:
V0.2.1}. Zenodo. \url{https://doi.org/10.5281/ZENODO.7775736}.

\bibitem[\citeproctext]{ref-moyner2024nonlinear}
Møyner, Olav, Atgeirr F Rasmussen, Øystein Klemetsdal, Halvor M Nilsen,
Arthur Moncorgé, and Knut-Andreas Lie. 2024. {``Nonlinear
Domain-Decomposition Preconditioning for Robust and Efficient
Field-Scale Simulation of Subsurface Flow.''} \emph{Computational
Geosciences} 28 (2): 241--51.

\bibitem[\citeproctext]{ref-nordbotten2011geological}
Nordbotten, Jan M, and Michael Anthony Celia. 2011. {``Geological
Storage of CO\(_{2}\): Modeling Approaches for Large-Scale
Simulation.''} In \emph{Geological Storage of CO 2: Modeling Approaches
for Large-Scale Simulation}. John Wiley; Sons.

\bibitem[\citeproctext]{ref-oghenekohwo2017low}
Oghenekohwo, Felix, Haneet Wason, Ernie Esser, and Felix J Herrmann.
2017. {``Low-Cost Time-Lapse Seismic with Distributed Compressive
Sensing---Part 1: Exploiting Common Information Among the Vintages.''}
\emph{Geophysics} 82 (3): P1--13.

\bibitem[\citeproctext]{ref-orozco2024probabilistic}
Orozco, Rafael, Felix J Herrmann, and Peng Chen. 2024. {``Probabilistic
Bayesian Optimal Experimental Design Using Conditional Normalizing
Flows.''} \emph{arXiv Preprint arXiv:2402.18337}.

\bibitem[\citeproctext]{ref-orozco2024aspireiterativeamortizedposterior}
Orozco, Rafael, Ali Siahkoohi, Mathias Louboutin, and Felix J. Herrmann.
2024. {``ASPIRE: Iterative Amortized Posterior Inference for Bayesian
Inverse Problems.''} \url{https://arxiv.org/abs/2405.05398}.

\bibitem[\citeproctext]{ref-orozco2023adjoint}
Orozco, Rafael, Ali Siahkoohi, Gabrio Rizzuti, Tristan van Leeuwen, and
Felix J Herrmann. 2023. {``Adjoint Operators Enable Fast and Amortized
Machine Learning Based Bayesian Uncertainty Quantification.''} In
\emph{Medical Imaging 2023: Image Processing}, 12464:357--67. SPIE.

\bibitem[\citeproctext]{ref-orozco2023invertiblenetworks}
Orozco, Rafael, Philipp Witte, Mathias Louboutin, Ali Siahkoohi, Gabrio
Rizzuti, Bas Peters, and Felix J. Herrmann. 2024b.
{``InvertibleNetworks.jl: A Julia Package for Scalable Normalizing
Flows.''} \emph{Journal of Open Source Software} 9 (99): 6554.
\url{https://doi.org/10.21105/joss.06554}.

\bibitem[\citeproctext]{ref-Orozco2024}
---------. 2024a. {``InvertibleNetworks.jl: A Julia Package for Scalable
Normalizing Flows.''} \emph{Journal of Open Source Software} 9 (99):
6554. \url{https://doi.org/10.21105/joss.06554}.

\bibitem[\citeproctext]{ref-orr2009onshore}
Orr Jr, Franklin M. 2009. {``Onshore Geologic Storage of CO\(_{2}\).''}
\emph{Science} 325 (5948): 1656--58.

\bibitem[\citeproctext]{ref-page2020global}
Page, Brad, Guloren Turan, Alex Zapantis, Jamie Burrows, Chris Consoli,
Jeff Erikson, Ian Havercroft, et al. 2020. {``The Global Status of CCS
2020: Vital to Achieve Net Zero.''}

\bibitem[\citeproctext]{ref-ipcc2018global}
Panel on Climate Change), IPCC (Intergovernmental. 2018. \emph{Global
Warming of 1.5° c. An IPCC Special Report on the Impacts of Global
Warming of 1.5° c Above Pre-Industrial Levels and Related Global
Greenhouse Gas Emission Pathways, in the Context of Strengthening the
Global Response to the Threat of Climate Change, Sustainable
Development, and Efforts to Eradicate Poverty}. ipcc Geneva.

\bibitem[\citeproctext]{ref-papamakarios2016fast}
Papamakarios, George, and Iain Murray. 2016. {``Fast
\(\varepsilon\)-Free Inference of Simulation Models with Bayesian
Conditional Density Estimation.''} \emph{Advances in Neural Information
Processing Systems} 29.

\bibitem[\citeproctext]{ref-nf}
Papamakarios, George, Eric Nalisnick, Danilo Jimenez Rezende, Shakir
Mohamed, and Balaji Lakshminarayanan. 2021. {``Normalizing Flows for
Probabilistic Modeling and Inference.''} \emph{J. Mach. Learn. Res.} 22
(1).

\bibitem[\citeproctext]{ref-pruess2011}
Pruess, Karsten, and Jan Nordbotten. 2011. {``Numerical Simulation
Studies of the Long-Term Evolution of a CO\(_{2}\) Plume in a Saline
Aquifer with a Sloping Caprock.''} \emph{Transport in Porous Media} 90
(1): 135--51. \url{https://doi.org/10.1007/s11242-011-9729-6}.

\bibitem[\citeproctext]{ref-qu2024uncertaintyquantificationseismicinversion}
Qu, Luping, Mauricio Araya-Polo, and Laurent Demanet. 2024.
{``Uncertainty Quantification in Seismic Inversion Through Integrated
Importance Sampling and Ensemble Methods.''}
\url{https://arxiv.org/abs/2409.06840}.

\bibitem[\citeproctext]{ref-radev2020bayesflow}
Radev, Stefan T, Ulf K Mertens, Andreas Voss, Lynton Ardizzone, and
Ullrich Köthe. 2020. {``BayesFlow: Learning Complex Stochastic Models
with Invertible Neural Networks.''} \emph{IEEE Transactions on Neural
Networks and Learning Systems} 33 (4): 1452--66.

\bibitem[\citeproctext]{ref-ramgraber2023ensembleb}
Ramgraber, Maximilian, Ricardo Baptista, Dennis McLaughlin, and Youssef
Marzouk. 2023. {``Ensemble Transport Smoothing. Part II: Nonlinear
Updates.''} \emph{Journal of Computational Physics: X} 17: 100133.

\bibitem[\citeproctext]{ref-rasmussen2021open}
Rasmussen, Atgeirr Flø, Tor Harald Sandve, Kai Bao, Andreas Lauser,
Joakim Hove, Bård Skaflestad, Robert Klöfkorn, et al. 2021. {``The Open
Porous Media Flow Reservoir Simulator.''} \emph{Computers \& Mathematics
with Applications} 81: 159--85.
https://doi.org/\url{https://doi.org/10.1016/j.camwa.2020.05.014}.

\bibitem[\citeproctext]{ref-raza2016integrity}
Raza, Arshad, Raoof Gholami, Mohammad Sarmadivaleh, Nathan Tarom, Reza
Rezaee, Chua Han Bing, Ramasamy Nagarajan, Mohamed Ali Hamid, and Henry
Elochukwu. 2016. {``Integrity Analysis of CO\(_{2}\) Storage Sites
Concerning Geochemical-Geomechanical Interactions in Saline Aquifers.''}
\emph{Journal of Natural Gas Science and Engineering} 36: 224--40.

\bibitem[\citeproctext]{ref-rezende2016variational}
Rezende, Danilo, and Shakir Mohamed. 2015. {``Variational Inference with
Normalizing Flows.''} In \emph{International Conference on Machine
Learning}, 1530--38. PMLR.

\bibitem[\citeproctext]{ref-ringrose2020store}
Ringrose, Philip. 2020. \emph{How to Store CO\(_{2}\) Underground:
Insights from Early-Mover CCS Projects}. Vol. 129. Springer.

\bibitem[\citeproctext]{ref-ringrose2023storage}
---------. 2023. \emph{Storage of Carbon Dioxide in Saline Aquifers:
Building Confidence by Forecasting and Monitoring}. Society of
Exploration Geophysicists.

\bibitem[\citeproctext]{ref-ringrose2016reservoir}
Ringrose, Philip, and Mark Bentley. 2016. \emph{Reservoir Model Design}.
Vol. 2. Springer.

\bibitem[\citeproctext]{ref-ronneberger2015u}
Ronneberger, Olaf, Philipp Fischer, and Thomas Brox. 2015. {``U-Net:
Convolutional Networks for Biomedical Image Segmentation.''} In
\emph{Medical Image Computing and Computer-Assisted Intervention--MICCAI
2015: 18th International Conference, Munich, Germany, October 5-9, 2015,
Proceedings, Part III 18}, 234--41. Springer.

\bibitem[\citeproctext]{ref-rozet2023score}
Rozet, François, and Gilles Louppe. 2023. {``Score-Based Data
Assimilation.''} \emph{Advances in Neural Information Processing
Systems} 36: 40521--41.

\bibitem[\citeproctext]{ref-rubio2023transport}
Rubio, Paul-Baptiste, Youssef Marzouk, and Matthew Parno. 2023. {``A
Transport Approach to Sequential Simulation-Based Inference.''}
\emph{arXiv Preprint arXiv:2308.13940}.

\bibitem[\citeproctext]{ref-settgast2018geosx}
Settgast, Randolph R, JA White, BC Corbett, A Vargas, C Sherman, P Fu, C
Annavarapu, et al. 2018. {``Geosx Simulation Framework.''} Lawrence
Livermore National Lab.(LLNL), Livermore, CA (United States).

\bibitem[\citeproctext]{ref-si2024latentensflatentensemblescore}
Si, Phillip, and Peng Chen. 2024. {``Latent-EnSF: A Latent Ensemble
Score Filter for High-Dimensional Data Assimilation with Sparse
Observation Data.''} \url{https://arxiv.org/abs/2409.00127}.

\bibitem[\citeproctext]{ref-siahkoohi2020weak}
Siahkoohi, Ali, Gabrio Rizzuti, and Felix J Herrmann. 2020. {``Weak Deep
Priors for Seismic Imaging.''} In \emph{SEG Technical Program Expanded
Abstracts 2020}, 2998--3002. Society of Exploration Geophysicists.

\bibitem[\citeproctext]{ref-siahkoohi2023reliable}
Siahkoohi, Ali, Gabrio Rizzuti, Rafael Orozco, and Felix J Herrmann.
2023a. {``Reliable Amortized Variational Inference with Physics-Based
Latent Distribution Correction.''} \emph{Geophysics} 88 (3): R297--322.

\bibitem[\citeproctext]{ref-siahkoohi2022ravi}
Siahkoohi, Ali, Gabrio Rizzuti, Rafael Orozco, and Felix J. Herrmann.
2023b. {``Reliable Amortized Variational Inference with Physics-Based
Latent Distribution Correction.''} \emph{Geophysics} 88 (3).
\url{https://doi.org/10.1190/geo2022-0472.1}.

\bibitem[\citeproctext]{ref-spantini2022coupling}
Spantini, Alessio, Ricardo Baptista, and Youssef Marzouk. 2022.
{``Coupling Techniques for Nonlinear Ensemble Filtering.''} \emph{SIAM
Review} 64 (4): 921--53.

\bibitem[\citeproctext]{ref-stacey2017validation}
Stacey, Robert W, and Michael J Williams. 2017. {``Validation of ECLIPSE
Reservoir Simulator for Geothermal Problems.''} \emph{GRC Transactions}
41: 2095--2109.

\bibitem[\citeproctext]{ref-stolk2009inverse}
Stolk, CC, MV de Hoop, and TJPM Op't Root. 2009. {``Inverse Scattering
of Seismic Data in the Reverse Time Migration (RTM) Approach:
Proceedings of the Project Review, Geo-Mathematical Imaging Group of
Purdue University, 91--108.''}

\bibitem[\citeproctext]{ref-STRANDLI20144473}
Strandli, Christin W., Edward Mehnert, and Sally M. Benson. 2014.
{``CO\(_{2}\) Plume Tracking and History Matching Using Multilevel
Pressure Monitoring at the Illinois Basin -- Decatur Project.''}
\emph{Energy Procedia} 63: 4473--84.
https://doi.org/\url{https://doi.org/10.1016/j.egypro.2014.11.483}.

\bibitem[\citeproctext]{ref-TANG2021103488}
Tang, Hewei, Pengcheng Fu, Christopher S. Sherman, Jize Zhang, Xin Ju,
François Hamon, Nicholas A. Azzolina, Matthew Burton-Kelly, and Joseph
P. Morris. 2021. {``A Deep Learning-Accelerated Data Assimilation and
Forecasting Workflow for Commercial-Scale Geologic Carbon Storage.''}
\emph{International Journal of Greenhouse Gas Control} 112: 103488.
https://doi.org/\url{https://doi.org/10.1016/j.ijggc.2021.103488}.

\bibitem[\citeproctext]{ref-tatsis2022sequential}
Tatsis, Konstantinos E, Vasilis K Dertimanis, and Eleni N Chatzi. 2022.
{``Sequential Bayesian Inference for Uncertain Nonlinear Dynamic
Systems: A Tutorial.''} \emph{arXiv Preprint arXiv:2201.08180}.

\bibitem[\citeproctext]{ref-um2023real}
Um, Evan Schankee, David Alumbaugh, Youzuo Lin, and Shihang Feng. 2023.
{``Real-Time Deep-Learning Inversion of Seismic Full Waveform Data for
CO\(_{2}\) Saturation and Uncertainty in Geological Carbon Storage
Monitoring.''} \emph{Geophysical Prospecting} 72 (Machine learning
applications in geophysical exploration and monitoring): 199--212.

\bibitem[\citeproctext]{ref-virieux2009overview}
Virieux, Jean, and Stéphane Operto. 2009. {``An Overview of
Full-Waveform Inversion in Exploration Geophysics.''} \emph{Geophysics}
74 (6): WCC1--26.

\bibitem[\citeproctext]{ref-wangconsis2024}
Wang, Pengfei, Jidong Yang, Jianping Huang, Jiaxing Sun, and Chong Zhao.
2024. {``{Consistency of the inverse scattering imaging condition, the
energy norm imaging condition and the impedance kernel in acoustic and
elastic reverse-time migration}.''} \emph{Journal of Geophysics and
Engineering} 21 (2): 614--33. \url{https://doi.org/10.1093/jge/gxae022}.

\bibitem[\citeproctext]{ref-1284395}
Wang, Zhou, A. C. Bovik, H. R. Sheikh, and E. P. Simoncelli. 2004a.
{``Image Quality Assessment: From Error Visibility to Structural
Similarity.''} \emph{IEEE Transactions on Image Processing} 13 (4):
600--612. \url{https://doi.org/10.1109/TIP.2003.819861}.

\bibitem[\citeproctext]{ref-wang2004image}
Wang, Zhou, Alan C Bovik, Hamid R Sheikh, and Eero P Simoncelli. 2004b.
{``Image Quality Assessment: From Error Visibility to Structural
Similarity.''} \emph{IEEE Transactions on Image Processing} 13 (4):
600--612.

\bibitem[\citeproctext]{ref-wason2017low}
Wason, Haneet, Felix Oghenekohwo, and Felix J Herrmann. 2017.
{``Low-Cost Time-Lapse Seismic with Distributed Compressive
Sensing---Part 2: Impact on Repeatability.''} \emph{Geophysics} 82 (3):
P15--30.

\bibitem[\citeproctext]{ref-whitmore2012applications}
Whitmore, ND, and Sean Crawley. 2012. {``Applications of RTM Inverse
Scattering Imaging Conditions.''} In \emph{SEG International Exposition
and Annual Meeting}, SEG--2012. SEG.

\bibitem[\citeproctext]{ref-witte2018alf}
Witte, Philipp A., Mathias Louboutin, Navjot Kukreja, Fabio Luporini,
Michael Lange, Gerard J. Gorman, and Felix J. Herrmann. 2019. {``A
Large-Scale Framework for Symbolic Implementations of Seismic Inversion
Algorithms in Julia.''} \emph{Geophysics} 84 (3): F57--71.
\url{https://doi.org/10.1190/geo2018-0174.1}.

\bibitem[\citeproctext]{ref-InvertibleNetworks}
Witte, Philipp, Mathias Louboutin, Rafael Orozco, grizzuti, Ali
Siahkoohi, Felix Herrmann, Bas Peters, Páll Haraldsson, and Ziyi Yin.
2023. {``Slimgroup/InvertibleNetworks.jl: V2.2.5.''} Zenodo.
\url{https://doi.org/10.5281/zenodo.7850287}.

\bibitem[\citeproctext]{ref-yin2022TLEdgc}
Yin, Ziyi, Huseyin Tuna Erdinc, Abhinav Prakash Gahlot, Mathias
Louboutin, and Felix J. Herrmann. 2023. {``Derisking Geological Carbon
Storage from High-Resolution Time-Lapse Seismic to Explainable Leakage
Detection.''} \emph{The Leading Edge} 42 (1): 6976.
\url{https://doi.org/10.1190/tle42010069.1}.

\bibitem[\citeproctext]{ref-yin2024time}
Yin, Ziyi, Mathias Louboutin, Olav Møyner, and Felix J Herrmann. 2024.
{``Time-Lapse Full-Waveform Permeability Inversion: A Feasibility
Study.''} \emph{The Leading Edge} 43 (8): 544--53.

\bibitem[\citeproctext]{ref-yin2024wiser}
Yin, Ziyi, Rafael Orozco, and Felix J Herrmann. 2024. {``WISER:
Multimodal Variational Inference for Full-Waveform Inversion Without
Dimensionality Reduction.''} \emph{arXiv Preprint arXiv:2405.10327}.

\bibitem[\citeproctext]{ref-yin2023solving}
Yin, Ziyi, Rafael Orozco, Mathias Louboutin, and Felix J Herrmann. 2023.
{``Solving Multiphysics-Based Inverse Problems with Learned Surrogates
and Constraints.''} \emph{Advanced Modeling and Simulation in
Engineering Sciences} 10 (1): 14.

\bibitem[\citeproctext]{ref-yin2024wise}
---------. 2024. {``WISE: Full-Waveform Variational Inference via
Subsurface Extensions.''} \emph{Geophysics} 89 (4): 1--31.

\bibitem[\citeproctext]{ref-ZHANG20143000}
Zhang, Guanru, Peng Lu, and Chen Zhu. 2014. {``Model Predictions via
History Matching of CO\(_{2}\) Plume Migration at the Sleipner Project,
Norwegian North Sea.''} \emph{Energy Procedia} 63: 3000--3011.
https://doi.org/\url{https://doi.org/10.1016/j.egypro.2014.11.323}.

\bibitem[\citeproctext]{ref-zhang2023optimized}
Zhang, Yijun, Ziyi Yin, Oscar López, Ali Siahkoohi, Mathias Louboutin,
Rajiv Kumar, and Felix J Herrmann. 2023. {``Optimized Time-Lapse
Acquisition Design via Spectral Gap Ratio Minimization.''}
\emph{Geophysics} 88 (4): A19--23.

\bibitem[\citeproctext]{ref-zoback2012earthquake}
Zoback, Mark D, and Steven M Gorelick. 2012. {``Earthquake Triggering
and Large-Scale Geologic Storage of Carbon Dioxide.''} \emph{Proceedings
of the National Academy of Sciences} 109 (26): 10164--68.

\end{CSLReferences}
\end{document}